\documentclass[10pt,leqno]{article}
\usepackage{times}
\usepackage{mathrsfs}
\usepackage{latexsym,amsopn,amsmath,amssymb}
\usepackage{amsfonts,amsbsy,amscd,stmaryrd}
\long\def\symbolfootnote[#1]#2{\begingroup%
    \def\thefootnote{\fnsymbol{footnote}}\footnote[#1]{#2}\endgroup}
\def\thesection{\arabic{section}}
\renewcommand{\theequation}{\arabic{section}.\arabic{equation}}

\newcounter{PAR}[section]
\def\thePAR{\thesection.\arabic{PAR}}
\def\PAR{\par\addvspace{2 ex} \noindent
\refstepcounter{PAR}{\bf \thePAR.} 
}

\newlength{\leftcolnumber}     
\newlength{\rightcoltext}      

\protect\newlength{\thname}
\font\tenmsb=msbm10 at 11pt
\font\sevenmsb=msbm7 at 8pt
\font\fivemsb=msbm5 at 6pt
\newfam\msbfam
\textfont\msbfam=\tenmsb
\scriptfont\msbfam=\sevenmsb
\scriptscriptfont\msbfam=\fivemsb

\def\CC{\mathbb C}

\def\ZZ{\mathbb Z}

\newcommand{\mbc}{\mathbb{C}}

\newcommand{\mbf}{\mathbb{F}}
\newcommand{\mbg}{\mathbb{G}}
\newcommand{\mbe}{\mathbb{E}}

\newcommand{\mbp}{\mathbb{P}}
\newcommand{\mbr}{\mathbb{R}}

\newcommand{\mbz}{\mathbb{Z}}

\newcommand{\bfe}{{\bf E}}
\newcommand{\ca}{\mathcal{A}}
\newcommand{\cb}{\mathcal{B}}
\newcommand{\cc}{\mathcal{C}}

\newcommand{\cf}{\mathcal{F}}
\newcommand{\cg}{\mathcal{G}}
\newcommand{\ch}{\mathcal{H}}

\newcommand{\cj}{\mathcal{J}}
\newcommand{\ck}{\mathcal{K}}
\newcommand{\cl}{\mathcal{L}}
\newcommand{\cm}{\mathcal{M}}

\newcommand{\cp}{\mathcal{P}}

\newcommand{\cs}{\mathcal{S}}
\newcommand{\cv}{\mathcal{V}}
\def\rond{\mathscr}
\newcommand{\ra}{\rond{A}}
\newcommand{\rb}{\rond{B}}
\newcommand{\rc}{\rond{C}}

\newcommand{\rg}{\rond{G}}

\newcommand{\ri}{\rond{I}}
\newcommand{\rj}{\rond{J}}
\newcommand{\rk}{\rond{K}}

\newcommand{\rt}{\rond{T}}
\newcommand{\rv}{\rond{V}}

\newcommand{\gof}{\goth f}
\newcommand{\gog}{\goth g}

\newcommand{\ind}{\mbox{\bf 1}}
\font\elevencmb=cmb10 at 11pt
\font\eightcmb=cmb10 at 8pt
\font\sixcmb=cmb10 at 6pt
\newfam\cmbfam
\textfont\cmbfam=\elevencmb
\scriptfont\cmbfam=\eightcmb
\scriptscriptfont\cmbfam=\sixcmb

\font\teneuf=eufm10 at 12pt
\font\seveneuf=eufm7 at 8pt
\font\fiveeuf=eufm5 at 6pt
\newfam\euffam
\textfont\euffam=\teneuf
\scriptfont\euffam=\seveneuf
\scriptscriptfont\euffam=\fiveeuf
\def\goth#1{{\teneuf\fam\euffam#1}}
\newfont{\secgoth}{eufm10 at 16pt}
\font\tenrsf=rsfs10 at 10pt
\font\sevenrsf=rsfs7 at 8pt
\font\fiversf=rsfs5 at 6pt
\newfam\rsffam
\textfont\rsffam=\tenrsf
\scriptfont\rsffam=\sevenrsf
\scriptscriptfont\rsffam=\fiversf
\def\rond#1{{\tenrsf\fam\rsffam#1}}
\newfont{\secrond}{rsfs10 at 14pt}

\font\tenmsa=msam10 at 11pt
\font\sevenmsa=msam8
\font\fivemsa=msam6
\newfam\msafam
\textfont\msafam=\tenmsa
\scriptfont\msafam=\sevenmsa
\scriptscriptfont\msafam=\fivemsa

\newlength{\rightcolbox}
\newlength{\leftcoltext} 
\def\ve{\varepsilon}
\def\vk{\varkappa}
\def\vphi{\varphi}
\def\vf{\varphi}

\def\cchi{\raisebox{.40 ex}{$\chi$}} 
\def\[[{[\hspace{-.5mm}[}
\def\]]{]\hspace{-.5mm}]}
\def\Left[[{\left[\hspace{-1mm}\left[}
\def\Right]]{\right]\hspace{-1mm}\right]}
\def\qed{\hfill \vrule width 8pt height 9pt depth-1pt \par\medskip}

\def\norm#1{\left\Vert #1\right\Vert }
\def\nnorm#1{|{\!|}{\!|} #1 |{\!|}{\!|}}

\def\proof{\par\addvspace{1 ex}\noindent \hbox{\bf Proof: }}

\newcommand{\rarrow}{\rightarrow}

\def\braket#1{\langle #1 \rangle}
\protect

\def\nin{\notin}
\def\wtilde#1{\widetilde{#1\,}}

\def\what#1{\widehat{ #1\,}}
\def\se{\sigma_{\rm ess}}
\def\pprod{\textstyle \prod}

\def\ccup{\raisebox{.25 ex}{$\textstyle\bigcup$}}
\def\ccap{\raisebox{.45 ex}{$\textstyle\bigcap$}}

\newcommand{\e}{\mbox{e}}

\protect
\protect
\protect
\protect
\protect
\protect
\protect
\protect
\protect
\protect
\protect
\protect
\protect

\def\Slim#1{\mbox{\rm s\,-}\!\lim_{#1}}

\protect\newcommand{\supp}{\mbox{\rm supp}\,}
\protect
\newcommand{\be}{\begin{equation}}
\newcommand{\ee}{\end{equation}}
\newcommand{\bea}{\begin{eqnarray}}
\newcommand{\eea}{\end{eqnarray}}

\def\@begintheorem#1#2{\it \trivlist \item[\hskip
\labelsep{\bf #1\ #2}]}
\def\@endtheorem{\endtrivlist}
\newtheorem{theorem}{Theorem}[section]
\newtheorem{lemma}[theorem]{Lemma}
\newtheorem{proposition}[theorem]{Proposition}
\newtheorem{corollary}[theorem]{Corollary}

\newtheorem{definition}[theorem]{Definition}
\newtheorem{remark}[theorem]{Remark}
\newtheorem{remarks}[theorem]{Remarks}
\newtheorem{example}[theorem]{Example}
\newtheorem{examples}[theorem]{Examples}
\renewcommand{\thefootnote}{(\alph{footnote})}
\def\co{{\mathcal C}_{0}}

\def\Cc{{\mathcal C}_{\mbox{\rm \scriptsize c}}}

\def\Cu0#1{{\mathcal C}_{\mbox{\rm \scriptsize u}}^0({#1})}
\newcommand{\dd}{\,\mbox{\rm d}}

\begin{document}
\title{ Localizations at infinity and essential spectrum of quantum
Hamiltonians: I. General theory}
\author{by \medskip \\ Vladimir GEORGESCU\thanks{CNRS (UMR 8088) and
    $^\star$Department of Mathematics, University of Cergy-Pontoise,
    2 avenue Adolphe Chauvin, 95302 Cergy-Pontoise Cedex, France;
    e-mail: \texttt{Vladimir.Georgescu@math.u-cergy.fr {\rm and}
      Andrei.Iftimovici@math.u-cergy.fr} } $^\star$ and Andrei
  IFTIMOVICI$^\star$}
\date{}
\maketitle
\vspace{-0.5cm}
\begin{abstract}\noindent
We isolate a large class of self-adjoint operators $H$
whose essential spectrum is determined by their behavior at
$x\sim\infty$ and we give a canonical representation of $\se(H)$ in
terms of spectra of limits at infinity of translations of $H$.
\end{abstract} 
\tableofcontents
\vspace{0.5cm}
\pagebreak
%
\section{Introduction}             \label{s:1}
\protect\setcounter{equation}{0}
In this paper we continue the investigation of the spectral
properties of quantum Hamiltonians with $C^*$-algebra methods on the
lines of our previous work \cite{GI0}. More precisely, our aim is
to study the essential spectrum of general classes of (unbounded)
operators in $L^2(X)$, where $X$ is a locally compact non-compact
abelian group, by using crossed product techniques.
For some historical remarks and comparison with other recently
obtained results, see Subsections \ref{ss:Ie} and \ref{ss:In}.

\PAR                      \label{ss:I1}
We set $\rb(X)=B(L^2(X))$ and we denote by $U_x$ the operator of
translation by $x\in X$ and by $V_k$ the operator of multiplication
by the character $k\in X^*$ (our notations, although rather
standard, are summarized in Section \ref{s:p}). We
define\,\symbolfootnote[2]{\ We make the following convention: if a
symbol like $T^{(*)}$ appears in a relation, then the relation must
hold for the operator $T$ and for its adjoint $T^*$.}
\begin{equation}\label{eq:01} 
\rc(X)=\{T\in \rb(X)\mid \lim_{ k\rarrow 0}\|[T, V_k]\|=0
\mbox{\, \rm and }
\lim_{ x\rarrow 0}\|(U_x - 1) T^{(*)}\|=0\}
\end{equation}
which is clearly a $C^*$-algebra of operators on $L^2(X)$ (without
unit if $X$ is not discrete).  Besides the norm topology on $\rc(X)$
we shall also consider on it the topology defined by the family of
seminorms $\|S\|_\theta=\|S\theta(Q)\|+\|\theta(Q)S\|$ with
$\theta\in\co(X)$ and we shall denote $\rc_s(X)$ the corresponding
topological space (see Remark \ref{re:simon}). Here $\theta(Q)$ is
the operator of multiplication by $\theta$ in $L^2(X)$.

Our main result is a description of the essential spectrum of the
operators $T\in\rc$ in terms of their ``localizations at infinity''.
We denote by $\delta X$ the set of all ultrafilters on $X$ finer than
the Fr\'echet filter (cf.\ page \pageref{fr}). If $A_i$ are subsets
of a topological space we denote $\overline\cup_{i\in I}A_i$  the
closure of their union.
\begin{theorem}\label{th:intro}
If $T\in\rc(X)$ is a normal operator, then for each $\vk\in\delta X$
the limit $\lim_{x\rarrow\vk}U_xTU_x^*=\vk.T$ exists in $\rc_s(X)$
and
\begin{equation}\label{eq:intro2}
\se(T)=\overline\ccup_{\vk}\sigma(\vk.T).
\end{equation}
\end{theorem}
Note that $x\rarrow\vk$ should be read ``$x$ tends to infinity along
the filter $\vk$''.  The limit operator $\vk.T$ will be called
\emph{localization at $\vk$ of $T$}.  Since an ultrafilter finer
than the Fr\'echet filter can be thought as a point on an ideal
boundary at infinity of $X$, the operators $\vk.T$ will also be
called \emph{localizations at infinity of $T$}.

\medskip

We are mainly interested in the essential spectrum of unbounded
self-adjoint operators $H$ ``affiliated'' to $\rc(X)$, but the
corresponding result is an immediate consequence of Theorem
\ref{th:intro}. We say that $H$ is affiliated to some $C^*$-algebra
$\ra$ of operators on $L^2(X)$ if $\vf(H)\in\ra$ for all
$\vf\in\co(\mbr)$ (for this it suffices to have $(H-z)^{-1}\in\ra$
for one $z\in\rho(H)$).  For technical reasons we have to consider
self-adjoint operators which are not necessarily densely defined
and, in order to avoid confusions with the standard terminology, we
shall call these more general objects \emph{observables}. A more
detailed presentation of this notion can be found in Subsection
\ref{ss:pd1}. For the moment we note only that an observable $H$ is
affiliated to $\rc(X)$ if and only if
\begin{equation}\label{eq:intro1}
\lim_{k\rarrow0}\|[V_k,(H-z)^{-1}]\|=0
\hspace{3mm} \mbox{\rm and} \hspace{3mm}
\lim_{x\rarrow0}\|(U_x-1)(H-z)^{-1}\|=0
\end{equation} 
for some $z\in\rho(H)$.  This follows from the fact that if
$T\in\rb(X)$ is normal, then
$$
\lim_{ x\rarrow 0}\|(U_x - 1) T\|=0 \Longrightarrow
\lim_{ x\rarrow 0}\|(U_x - 1) T^*\|=0.
$$

\begin{theorem}\label{th:introo}
  Let $H$ be an observable on $L^2(X)$ affiliated to $\rc(X)$.  Then
  for each $\vk\in\delta X$ the limit $\vk.H:=\lim_{x\rarrow\vk}x.H$
  exists in the following sense: there is an observable $\vk.H$
  affiliated to $\rc(X)$ such that
  $\lim_{x\rarrow\vk}U_x\vf(H)U_x^*=\vf(\vk.H)$ in $\rc_s(X)$ for
  all $\vf\in\co(\mbr)$. Moreover, we have
\begin{equation}\label{eq:introo}
\se(H)=\overline\ccup_{\vk}\sigma(\vk.H).
\end{equation}
\end{theorem}
Practically we are interested only in the case when $H$ is a
self-adjoint operator in the standard sense. However, even in this
case $\vk.H$ could be not densely defined and quite often we have
$\vk.H=\infty$ (i.e.\ the domain of $\vk.H$ is $\{0\}$). For example,
if $H$ has purely discrete spectrum, then $\vf(H)$ is a compact
operator and we clearly get $\vk.H=\infty$ for all $\vk$. Since
$\sigma(\infty)=\emptyset$, we then obtain $\se(H)=\emptyset$, as it
should be.

\begin{remark}\label{re:8}{\rm
    The observable $H$ should be thought as the Hamiltonian (energy
    observable) of a physical system. Thus (\ref{eq:introo}) says
    that the essential spectrum of the Hamiltonian $H$ can be
    computed in terms of the spectra of its localizations at
    infinity $\vk.H$.  We emphasize that this notion of infinity is
    determined by the position observable $Q$. In other terms, if $H$
    satisfies (\ref{eq:intro1}) then $\se(H)$ is given by its
    localizations in the region $Q=\infty$.  This property does not
    hold in many situations of physical interest (e.g.\ if magnetic
    fields which do not vanish at infinity are involved) because
    localizations at infinity with respect to other observables must
    be taken into account, see \cite{GI**}.  }\end{remark}

\begin{remark}\label{re:kk}{\rm
It will be clear from the proof of Theorem \ref{th:introo} (see
Lemma \ref{lm:linf} and Proposition \ref{pr:ker}) that
(\ref{eq:introo}) remains valid if $\vk$ runs over sets much smaller
than $\delta X$: it suffices to take $\vk\in\ck$ if
$\ck\subset\delta X$ has the property: if $\vf$ is a bounded
uniformly continuous function on $X$ and
$\lim_{x\rarrow\vk}\vf(x+y)=0$ for all $y\in X,\vk\in\ck$, then
$\vf\in\co(X)$. 
}\end{remark}

\begin{remark}\label{re:stab}{\rm
We mention the following immediate consequence of (\ref{eq:introo}):
\emph{ if two observables affiliated to $\rc(X)$ have the same
localizations at infinity, then they have the same essential
spectrum}. If the difference of the resolvents is a compact
operator, then clearly they have the same localizations at infinity,
but the converse is far from being true (e.g.\ see the example on
page 531 from \cite{GI0}, where the essential spectrum is
independent of the details of the shape of the function $\omega$).
On the other hand, one may find in \cite{GG2} criteria which ensure
the compactness of the difference of the resolvents of two
self-adjoint operators under rather weak conditions, e.g.\ an
example from \cite[p.\,26 ]{GG2} is a general version of
\cite[Proposition 4.1]{LaS}.  }\end{remark}

\begin{remark}\label{re:fct}{\rm
    The following remark is useful in applications: if $H$ is an
    observable affiliated to $\rc(X)$ and if
    $\theta:\sigma(H)\rarrow\mbr$ is a proper continuous function,
    then $\theta(H)$ is affiliated to $\rc(X)$ and we have
    $\vk.\theta(H)=\theta(\vk.H)$ for all $\vk\in\delta X$ (see page
    \pageref{fct}).  }\end{remark}

\begin{remark}\label{re:matx}{\rm
    As explained in \cite[p.\,520]{GI0}, all our results extend
    trivially to the case when $L^2(X)$ is replaced with the space
    of $L^2$ functions with values in a Hilbert space $\bfe$: it
    suffices to replace the algebra $\ca$ with $\ca\otimes K(\bf
    E)$.  For example, Theorem \ref{th:introo} remains valid without
    any change if $L^2(X)$ is replaced by $L^2(X;\bfe)$, where $\bf
    E$ is finite dimensional, and $\rc(X)$ is defined exactly as
    before.  Thus in applications we can consider differential
    operators with matrix valued coefficients, like Dirac operators.
  }\end{remark}

\PAR                      \label{ss:Ie}
We give here the simplest applications of Theorems \ref{th:intro}
and \ref{th:introo}, a more detailed study and more general examples
can be found in Section \ref{s:opaf}.

Assume first that $X$ is discrete. Note that in the particularly
important case $X=\mbz^n$ Theorem \ref{th:intro} has been proved in
\cite{RRS} (with a slightly different formulation and with quite
different methods). Now we have
\begin{equation}\label{eq:disc}
\rc(X)=\{T\in \rb(X)\mid \lim_{ k\rarrow 0}\|[T, V_k]\|=0\}.
\end{equation}
Since $V_k^*U_xV_k=k(x)U_x$ we see that each operator of the form
$T=\sum_{a\in X}\vf_a(Q)U_a$, with $\vf_a\in\ell_\infty(X)$ and
$\vf_a\neq0$ only for a finite number of $a$, belongs to $\rc(X)$.
Clearly, we have $\vk.T=\sum_{a\in X}(\vk.\vf_a)(Q)U_a$
where the function $\vk.\vf_a\in\ell_\infty(X)$ is defined by
$(\vk.\vf_a)(y)=\lim_{x\rarrow\vk}\vf(x+y)$. The Jacobi and CMV
operators considered in \cite{LaS} are particular cases of such
operators $T$.

Now we give three examples in the case $X=\mbr^n$. We start with the
Schr\"odinger operator. We denote by $\ch^s$ the Sobolev space of
order $s\in\mbr$ associated to $L^2(\mbr^n)$. Note that $\Delta$ is
the positive Laplacian. From Proposition \ref{pr:3.3} we get:

\begin{proposition}\label{pr:schr}
Let $W$ be a continuous symmetric sesquilinear form on $\ch^1$ such
that:\\ 
(1) $W\geq-\mu\Delta-\nu$ as forms on $\ch^1$ for some
numbers $\mu<1$ and $\nu>0$,\\ 
(2) $\lim_{k\rarrow0}\|[V_k,W]\|_{\ch^1\rarrow\ch^{-1}}=0$.\\ 
Let $H_0$ be the self-adjoint operator associated to the form sum
$\Delta+W$ and let $V$ be a real function in $L^1_{\rm loc}(\mbr^n)$
such that its negative part is relatively bounded with respect to
$H_0$ with relative bound $<1$. Then the self-adjoint operator
$H=H_0+V(Q)$ (form sum) is affiliated to $\rc(\mbr^n)$, hence the
conclusions of Theorem \ref{th:introo} hold for it.
\end{proposition}

This can be extended to a general class of hypoelliptic operators,
cf.\ Proposition \ref{p:3.3}. We present below a very particular
case. 

\begin{proposition}\label{pr:hyp}
Let $h:\mbr^n\rarrow\mbr$ be of class $C^m$ for some $m\geq1$ and
such that:\\ 
(1) $\lim_{k\rarrow\infty}h(k)=+\infty$,\\ 
(2) the derivatives of order $m$ of $h$ are bounded,\\
(3) $\sum_{|\alpha|\leq m}|h^{(\alpha)}(k)|\leq C(1+|h(k)|)$.\\
Let $\cg=D(|h(P)|^{1/2})$ be the form domain of the operator $h(P)$
and assume that $W$ is a symmetric continuous form on
$\cg$ such that:\\
(4) $W\geq-\mu h(P)-\nu$ as forms on $\cg$ for some
numbers $\mu<1$ and $\nu>0$,\\ 
(5) $\lim_{k\rarrow0}\|[V_k,W]\|_{\cg\rarrow\cg^*}=0$.\\
Let $H_0=h(P)+W$ (form sum) and let  $V\in L^1_{\rm loc}(\mbr^n)$
real such that its negative part is relatively bounded with respect
to $H_0$ with relative bound $<1$. Then the self-adjoint operator
$H=H_0+V(Q)$ (form sum) is affiliated to $\rc(\mbr^n)$, hence the
conclusions of Theorem \ref{th:introo} hold for it.
\end{proposition}

\begin{remark}\label{re:smoo}{\rm
If $X$ is an arbitrary group, $h:X\rarrow\mbr$ is continuous and
satisfies $|h(k)|\rarrow\infty$ as $k\rarrow\infty$, and if $V\in
L^\infty(X)$, then obviously $h(P)+V(Q)$ is affiliated to $\rc(X)$
and so we can apply Theorem \ref{th:introo}. In order to cover
unbounded $V$ without much effort a quite weak regularity
condition on $h$ is sufficient, see Proposition \ref{p:3.3}
and especially relation (\ref{e:3.x}). We shall not try to optimize
on this here. 
}\end{remark}

Finally, we consider a Dirac operator $D$.  Let
$\ch=L^2(\mbr^n;\bfe)$ for some finite dimensional Hilbert space
$\bfe$. We only need to know that $D$ is a symmetric first order
differential operator with constant coefficients acting on
$\bfe$-valued functions and which is realized as a self-adjoint
operator on $\ch$ such that the domain of $|D|^{1/2}$ is the Sobolev
space $\ch^{1/2}$. Now from Corollary \ref{co:ts} we get:

\begin{proposition}\label{pr:dirac}
Let $W$ be a continuous symmetric form on $\ch^{1/2}$ such that:\\
(1) $\pm W\leq \mu|D|+\nu$ as forms on $\ch^{1/2}$ for some numbers
$\mu<1$ and $\nu>0$,\\ 
(2) $\lim_{k\rarrow0}\|[V_k,W]\|_{\ch^{1/2}\rarrow\ch^{-1/2}}=0$.\\
Then the self-adjoint operator $H=D+W$, defined as explained on page
\pageref{dirac}, is affiliated to $\rc(\mbr^n)$, hence the
conclusions of Theorem \ref{th:introo} hold for it.
\end{proposition}
Observe that condition (2) is trivially satisfied if $W$ is the
operator of multiplication by an operator valued function
$W:\mbr^n\rarrow B(\bfe)$.

\begin{remark}\label{re:2ord}{\rm
We emphasize that the conditions on the perturbation $W$ in
Propositions \ref{pr:schr}-\ref{pr:dirac} is such that $W$ can
contain terms of the same order as $\Delta,h(P)$ or $D$
respectively. For example, operators of the form
$$
-\textstyle\sum_{j,k} \partial_j a_{jk}\partial_k +
\mbox{ singular lower order terms }
$$ with $a_{jk}\in L^\infty$ such that the matrix $(a_{jk}(x))$ is
bounded from below by a strictly positive constant are already
covered by Proposition \ref{pr:schr}.  See Example \ref{ex:ell} for
much more general results. These examples may be combined with the
Remark \ref{re:fct} to cover functions of operators, e.g.\
$\sqrt{H}$ if $H\geq0$.
}\end{remark}

\PAR                      \label{ss:I2}
Crossed products of $C^*$-algebras by the action of $X$ play a
fundamental r\^ole in our proof of Theorem \ref{th:intro} but we
have to stress that they are important for two distinct reasons.
First, they are in a natural sense $C^*$-algebras of
\emph{energy}\,\symbolfootnote[3]{\ We emphasize ``energy'' because
  algebras of observables and crossed products were frequently used
  in various domains of the quantum theory in the last 50 years, but
  with different meanings and scopes than here.}  observables (or
quantum Hamiltonians), and hence they allow one to organize the
Hamiltonians in classes each having some specific properties, e.g.\ 
the essential spectrum of the operators in a class is given by a
``canonical'' formula specific to that class (see (\ref{eq:imp})).
On the other hand, crossed products are very efficient at a
technical level, their use allows one to solve a non-abelian problem
by abelian means: the problem of computing the quotient of a
non-commutative $C^*$-algebra $\ra\subset \rb(X)$ with respect to
the ideal $\rk(X)\equiv K(L^2(X))$ is reduced to that of computing
$\ca/\co(X)$ where $\ca$ is a $C^*$-algebra of bounded uniformly
continuous functions on $X$.

The first reason mentioned above will be clarified by the later
developments, but one may observe already now that the decomposition
(\ref{eq:intro2}) is far from efficient.  Indeed,
its extreme redundancy\label{red} becomes clear when we realize that
many $\vk$ give the same $\vk.H$
(e.g.\ if the filters $\vk$ and $\cchi$ have the same envelope then
$\cchi.T=\vk.T$, see page \pageref{env}) and many more give the same
$\sigma(\vk.H)$ (e.g. $\cchi.T=U_x\vk.TU_x^*$ if $\cchi$ is the
translation by $x\in X$ of $\vk$).

Thus at a qualitative level (\ref{eq:intro2}) is
not very significant, it does not say much about $\se(H)$, at
least when compared with the $N$-body situation where the HVZ
theorem has such a nice physical interpretation that you can predict
it and believe it without proof.

In order to partially remediate this drawback we consider smaller
classes of Hamiltonians. The following framework, introduced in
\cite{GI0}, gives us more specific information about $\se(H)$.
Let $\cc(X)$ be the $C^*$-algebra of all bounded uniformly continuous
functions on $X$ and $\cc_\infty(X)$ that of continuous
functions which have a limit at infinity (in the usual sense).

\begin{definition}\label{df:aqh}
An \emph{algebra of interactions $\ca$} on $X$ is a $C^*$-subalgebra
of\, $\cc(X)$ which is stable under translations and which contains
$\cc_\infty(X)$. The \emph{$C^*$-algebra of quantum Hamiltonians
of type $\ca$} is the norm closed linear space $\ra\equiv\ca\rtimes
X\subset \rb(X)$ generated by the operators of the form
$\varphi(Q)\psi(P)$ with $\varphi\in\ca$ and $\psi\in\co(X^*)$.
\end{definition}
We have denoted $\vf(Q)$ the operator of multiplication by $\vf$ in
$L^2(X)$ and $\psi(P)$ becomes multiplication by $\psi$ after a
Fourier transformation. The Propositions \ref{pr:och} and
\ref{pr:da} explain why we think of $\ra$ as a $C^*$-algebra of
Hamiltonians.  For example, if $X=\mbr^n$, the self-adjoint
operators of the form $\Delta+\sum_{k=1}^na_k(x)\partial_k+a_0(x)$
with $a_j\in\ca^\infty$ (functions in $\ca$ with all derivatives
in $\ca$) generate $\ra$.  It turns out that $\ra$ is canonically
isomorphic with the crossed product of $\ca$ by the natural action
of $X$, which explains the notation $\ca\rtimes X$ and the relevance
of crossed products in our context.

\begin{remark}\label{re:sb}{\rm
Note that the definition and the quoted propositions tend to give
the impression that the algebra $\ra$ is rather small. But this is
wrong, $\ra$ \emph{is much larger than expected}. For example,
$\cc(X)\rtimes X=\rc(X)$ and we shall see in Section \ref{s:opaf}
that the set of self-adjoint operators affiliated to $\rc(X)$ is
very large.  Other examples are the $N$-body algebra and the
``bumps'' algebras. In fact, we may summarize our approach as
follows: we first isolate a class of \emph{elementary Hamiltonians},
these being the simplest operators we would like to study, but our
results concern all the operators affiliated to the $C^*$-algebra
they generate, which happens to be a crossed product and is very
rich.  }\end{remark}

In order to state the next consequence of Theorem \ref{th:intro} we
have to introduce some new notations.  Let $\sigma(\ca)$ be the
space of characters of the abelian $C^*$-algebra $\ca$. Then
$\sigma(\ca)$ is a compact topological space which contains $X$ as
an open dense subset, so $\delta(\ca)=\sigma(\ca)\setminus X$ is a
compact space.  We shall adopt the following abbreviation:
$H\in'\ra$ means that $H$ is either a normal element of the algebra
$\ra$ or an observable affiliated to $\ra$.  If $H$ is an observable
affiliated to $\ra$ then $U_xHU_x^*$ is also an observable
affiliated to $\ra$ and we have $\vf(U_xHU_x^*)=U_x\vf(H)U_x^*$ for
$\vf\in\co(\mbr)$. By ``continuity'' of a map $\sigma(\ca)\ni
\vk\mapsto \vk.H\in'\rc_s(X)$ whose values are observables we mean
that $\sigma(\ca)\ni \vk\mapsto \vf(\vk.H)\in\rc_s(X)$ is continuous
for all $\vf\in\co(\mbr)$.

\begin{theorem}\label{th:imp}
If $H\in'\ra$ then the map $X\ni x\mapsto U_xHU_x^*\in'\ra$ extends
to a continuous map $\sigma(\ca)\ni \vk\mapsto \vk.H\in'\rc_s(X)$
and we have
\begin{equation}\label{eq:imp}
\se(H)=\overline\ccup_{\vk\in\delta(\ca)}\sigma(\vk.H).
\end{equation}
\end{theorem}
{\bf Remark:} To see the connection between this and Theorems
\ref{th:intro} and \ref{th:introo}, we recall that an ultrafilter
finer than the Fr\'echet filter is the same thing as a character
$\vk$ of the algebra of all bounded functions on $X$ such that
$\vk(\vf)=0$ if $\vf\in\co(X)$ (see Subsection \ref{ss:pd5}).
Moreover, if $\cchi$ is a second such ultrafilter and
$\vk(\vf)=\cchi(\vf)$ for all $\vf\in\cc(X)$, then $\vk.H=\cchi.H$
for all $H\in'\rc(X)$, thus the union in (\ref{eq:intro2}) and
(\ref{eq:introo}) may be taken in fact over $\vk\in\delta(\cc(X))$.
We emphasize that, although Theorem \ref{th:imp} seems stronger than
Theorems \ref{th:intro} and \ref{th:introo}, it is in fact an
immediate consequence of Theorem \ref{th:intro} (just ``abstract
nonsense'', see Subsection \ref{ss:LIT} for details).  Note also
that (\ref{eq:imp}) is a canonical decomposition of the essential
spectrum of $H$, all the objects in the formula being canonically
associated to $\ca$. The representation (\ref{eq:imp}) is further
discussed in Subsection \ref{ss:LIT}, see page \pageref{disc}.

\begin{remark}\label{re:info}{\rm
    We mention that, by using a more involved algebraic formalism as
    in \cite{GI0}, one can obtain partial, but often relevant,
    information concerning the essential spectrum of $H$ as
    follows. Let $\cj$ be an $X$-ideal such that
    $\co(X)\subset\cj\subset\ca$ and let $\rj=\cj\rtimes X$ (we use
    here notations and results from \cite{GI0}). Then
    $\rk(X)\subset\rj\subset\ra$ and $\rj$ is an ideal in $\ra$, so
    the image $H_\cj$ of $H$ is well defined as observable affiliated
    to the quotient algebra $\ra/\rj$. By using the natural
    surjection $\ra/\rk(X)\rarrow \ra/\rj$ we clearly get
    $\sigma(H_\cj)\subset\se(H)$. In this argument $\rj$ need not be
    a crossed product, but if it is, we can use $\ra/\rj\cong
    (\ca/\cj)\rtimes X$ to get a concrete representation of $H_\cj$.
    }\end{remark}

\PAR                      \label{ss:In}
This subsection is devoted to some historical comments and a
discussion of some related results from the literature.

Theorem \ref{th:intro} was announced in the preprints
\cite{Ift,GI*}, see Theorems 1.3 and 4.2 in \cite{Ift} and Theorem
4.1 and Corollary 4.2 in \cite{GI*}. In fact, the theorem was stated
in a stronger form, namely we assert that the union in
(\ref{eq:intro2}) is already closed. Moreover, some nontrivial
applications are stated at page 149 of \cite{GIc}.  The closedness
of the union in (\ref{eq:intro2}) as well as more explicit
applications of Theorem \ref{th:intro} will be discussed in the
second part of this paper.  However we show here that the union in
(\ref{eq:imp}) is closed for some special algebras $\ca$ when the
result is far from obvious (Section \ref{s:exloc}).

The main idea of the proof of Theorem \ref{th:intro} we had in mind
at that moment is presented at \cite[p.\ 30--31]{GI*} and it has to
be combined with the two main points of the algebraic approach we
used in that paper, namely:
\begin{itemize}
\item[(1)] If $\ch$ is a Hilbert space then the quotient algebra
$B(\ch)/K(\ch)$ is a $C^*$-algebra and, if $\what T$ is the
projection of $T\in B(\ch)$ in the quotient, then
$\se(T)=\sigma(\what T)$.
\item[(2)] We have $K(L^2(X))=\co(X)\rtimes X$ and if $\ca$ is an
algebra of interactions then
\begin{equation}\label{eq:mp}
\big(\ca\rtimes X\big)/\big(\co(X)\rtimes X\big)
\cong\big(\ca/\co(X)\big)\rtimes X.
\end{equation}
\end{itemize}
If $T\in\ca\rtimes X$ the isomorphism (\ref{eq:mp}) allows us to
reduce the computation of $\what T$ to an abelian problem and hence
to deduce $\what T\cong(\vk.T)_{\vk\in\delta(\ca)}\in
\prod_{\vk\in\delta(\ca)}\rc(X)$.

The preceding strategy requires a lot of abstract machinery and is
not adapted to a purely Hilbert space setting. For example, the
isomorphism (\ref{eq:mp}) is a consequence of the fact that the
functor $\ca\mapsto\ca\rtimes X$ transforms short exact sequences in
short exact sequences, an assertion which does not even make sense
if we fix the Hilbert space on which the algebras are realized.

Instead, in the present paper we decided to avoid step (2) of
this strategy and to base our arguments on a beautiful
theorem due to M.\,B.\,Landstad \cite{Lad} which gives an intrinsic
characterization of crossed products. We feel that this makes the
argument more elementary and gives deeper insight into the matters
treated here. In fact, we could now avoid completely going out from
the purely Hilbert space setting (in particular, forget about the
step (1) above), but this does not seem to us a natural attitude and
we finally decided to adopt a median approach.

\medskip

It is remarkable that $\rc(X)$ as defined in (\ref{eq:01}) is
precisely the crossed product $\cc(X)\rtimes X$.  Initially, this
fact was proved by direct methods in the case $X=\mbr^n$ in
\cite{DaG2} (because of this Corollary 4.2 from \cite{GI*} was stated
only for $X=\mbr^n$).  The general case follows in fact immediately
from Landstad's theorem.

\medskip

We make now some comments concerning other papers with goals similar
to ours.  We note first that, in the particular case $X=\mbz^n$,
Rabinovich, Roch and Silberman \cite{RRS} discovered Theorem
\ref{th:intro} before us and proved it with no $C^*$-algebra
techniques (in Remark \ref{re:gog} we explain why (\ref{eq:disc}) is
just their algebra of ``band dominated operators''). It seems that
they realized the fact that their algebra in the case $X=\mbz$ is a
crossed product only in \cite{RRR} (this fact is a particular case
of \cite[Theorem 4.1]{GI0}). In \cite{RRS} and in subsequent works
\cite{Rab,RRS,RRS2} (see also \cite{RRS2} for references to earlier
papers) these authors use a discretization technique in order to
treat perturbations of pseudo-differential operators in
$L^2(\mbr^n)$.  They get relations like (\ref{eq:introo}) and show
that in some situations the union is already closed.  Moreover, in
Chapter 7 of \cite{RRS2} they present an abstract version of their
approach (in particular they consider groups more general than
$\mbz^n$) which seems to us complementary to our approach and
relevant in contexts like that of \cite{Gol}. We learned about these
works quite recently thanks to a correspondence with Barry Simon who
sent us a copy of the paper \cite{Rab}; this explains why the above
references were not included in our previous works on this topic.

We discuss now the relation between our paper and the article
\cite{Ro3} (this reference was pointed out to us by one of the
referees). We shall do it in some detail because $C^*$-algebra
techniques are emphasized in \cite{Ro3}. The purpose of Roe is to
extend the results of Rabinovich, Roch and Silberman to nonabelian
groups. He considers a finitely generated discrete (nonabelian)
group $\Gamma$ and defines $A$ as the $C^*$-algebra of operators on
$\ell_2(\Gamma)$ generated by $\ell_\infty(\Gamma)$ and by the right
translation operators $R_\gamma$ (this is a natural extension of the
procedure introduced in \cite{RRS}). Then, denoting $L_\gamma$ the
left translation operators, he shows that for each $T\in A$ the map
$\gamma\mapsto L_\gamma T L_\gamma^*$ extends to a $*$-strongly
continuous map $\beta\Gamma\to A$, where $\beta\Gamma$ is the
Stone-\v{C}ech compactification of $\Gamma$ (the space of characters
of $\ell_\infty(\Gamma)$).  The restriction to
$\delta\Gamma=\beta\Gamma\setminus\Gamma$ of this map is the
\emph{symbol} of $T$ and the main result of \cite{Ro3} is that for
exact groups (in the $C^*$-algebra sense) an operator $T\in A$ has
symbol equal to zero if and only if $T$ is compact.

On \cite[p.\ 30-31]{GI*}, where we describe the main ideas of the
proof of Theorem \ref{th:introo}, we introduce the notion of regular
operator on $L^2(X)$ for $X$ an abelian locally compact group (and
in a more general context in the footnote on \cite[p.\ 31]{GI*}): we
say that a bounded operator $T$ on $L^2(X)$ is regular if
$\{U_xT^{(*)}U_x^*\mid x\in X\}$ are strongly relatively compact
sets. Then we note that for such operators the map $x\mapsto
U_xTU_x^*$ extends to a strongly continuous map $\beta
X\ni\vk\mapsto T_\vk\in B(L^2(X))$ (this time the Stone-\v{C}ech
compactification $\beta X$ involves the topology of $X$) and call
the values $T_\vk$ with $\vk\in\delta X$ localizations at infinity
of $T$. We show that the elements of $\cc(X)\rtimes X$ are regular
and from the arguments on page 31 it is rather obvious that their
localizations at infinity belong to the same algebra $\cc(X)\rtimes
X$. This is more explicitly stated and proved in \cite[Lemma
3.10]{GIc} (which is Lemma 3.9 in the preprint version and Lemma
\ref{lm:sop} here).  All this can also be done at the level of the
algebra $\cc(X)$ and at the bottom of \cite[p.\ 31]{GI*} we say that
if $\varphi\in\cc(X)$ and all its localizations at infinity are
zero, then $\varphi\in\co(X)$ (this is easy to prove, cf.\ Lemma
\ref{lm:linf} here) and finish by saying that this remains true
after taking crossed products (which is not obvious but can be
deduced from \cite[Theorem 3.4]{GI*} or \eqref{eq:mp} here; as we
said before, in this paper we prefer to use Landstad's theorem at
this last step).

We emphasize that although the starting points of \cite{GI*} and
\cite{Ro3} (in particular the relevance of the Stone-\v{C}ech
compactification) are similar, the proofs of the main fact (that the
kernel of the symbol map, in the terminology of \cite{RRS}, is just
the compacts) are of a quite different nature.  Indeed, Roe mentions
that $A$ is the reduced crossed product
$L^\infty(\Gamma)\rtimes_r\Gamma$ but never uses this fact, cf.\ the
proof of \cite[Proposition 3.3]{Ro3}. On the other hand, the crossed
product structure and relations like \eqref{eq:mp} are the heart of
our approach (and we expect that \eqref{eq:mp} is also true under
Roe's conditions).

The methods used by Roe also seem relevant for the solution of a
problem left open (but not explicitly stated) in \cite{Gol}. The
space $\Gamma$ considered there is a tree, which is a finitely
generated \emph{mono\"id}. The natural object in this case is the
$C^*$-algebra generated by the right translations and by
$\ell_\infty(\Gamma)$, the localizations at infinity being given by
left translations. Due to obvious technical difficulties the algebra
considered in \cite{Gol} is much smaller: it is generated by the
Laplacian (which is a certain polynomial in the right translations)
and by the functions in $\ell_\infty(\Gamma)$ which extend
continuously to the hyperbolic compactification of $\Gamma$ (see
Sections 3.4 and 3.5 and Theorem 5.1 and its proof in \cite{Gol};
the references are to the preprint version). A larger algebra,
associated to the analogue of the slowly oscillating functions on
$\Gamma$, is considered in \cite{GG1}, where the problem is treated
by very different techniques. It would be interesting to know if the
techniques from \cite[Section 3]{Ro3} can be adapted to solve the
most general situation.

\medskip

Y.\,Last and B.\,Simon obtained in \cite{LaS} relations like
(\ref{eq:introo}) for large classes of Schr\"odinger operators on
$\mbr^n$ and their discrete versions (Jacobi or CMV operators).
Their proofs involve ``classical'' geometrical methods (localization
with the help of a partition of unity).

\medskip

We have to emphasize that many people working on pseudo-differential
operators have been led to consider $C^*$-algebras generated by such
operators and to describe their quotients with respect to the ideal
of compact operators: in fact, this is one of the most efficient
ways to define the symbol of an operator (see \cite{CMS} for
example).  Much more specific and relevant with respect to our goals
is the work of H.\,O.\,Cordes (see \cite{Cor} for a review). For
example, the $C^*$-algebra generated by a hypoelliptic operator and
by the algebra of slowly oscillating functions and the computation
of its quotient with respect to the compacts seem to have been
considered for the first time in M.\,Taylor's thesis (see
\cite[Theorem 1]{Tay}).  For more recent work on these lines, we
refer to \cite{Nis}.

A rather different class of ``$C^*$-algebras of Hamiltonians''
appears in the work of J.\,Bellissard on solid state physics
\cite{Be1,Be2}: he fixes a Hamiltonian $H$ and considers the
$C^*$-algebra generated by its translates. These algebras do not
contain compact operators in general, so the techniques we use do
not seem relevant in his setting. A more detailed discussion of the
connection between the approach of Bellissard and ours can be found
in \cite{GI0}.

\medskip

The origin of our approach can be traced back to the algebraic
treatment of the $N$-body problem from \cite{BoG1,BoG2} (where the
HVZ theorem and the Mourre estimate are proved in an abstract graded
$C^*$-algebra framework for a very general class of $N$-body
Hamiltonians). The r\^ole of the crossed products was pointed out in
\cite{GI*,GI**,GI0} and a treatment of the $N$-body problem along
these lines is presented in \cite{DaG1,DaG2,DaG3}. Various
applications and extensions of the crossed product technique can be
found in \cite{AMP,Ma2,MPR,Ric,Rod} and references therein.

\medskip

Our interest in localizations at infinity of a Hamiltonian was
initially motivated by our desire to go beyond the $N$-body problem
and to consider general (phase space) anisotropic systems \cite{GI4,
Ift}. Indeed, in the $N$-body case there is a lot of supplementary
structure which makes the theory simple and beautiful (cf.\
Subsection \ref{ss:nbody}), but this structure has no analogue in
other types of anisotropy. We first found that the $C^*$-algebra
techniques are quite well adapted to the study of Hamiltonians with
Klaus type potentials, see \cite{GI*,GI0} and also Subsection
\ref{ss:klaus} here for a treatment in the spirit of Theorem
\ref{th:intro}. We finally realized that the relation (\ref{eq:mp}),
which is the main point of the algebraic approach that we used,
predicts in fact the description (\ref{eq:introo}) of the essential
spectrum of $H$ in terms of its localizations at infinity. 

 The paper \cite{HeM} played an important r\^ole in our
understanding of this fact. Indeed, B.\,Helffer and A.\,Mohamed
prove there that the essential spectrum of a magnetic Hamiltonian
$(P-A)^2+V$ is the closure of the union of the
spectra of some limit Schr\"odinger operators. Their proof is based
on hypoellipticity techniques and the result is already interesting
if the magnetic field is not present. The class of potentials they
consider is quite large, but the function $V$ has to be bounded
from below and to satisfy some regularity conditions.  These
assumptions imply that the limit operators have only polynomial
electric and magnetic potentials, which is easily explained in our
framework, see \cite[Proposition 3.13]{GIc}.

\PAR                      \label{ss:pp}
{\bf Plan of the paper.}  Our purpose being to emphasize not only the
power but also the simplicity of the $C^*$-algebra techniques, 
we made an effort to make the paper essentially self-contained and
easy to read by people working in the spectral theory of quantum
Hamiltonians and with little background in $C^*$-algebras.

We could have written a much shorter paper but which would have been
accessible mostly to people with no interest in spectral theory.
Instead, we have chosen to present in some detail most of the tools
which are not standard among those interested in the subject.  In
particular we give in an Appendix a simple and self-contained proof
of Landstad's theorem (Theorem \ref{th:Land}) which plays an
important r\^ole in our arguments. 

In Section \ref{s:p} we introduce our notations and make a
r\'esum\'e of what we need concerning (ultra)filters and their
relation with the characters of some abelian $C^*$-algebras. In
Section \ref{s:B} we introduce crossed products in the version we
need and we point out several useful consequences of Landstad's
theorem. This replaces the much more abstract arguments from
\cite{GI*,GI0}, since we remain in a purely Hilbert space setting,
but also gives stronger and more explicit results in applications.
Section \ref{s:opaf} is devoted to criteria of affiliation to the
algebra $\rc(X)$, we show there that this algebra is much larger
than one would think at first sight. 

In Section \ref{s:LIF} we prove our main result, Theorem
\ref{th:ess}. Finally, in Section \ref{s:exloc} we consider three
algebras of quantum Hamiltonians, those which seem the most
interesting to us.  The first one $\rv(X)$ is generated by slowly
oscillating potentials and is the simplest non trivial algebra of
Hamiltonians since it is defined by the property that if $H$ is
affiliated to $\rv(X)$ then all its localizations at infinity are
free Hamiltonians (i.e.\ functions of the momentum). The second one
is the algebra associated to a \emph{sparse set} and it is
remarkable because the localizations at infinity of the Hamiltonians
affiliated to it are two-body Hamiltonians and thus their essential
spectrum has a quite interesting structure. The third one is, of
course, the $N$-body algebra, or rather a more general and
geometrically natural algebra that we call \emph{Grassmann algebra},
an object of a remarkable simplicity, richness and beauty. The final
subsections are devoted to some remarks of a different nature on the
localizations at infinity of Hamiltonians of the form $h(P)+v(Q)$
with $v(Q)$ relatively bounded with respect to $h(P)$.

\bigskip

\noindent{\bf Acknowledgments:} It is a pleasure to thank Barry
Simon for stimulating correspondence and for sending us a copy of
the paper \cite{Rab}. We are also indebted to Fran\c coise Piquard
and George Skandalis for helpful discussions and to Steffen Roch
for pointing out to us an erroneous assertion in the first version
of this paper, cf.\ Remark \ref{re:roch}. Finally, we are grateful
to the referees, their comments and critics allowed us to eliminate
several errors form the first version of this paper and to
significantly improve the general presentation. 

\section{Preliminaries} \label{s:p}
\protect\setcounter{equation}{0}
In this section we describe our notations and recall facts
needed in the rest of the paper.
\PAR                      \label{ss:pd2}
If $X$ is a locally compact topological space then $\cc_\infty(X)$
is the $C^*$-algebra of continuous functions which have a limit at
infinity and $\co(X)$ is the subalgebra of functions which converge
to zero at infinity; thus $\cc_\infty(X)=\mbc+\co(X)$. Let $\Cc(X)$
be the subalgebra of functions with compact support. If $\ra$ is a
$C^*$-algebra then we similarly define $\co(X;\ra)$ for example,
which is also a $C^*$-algebra. If $X$ is compact we set
$\cc(X;\ra)=\co(X;\ra)$ and $\cc(X)=\cc(X;\mbc)$, which does not
conflict with the notation (\ref{eq:buc}) because the continuous
functions on $X$ are uniformly continuous.
The characteristic function of a set $S\subset X$ is denoted
$\ind_S$.

In order to facilitate the reading of the paper we tried to respect
as much as possible the following notational conventions.  For
abelian algebras (abstract as well as concrete ones, like function
algebras) we use ``mathcal'' fonts, like $\ca,\cc$. For nonabelian
algebras we use ``mathscr'' fonts, like $\ra,\rc$. Moreover, the
crossed product of an abelian algebra $\ca$ by the action of some
group is denoted $\ra$. For other mathematical objects we use either
greek letters or ``mathcal'' fonts with one exception: filters are
often denoted by small gothic letters like $\gof,\gog$.  However,
ultrafilters are generally denoted $\vk$ because we think of
them as ``points at infinity'' of the space $X$ whose points are
denoted $x$.

\PAR                      \label{ss:pd1}
If $\ch$ is a Hilbert space then $B(\ch)$ and $K(\ch)$ are the
$C^*$-algebras of bounded and compact operators on $\ch$
respectively.  The resolvent set, the spectrum and the essential
spectrum of an operator $S$ are denoted $\rho(S)$, $\sigma(S)$ and
$\se(S)$ respectively.

By morphism between two $C^*$-algebras we understand
$*$-homomorphism.  An ideal in a $C^*$-algebra is assumed to be
closed and two-sided.

An \emph{observable} is a linear operator $H:D(H)\rarrow \ch$ such
that $HD(H)\subset\ck$, where $\ck$ is the closure of $D(H)$ in
$\ch$, and such that $H$ when considered as operator in $\ck$ is
self-adjoint in the usual sense.  A trivial observable which,
however, is quite important, is the unique observable whose domain
is equal to $\{0\}$; we shall denote it $\infty$. One has to think
that $H$ is equal to $\infty$ on $\ck^\bot$ and for this reason we
set $\vf(H)=0$ on $\ck^\bot$ if $\vf\in\co(\mbr)$. Note that we keep
the notation $(H-z)^{-1}$ for the resolvent of $H$ in $\ch$ but
$(H-z)^{-1}=0$ on $\ck^\bot$.

If $\rc\subset B(\ch)$ is any $C^*$-subalgebra then an observable
$H$ is said to be {\em affiliated} to $\rc$ if $(H-z)^{-1}\in\rc$
for some $z\in\rho(H)$. Then $\varphi(H)\in\rc$ for all $\varphi\in
\co (\mbr)$.

It is theoretically much more convenient to define an observable
affiliated to $\rc$ as a morphism $H:\co(\mbr)\rarrow \rc$ and then
to set $H(\vf)\equiv\vf(H)$.  We refer to \cite[p.\ 522--523]{GI0}
for a r\'esum\'e of what we need and also to \cite{DaG3} for
comments on this notion which should not be confused with that
introduced by S.\ Baaj and S.\ L.\ Woronowicz
(in \cite[Sec.\ 8.1]{ABG} one can find a systematic presentation of
this point of view). 

We recall two definitions which make the transition from Theorem
\ref{th:intro} to Theorem \ref{th:introo} trivial. The
\emph{spectrum} of the observable $H$ is the set
\begin{equation}\label{eq:sp}
\sigma(H)=\{\lambda\in\mbr\mid \vf\in\co(\mbr)\mbox{ and }
\vf(\lambda)\neq0\Rightarrow \vf(H)\neq0\}.
\end{equation}
Let $\rk=\rc\cap K(\ch)$, this is an ideal in $\rc$. Then the 
\emph{essential spectrum} of $H$ is the set
\begin{equation}\label{eq:essp}
\se(H)=\{\lambda\in\mbr\mid \vf\in\co(\mbr)\mbox{ and }
\vf(\lambda)\neq0\Rightarrow \vf(H)\nin\rk\}.
\end{equation}
We also note that any morphism $\pi:\rc\rarrow\rb$ between two
$C^*$-algebras extends in a trivial way to a map between
observables affiliated to $\rc$ to observables affiliated to $\rb$.
Indeed, it suffices to define $\pi(H)$ by the condition
$\vf(\pi(H))=\pi(\vf(H))$. 
For example, if $\pi:\rc\rarrow\rc/\rk$ is the canonical morphism of
$\rc$ onto the quotient algebra $\rc/\rk$, we have
$\se(H)=\sigma(\pi(H))$.

Finally, we mention one more immediate consequence of the definition
of an observable in terms of morphisms, cf.\ \cite[p.\ 370]{ABG}.
We shall use the easily proven fact that $\vf(H)$ depends only on
the restriction of $\vf$ to the closed real set $\sigma(H)$.
Let $\theta:\sigma(H)\rarrow\mbr$ be continuous and proper (i.e.\
$|\theta(\lambda)|\rarrow\infty$ if $|\lambda|\rarrow\infty$). Then
the observable $\theta(H)$ is well defined by the rule
$\vf(\theta(H))=(\vf\circ\theta)(H)$ for $\vf\in\co(\mbr)$ (if
$H$ is a self-adjoint operator then $\theta(H)$ is just the
operator defined by the usual functional calculus). Clearly:
\emph{if $H$ is affiliated to $\rc$, the observable $\theta(H)$ is
  also affiliated to $\rc$}. \label{fct}

\PAR                      \label{ss:pd3}
We describe now objects and notations from the harmonic
analysis on groups. Everything we need can be found in \cite{F} or
\cite{FeD}; see also \cite{W}.

Let $X$ be an abelian locally compact group (with the operation
denoted additively) equipped with a Haar measure $\dd x$.  We
abbreviate ${\rb}(X)=B(L^2(X)),\ {\rk}(X)=K(L^2(X))$ and note that
these are $C^*$-algebras depending on $X$ and not on the choice of
the Haar measure. Other such $C^*$-algebras are $L^\infty(X)$,
$\cc_\infty(X)$, $\co(X)$ and the $C^*$-algebra of bounded
uniformly continuous functions on $X$, which plays the most
important r\^ole in what follows:
\begin{equation}\label{eq:buc}
\cc(X)=\{\varphi:X\rarrow\mbc\mid \varphi \mbox{ is bounded and
  uniformly continuous } \}.
\end{equation}
In order to avoid ambiguities, if $\vphi $ is a measurable function
on $X$ then we denote $\vphi (Q)$ the operator of multiplication by
$\vphi$ in $L^2(X)$ (the symbol $Q$ has no operator meaning).
By using this map we identify the algebra $L^\infty(X)$ and its
$C^*$-subalgebras with $C^*$-subalgebras of ${\rb}(X)$, in
particular we always embed
\begin{equation}\label{eq:cemb}
\co(X)\subset\cc_\infty(X)\subset\cc(X)\subset\rb(X).
\end{equation}
Note that the $C^*$-algebra $\ell_\infty(X)$ of all bounded
functions on $X$ cannot be embedded in $\rb(X)$ (neither can the
$C^*$-algebra $\cb(X)$ of bounded Borel functions).

\medskip

Let $X^*$ be the set of characters of $X$ (continuous homomorphisms
$k:X\rarrow\mathbb C$ with $|k(x)|=1$) equipped with the locally
compact group structure defined by the operation of multiplication
and the topology of uniform convergence on compact sets.  We denote
the operation in $X^*$ additively and its neutral element by $0$, as
in \cite[ch.II, \S5]{W} (this convention looks rather strange if
$X=\ZZ^n$, for example). If $X$ is a real finite dimensional vector
space then $X^*$ is identified with the vector space dual to $X$ as
follows: let $\braket{\cdot\,,\cdot\,} :X\times X^*\rarrow\mbr$ be
the canonical bilinear map and take $k(x)=\e^{i\langle x,k\rangle}$.
In fact, the field of real numbers can be replaced here by an
arbitrary non-discrete locally compact field, see \cite[page 91]{F})
and \cite[ch.II, \S5]{W}.  We recall that the dual group $(X^*)^*$
of $X^*$ is identified with $X$, each $x\in X$ being seen as a
character of $X^*$ through the formula $x(k)=k(x)$.

The Fourier transform of $u \in L^1(X)$ is the function 
$\cf u\equiv\what{u}:X^* \rarrow {\mbc}$ given by $\what{u}(k)=
\int_X \overline{k(x)}u(x)\dd x$.
We equip $X^*$ with the unique Haar measure $\dd k$ such that $\cf$
induces a unitary map $\cf:L^2(X)\rarrow L^2(X^*)$.  From 
$\cf^{-1}=\cf^*$ we get $(\cf^{-1}v)(x)=
\int_{X^*}k(x)v(k)\dd k$ for $v\in L^2(X^*)$. By taking into account
the identification $X^{**}=X$, the Fourier transform of $\psi\in
L^1(X^*)$ and the Fourier inversion formula are
\begin{equation}\label{eq:ft}
\what\psi(x)=\int_{X^*}\overline{k(x)}\psi(k)\dd k
\hspace{3mm} \mbox{and} \hspace{3mm}
\psi(k)=\int_{X}k(x)\what\psi(x)\dd x.
\end{equation} 
For each measurable $\psi:X^*\rarrow\mbc$ we define the operator
$\psi(P)$ on $L^2(X)$ by $\psi(P)=\cf^*M_\psi\cf$, where $M_\psi$ is
the operator of multiplication by $\psi$ in $L^2(X^*)$.  In
particular, the restriction to $L^\infty (X^*)$ of the map
$\psi\mapsto\psi(P)$ is injective and gives us $C^*$-subalgebras
\begin{equation}\label{eq:0*}
\co(X^*)\subset L^\infty (X^*)\subset{\rb}(X).
\end{equation}

Let $\{U_x\}_{x\in X}$ and $\{V_k\}_{k\in X^*}$ be the strongly
continuous unitary representations of $X$ and $X^*$ in $L^2(X)$
defined by $(U_x f)(y) = f(x+y)$ and $(V_k f)(y) = k(y)f(y)$
respectively.  Note that $U_x$ and $V_k$ satisfy the canonical
commutation relations 
\begin{equation}\label{eq:ccr}
U_x V_k =k(x) V_k U_x.
\end{equation}
Observe that we have $U_x=x(P)$ if $x\in X$ is identified with the
function $k\mapsto k(x)$ and similarly $V_k=k(Q)$. Also, we have,
cf.\ (\ref{eq:ft}):
\begin{equation}\label{eq:fr}
\psi(P)=\int_X U_x\what{\psi}(x)\dd x
\hspace{3mm}\mbox{if}\hspace{3mm} \what{\psi}\in L^1(X).
\end{equation}

\PAR                      \label{ss:pd4}
We summarize here some facts we need concerning filters, cf.\
\cite{Bo,HiSt,Sam}.

A \emph{filter} on $X$ is a family $\gof$ of subsets of $X$ which
does not contain the empty set, is stable under finite
intersections, and has the property: $G\supset F\in\goth
f\Rightarrow G\in\gof$ (the empty set is a filter!).  If $Y$ is a
topological space and $\theta:X\rarrow Y$ is any map, then
$\lim_{\gof} \theta=y$ means that $\theta^{-1}(V)\in\gof$ if $V$ is
a neighborhood of $y$. We shall often write $\lim_{x\rarrow\gof}
\theta(x)$ instead of $\lim_{\goth f}\theta$ for reasons which will
become clear later on.

If $\gof,\gog$ are filters and $\gof\subset\gog$ then
$\gog$ is said to be \emph{finer} than $\gof$. An
\emph{ultrafilter} is a maximal element in the set of all filters on
$X$ for this order relation. If $x\in X$ then the family of sets
which contain $x$ is the ultrafilter determined by $x$.
A filter $\gof$ is an ultrafilter if and only if for each $A\subset
X$ one has $A\in\gof$ or $A^c\equiv X\setminus A\in\gof$. 

Ultrafilters are important because of the following property: if
$\gof$ is an ultrafilter and $\theta:X\rarrow Y$ is an arbitrary map
with values in a compact space $Y$, then $\lim_\gof \theta$ exists.
This fact will become clear after the explanations in Subsection
\ref{ss:mf}. 

The space $\gamma X$ of all ultrafilters on $X$ is a compact space
for the topology defined as follows: the map
$\gof\mapsto\{\vk\in\gamma X\mid\vk\supset\gof\}$ is a bijection
from the set of all filters on $X$ onto the set of all closed
subsets of $\gamma X$.  Thus one should think that \emph{a filter is
  a closed subset of} $\gamma X$.\label{filclo} Another description
of this topology will be given later on.  The compact topological
space $\gamma X$ is the \emph{discrete Stone-\v{C}ech
  compactification of $X$} and it is characterized by the following
universal property:\label{up} \emph{if $Y$ is a compact space then
  each map $\theta:X\rarrow Y$ has a unique extension to a
  continuous map $\theta:\gamma X\rarrow Y$}. Since this property is
important for us, we shall further discuss it in Subsection
\ref{ss:mf}, see page \pageref{upc}.

The set $X$ is identified with an open dense subset of $\gamma X$
(to $x\in X$ one associates the ultrafilter determined by $x$) and
the topology induced by $\gamma X$ on $X$ is the discrete topology.
However, the space $\gamma X\setminus X$ is much too large for our
purposes, the only ultrafilters of interest to us belong to the
compact subset of $\gamma X$ defined by
\begin{equation}\label{eq:gam}
\delta X=\{\vk\mid \vk\mbox{ is an ultrafilter finer than
  the Fr\'echet filter}\}.
\end{equation}
We call \emph{Fr\'echet filter} \label{fr} the filter consisting of
the sets with relatively compact complement (this is not quite
standard). This filter depends on the locally compact non compact
topology given on $X$. In view of the standard meaning of the
notation $\lim_{x\rarrow\infty}$ it is natural to denote by $\infty$
the Fr\'echet filter. As explained above, one should think of
$\infty$ as a certain compact subset of $\gamma X$ and then we have
in fact $\infty=\delta X$.

\PAR                      \label{ss:pd5}
We shall explain now the relation between filters and characters of
certain abelian $C^*$-algebras.  If $\ca$ is such an algebra we
denote $\sigma(\ca)$ the space of characters of $\ca$ (a character
is a non zero morphism $\ca\rarrow\mbc$) equipped with the weak$^*$
topology.  This is a locally compact topological space which is
compact if and only if $\ca$ is unital.

Let $\cb$ be a unital abelian $C^*$-algebra and let $\ca\subset\cb$
be a $C^*$-subalgebra which contains the unit of $\cb$. Then each
character of $\cb$ restricts to a character of $\ca$ and each
character of $\ca$ is obtained in this way. This gives a canonical
map $\pi:\sigma(\cb)\rarrow\sigma(\ca)$ which is continuous and
surjective and if we define in $\sigma(\cb)$ an equivalence relation
$\vk\sim\cchi$ by the condition $\vk(S)=\cchi(S)\ \forall S\in\ca$,
the compact topological space $\sigma(\ca)$ is just the quotient of
$\sigma(\cb)$ with respect to this relation.

In particular, \emph{a map $f:\sigma(\ca)\rarrow Y$ is continuous if
and only if $f\circ\pi:\sigma(\cb)\rarrow Y$is continuous, where $Y$
is an arbitrary topological space}.\label{fact}

\medskip

Let $\ell_\infty(X)$ be the $C^*$-algebra of all bounded functions
on $X$.  Then the space of all characters of $\ell_\infty(X)$ can be
identified with the space $\gamma X$ of all ultrafilters on $X$:
\begin{equation}\label{eq:gx}
\gamma X=\sigma(\ell_\infty(X)).
\end{equation}
Indeed, the map which associates to an ultrafilter $\gof$ the
character $\vf\mapsto\lim_{\gof}\vf$ is a homeomorphism and the
inverse map associates to the character $\vk$ the ultrafilter
$\gof=\{F\subset X\mid \vk(\ind_F)=1\}$.  From now on we shall
identify $\gof$ and $\vk$, so \emph{an ultrafilter is the same thing
  as a character of $\ell_\infty(X)$}, and we shall work with the
interpretation which is most suited to the context. We also set
$\vk(F)=\vk(\ind_F)$ for $F\subset X$. Then
\begin{equation}\label{eq:gamd}
\delta X=\{\vk\in\gamma X\mid \vk(K)=0\quad\forall K\subset X
\mbox{ compact }\}.
\end{equation} 

The algebras $\ca$ that we consider are unital subalgebras of
$\ell_\infty(X)$, thus their character spaces $\sigma(\ca)$ are
quotients of $\gamma X$. In other terms, we can view the characters
of $\ca$ as equivalence classes of ultrafilters: if $\vk$ is a
character of $\ca$, then there is an ultrafilter $\gof$
such that $\vk(\vf)=\lim_\gof\vf$ for all $\vf\in\ca$, and in fact
there are many such ultrafilters.

For the algebras which are of interest for us we always have
\begin{equation}\label{eq:acl}
\cc_\infty(X)\subset\ca\subset\cc(X)
\subset\ell_\infty(X)
\end{equation}
Then $X$ is identified with an open dense subset of $\sigma(\ca)$
and the topology induced by $\sigma(\ca)$ on $X$ coincides with the
initial topology, so $\sigma(\ca)$ is a compactification of the
locally compact space $X$. Thus
\begin{equation}\label{eq:da}
\delta(\ca)=\sigma(\ca)\setminus X=\{\vk\in\sigma(\ca)\mid
\vk(\vf)=0 \quad\forall\vf\in\co(X)\}
\end{equation}
is a compact subset of $\sigma(\ca)$, the boundary of $X$ in the
compactification $\sigma(\ca)$.  The \emph{uniform compactification}
$\beta_{\rm u}X$ of $X$ is defined by the largest algebra $\cc(X)$: 
\begin{equation}\label{eq:sc}
\beta_{\rm u}X=\sigma(\cc(X)),\hspace{2mm}
\delta_{\rm u}X=\beta_{\rm u}X\setminus X=\delta(\cc(X)).
\end{equation}
Later on we shall explicitly describe the equivalence relation in
$\gamma X$ which defines $\beta_{\rm u}$.

\medskip

We are interested only in the boundary $\delta(\ca)$ of $X$ in
$\sigma(\ca)$. We show now that this is a quotient of $\delta X$.
\begin{lemma}\label{lm:delta}
Let $\gof$ be an ultrafilter on $X$ and let $\vk$ be the
character of $\ca$ defined by  $\vk(\vf)=\lim_\gof\vf$. Then
$\vk\in\delta(\ca)$ if and only if $\gof\in\delta X$.
\end{lemma}
\proof If $\gof$ is an ultrafilter and $Y\subset X$ then there
are only two possibilities: either $Y\nin\gof$, and then
$X\setminus Y\in\gof$ hence $Y\cap Z= \emptyset$ for all
$Z\in\gof$, or $Y\in\gof$, and then the sets $Y\cap Z$ with
$Z\in\gof$ form an ultrafilter on $Y$.  If $\gof$ is not
finer than the Fr\'echet filter then there is a set with compact
complement $Y$ which does not belong to $\gof$, and so
$Y\in\gof$. Since any ultrafilter on a compact set is
convergent, we see that there is $y\in Y$ such that $\gof$
contains the filter of neighborhoods of $y$. But then clearly
$\lim_\gof\vf=\vf(y)$ for any continuous function $\vf$, hence
the character $\vk(\vf)=\lim_\gof\vf$ is just $y$ and does not
belong to $\delta(\ca)$. On the other hand, if $\gof\in\delta X$
then clearly $\vk\in\delta(\ca)$.
\qed

Thus the \emph{characters $\vk\in\delta(\ca)$ are equivalence
  classes of ultrafilters $\gof\in\delta X$}. In general, we do not
distinguish\label{dist} between a character and the elements of the
equivalence class of ultrafilters which define it. However, when
needed for the clarity of the argument, we shall use the map
$\delta$ which sends an element into its equivalence class.  More
precisely, from (\ref{eq:acl}) we see that there are canonical
surjections
\begin{equation}\label{eq:lca}
\delta X\rarrow\delta_uX\rarrow\delta(\ca)
\rarrow\{\infty\} 
\end{equation}
and all of them (and their compositions) will be denoted $\delta$.
Here $\infty$ is the Fr\'echet filter and we have
$\delta(\cc_\infty(X))=\{\infty\}$.

\PAR                      \label{ss:pd6}
The space $\beta_{\rm u}X$ is the quotient of $\gamma X$ given by an
equivalence relation that we describe now (see \cite[p.\ 121]{Sam}).
If $\gof$ is a filter then its \emph{envelope}\label{env} is the
filter $\goth f^\circ$ generated by the sets $F+V$ where $F\in\gof$
and $V$ belongs to the filter of neighborhoods of the origin
(observe that the sets $F+V$, with $V$ an open neighborhood of the
origin, are open and form a basis of $\gof^\circ$). Note that
$\gof\supset\gof^\circ$ and $(\gof^\circ)^\circ=\goth f^\circ$.  Two
filters are \emph{u-equivalent} (uniformly equivalent) if they have
the same envelope.

\emph{The quotient of $\gamma X$ with respect to this relation is
$\beta_{\rm u}X$}. We shall give a complete proof of this assertion
since in \cite{Sam} the $C^*$-algebra point of view is not explicitly
considered. The following simple fact will be useful for other
purposes too.

\begin{lemma}\label{lm:easy}
Let $\vf:X\rarrow\mbc$ be uniformly continuous and let $\gof$ be a
filter on $X$.\\
(1) $\lim_\gof\vf$ exists if and only if $\lim_{\gof^\circ}\vf$
exists and in this case they are equal.\\
(2) If $\lim_{x\rarrow\gof}\vf(x+y)\equiv\xi(y)$ exists for each
$y\in X$ then the limit exists locally uniformly in $y$ and $\xi$ is
a uniformly continuous function.
\end{lemma}
\proof To prove (1) it suffices to show that
$\lim_{\gof^\circ}\vf=0$ if $\lim_\gof\vf=0$.  For $\varepsilon>0$
let $F_\varepsilon$ be the set of points where
$|\varphi(x)|<\varepsilon$. We have $F_\varepsilon\in\gof$ and if we
choose a neighborhood $V$ of the origin such that
$|\varphi(x)-\varphi(y)|<\varepsilon$ if $x-y\in V$, then for $x\in
F_\varepsilon +V$ we have $|\varphi(x)|<2\varepsilon$ hence
$F_{2\varepsilon}\in\gof^\circ$.  

Now we prove (2). Set $\omega_V(\vf)=\sup_{y-z\in V}|\vf(y)-\vf(z)|$
if $V$ is a neighborhood of the origin. Then $\vf$ is uniformly
continuous if and only if for each $\varepsilon>0$ there is $V$ such
that $\omega_V(\vf)<\varepsilon$. Clearly
$\omega_V(\xi)\leq\omega_V(\vf)$, so $\xi$ is uniformly continuous.
Now let $V$ be open and let $K$ be a compact set. Then $K$ is
covered by the open sets $x+V$, $x\in K$, hence there is $Z\subset
K$ finite such that $K\subset\ccup_{z\in Z}(z+V)$. Thus for each
$y\in K$ there is $z\in Z$ such that $y\in z+V$ and then
$|\vf(x+y)-\vf(x+z)|\leq\omega_V(\vf)$ for all $x\in X$. Then we
have:
\begin{eqnarray*}
|\vf(x+y)-\xi(y)| &\leq&
|\vf(x+y)-\vf(x+z)|+|\vf(x+z)-\xi(z)|+|\xi(z)-\xi(y)|\\
&\leq& \omega_V(\vf)+|\vf(x+z)-\xi(z)|+\omega_V(\xi)\\
&\leq& 2\omega_V(\vf)+|\vf(x+z)-\xi(z)|.
\end{eqnarray*}
We choose $V$ such that $\omega_V(\vf)\leq\varepsilon/3$ and then we
fix $Z$ as above.  Since $Z$ is finite, there is $F\in\gof$ such
that $|\vf(x+z)-\xi(z)|\leq\varepsilon/3$ for all $x\in F$ and $z\in
Z$.  Finally, we get $|\vf(x+y)-\xi(y)|\leq\varepsilon$ for all
$x\in F$ and $y\in K$.  \qed

\begin{lemma}\label{lm:dif}
Assume $\gof =\gof^\circ$ where $\gof$ is a filter on $X$. Then for
each $F\in \gof$ there is an open subset $G\in \gof$ of $F$ and a
function $\theta \in \cc(X)$ such that $\theta = 1$ on $G$
and $\theta = 0$ on $F^{\rm c}\equiv X\setminus F$.
\end{lemma}
\proof Note first that the open sets from $\gof$ form a basis of
$\gof$. Clearly there is an open $G\in \gof$ and an open, relatively
compact neighborhood of the origin $U$ such that $G + (U-U)\subset
F$, so denoting $A=G-U$ we shall have $A+U\subset F$. We then set
$\theta = |U|^{-1}\ind_A\ast \ind_U$, so for each $x\in X$ we have
$\theta(x) = |U|^{-1}|A\cap (x-U)|$.  For $x \in G$, $x-U\subset A$
thus $\theta(x) = |U|^{-1}|x-U|=1$, and for $x\nin A+U$, $A\cap
(x-U)=\emptyset$ hence $\theta(x) = 0$. But $A+U\subset F$, thus
$\theta = 0$ on $F^{\rm c}$ too. Finally, from
$$\|u\ast v\|_{L^\infty}\leq\|u\|_{L^1}\|v\|_{L^\infty}
\quad\mbox{and}\quad
\|x.(u\ast v)-u*v\|_{L^\infty}\leq\|x.u-u\|_{L^1}\|v\|_{L^\infty},
$$
where $(x.u)(y)=u(y+x)$,
we get $L^1(X)\ast L^\infty(X)\subset\cc(X)$, hence $\theta \in
\cc(X)$. 
\qed

\begin{proposition}
Let $\vk,\cchi$ be ultrafilters on $X$.
Then $\vk^\circ=\cchi^\circ$ if and only if
$\vk(\vf)=\cchi(\vf)$
for each $\vf \in \cc(X)$.
\end{proposition}
\proof The ``only if'' part follows from $
\vk(\vf)=\lim_\vk\vf=\lim_{\vk^\circ} \vf= \lim_{\cchi^\circ} \vf =
\lim_{\cchi} \vf = \cchi(\vf), $ the second and the fourth equality
being consequences of Lemma \ref{lm:easy}.  Conversely, let
$\vk^\circ\neq\cchi^\circ$.  Then there is $F\in \vk^\circ$ such
that $F\nin\cchi^\circ \subset \cchi$, hence $F^{\rm c}\in \cchi$
because $\cchi$ is an ultrafilter. Let now $G$ and $\theta$ be as in
Lemma \ref{lm:dif}. Since $G\in\vk^\circ \subset \vk$ we have
$\vk(\ind_G)=1$, thus $\vk(\ind_{G^{\rm c}})=0$. Hence $\vk(\theta)
= \vk(\theta\ind_G)+\vk(\theta\ind_{G^{\rm c}})= 1 +
\vk(\theta)\vk(\ind_{G^{\rm c}})=1$.  On the other hand, $F^{\rm
  c}\in\cchi$ implies $\cchi(\ind_{F^{\rm c}})=1$ and $\theta
\ind_{F^{\rm c}}=0$, thus $0=\cchi(\theta \ind_{F^{\rm c}})=
\cchi(\theta)\cchi( \ind_{F^{\rm c}})=\cchi(\theta)$.  \qed
%

\section{ Crossed products} \label{s:B} 
\protect\setcounter{equation}{0}
In this section we recall some facts concerning crossed products and
point out some properties important for our later arguments. A
locally compact non compact abelian group $X$ is fixed in what
follows.

We shall say that a $C^\ast$-algebra $\ca$ is an {\em $X$-algebra\/}
if a homomorphism $\alpha :x \mapsto \alpha_x$ of $X$ into the group
of automorphisms of $\ca$ is given, such that for each $A\in\ca$ the
map $x \mapsto \alpha_x(A)$ is norm continuous\,\symbolfootnote[2]{\ 
  The terminology ``$C^*$-dynamical system'' used by some
  $C^*$-algebra theorists seems to us extremely confusing in our
  context, even if $X$ is $\mbr$ or $\mbz$, so we shall not use
  it.}.  An \emph{$X$-subalgebra} of $\ca$ is a $C^*$-subalgebra
that is left invariant by all the automorphisms $\alpha_x$. An
\emph{$X$-ideal} is an ideal stable under the $\alpha_x$.  If
$(\ca,\alpha)$ and $(\cb,\beta)$ are two $X$-algebras, a morphism
$\phi : \ca\rarrow\cb$ is called {\em $X$-morphism\/} if
$\phi[\alpha_x (A)] = \beta_x [\phi (A)]$ for all $x \in X$ and $A
\in\ca$.

We shall not need the abstract definition of the crossed product
$\ca\rtimes X$ of an $X$-algebra $\ca$ by the action of $X$.  We
mention only that $\ca\rtimes X$ is a $C^*$-algebra uniquely defined
modulo a canonical isomorphism by a certain universal property (see
\cite{Rae} for example) and that the correspondence
$\ca\mapsto\ca\rtimes X$ has certain functorial properties (see
\cite{GIc}) which play an important r\^ole in \cite{GI0} but will
not be used here. On the other hand, the following concrete
realization of $\ca\rtimes X$ for certain $\ca$ will be important.

There is a natural action of $X$ on $L^\infty(X)$ by translations
$(\tau_x\varphi)(y)=\varphi(y+x)$\label{tau} and it is clear that
$x\mapsto\tau_x\varphi\in L^\infty(X)$ is norm continuous if and
only if $\varphi\in\cc(X)$. Thus $\cc(X)$ becomes an $X$-algebra and
we will be interested only in crossed products $\ca\rtimes X$ with
$\ca$ an $X$-subalgebra of $\cc(X)$, i.e.\ a $C^*$-subalgebra stable
under translations. In many cases we shall slightly simplify the
writing and set $x.\vf=\tau_x\vf$.\label{act} Note that if
$\vf\in\cc(X)\cap L^2(X)$ we have $x.\vf=U_x\vf$ but
$(x.\vf)(Q)=U_x\vf(Q)U_x^*$. More generally, we shall use the
notations: 
\begin{equation}\label{eq:not}
x\in X,T\in\rb(X)\Longrightarrow x.T\equiv\tau_x(T)= U_xTU_x^*.
\end{equation}

The next definition describes $\ca\rtimes X$ in what we could call
the \emph{pseudo-differential operator representation}, or
$\Psi$DO-representation.
\begin{definition}\label{df:crp}
  If $\ca$ is an $X$-subalgebra of $\cc(X)$, the \emph{crossed
    product} $\ca\rtimes X\equiv\ra$ is the norm closed linear
  subspace of $\rb(X)$ generated by the operators of the form
  $\varphi(Q)\psi(P)$ with $\varphi\in\ca$ and $\psi\in\co(X^*)$.
\end{definition}
The fact that $\ra$ is a $C^*$-algebra follows from:
\begin{lemma}\label{lm:crp}
    If $\vphi\in\cc(X)$ and $\psi\in\co(X^*)$ then for each number
    $\varepsilon >0$ there are elements $x_1, ..., x_n \in X$ and
    functions $\psi_1, ..., \psi_n \in\co(X^*)$ such that:
\begin{equation}\label{eq:dag}
\|\psi(P) \varphi (Q) - 
\textstyle\sum_k \vphi(Q + x_k) \psi_k(P)\| < \varepsilon .
\end{equation}
\end{lemma}
For the proof, first approximate $\psi$ by functions such that
$\what\psi\in L^1(X)$ and then adapt the proof of \cite[Lemma
2.1]{DaG1}. We mention two results which explain why we think of
$\ra$ as a $C^*$-algebra of quantum Hamiltonians. The first one is
\cite[Proposition 4.1]{GI0}.
\begin{proposition}             \label{pr:och}
  Let $\ca$ be an $X$-subalgebra of\, $\cc(X)$ which contains the
  constants.  Let $h:X^*\rarrow\mbr$ be a continuous non-constant
  function such that $\lim_{k\rarrow \infty}|h(k)| = \infty$.  Then
  $\ca\rtimes X$ is the $C^*$-algebra
  generated\,\symbolfootnote[3]{\ If $\cs$ is a family of
  self-adjoint operators then the $C^*$-algebra generated by $\cs$
  is the smallest $C^*$-algebra of operators on ${\ch}$ to which is
  affiliated each $H\in\cs$.}  by the self-adjoint operators of the
  form $h(P+k)+v(Q)$, with $k\in X^*$ and $v\in\ca$ real.
\end{proposition}
The second one is \cite[Corollary 2.4]{DaG1}. Here we assume
$X=\mbr^n$ and denote $\ca^\infty$ the set of functions in $\ca$
such that all their derivatives exist and belong to $\ca$.
\begin{proposition}\label{pr:da}
  Let $h$ be a real elliptic polynomial of order $m$ on $X$ and let
  $\ca$ be as in Proposition \ref{pr:och}.  Then $\ca\rtimes X$ is
  the $C^*$-algebra generated by the self-adjoint operators of the
  form $h(P)+S$, where $S$ runs over the set of symmetric
  differential operators of order $<m$ with coefficients in
  $\ca^\infty$.
\end{proposition}

\begin{examples}\label{ex:exs1}{\rm
We shall point out now the simplest crossed products. The smallest
crossed product $\{0\}=\{0\}\rtimes X$ is, of course, of no interest.
\begin{itemize}
\item[{\rm(1)}]
The largest crossed product is
$\rc(X)=\cc(X)\rtimes X$, see Theorem \ref{th:dg}.
\item[{\rm(2)}]
The $C_0$ functions of momentum: $\co(X^*)=\mbc\rtimes X$.
\item[{\rm(3)}]
The algebra of compact operators:
$\rk(X) = \co(X)\rtimes X$.
\item[{\rm(4)}]
The \emph{two-body algebra}: 
$\rt(X):=\cc_\infty(X)\rtimes X=\co(X^*)+\rk(X)$.
\end{itemize}
The name of the fourth algebra is justified by Propositions
\ref{pr:och} and \ref{pr:da}. Indeed, if $X=\mbr^n$ then $\rt(X)$ is
the $C^*$-algebra generated by the self-adjoint operators of the
form $(P+k)^2+v(Q)$ with $k\in X$ and $v\in\Cc^\infty(X)$ is real,
or by those of the form $\Delta+\sum_{j=1}^na_j\partial_j+a_0$ where
$a_j$ are $C^\infty$ functions constant outside a compact.
}\end{examples}

\begin{remark}\label{re:ab}{\rm
Note that the only abelian crossed products are $\{0\}$ and
$\co(X^*)$. 
}\end{remark}

We have defined a map $\ca\mapsto\ca\rtimes X$ from the set of all
$X$-subalgebras of $\cc(X)$ into the set of $C^*$-subalgebras of
$\rb(X)$ which is obviously increasing. The following theorem, which
is an immediate consequence of a more general abstract result due to
M.B.\ Landstad, cf.\ \cite[Theorem 4]{Lad} or \cite{Ped}, says that
this map is injective and describes its range.
\begin{theorem}\label{th:Land}
A $C^*$-subalgebra $\ra$ of $\rb(X)$ is a crossed product if and
only if for each $A\in\ra$ the following two conditions are
satisfied:
\begin{itemize}
\item
If $k\in X^*$ then $V_k^*AV_k\in\ra$ and
$\lim_{k\rarrow0}\|V_k^*AV_k-A\|=0$,
\item
If $x\in X$ then $U_xA\in\ra$ and $\lim_{x\rarrow0}\|(U_x-1)A\|=0$.
\end{itemize} 
In this case, there is a unique $X$-subalgebra $\ca\subset\cc(X)$
such that $\ra=\ca\rtimes X$, and this algebra is given by
\begin{equation}\label{eq:Land}
\ca =\ra_\flat:=\{\varphi\in\cc(X)\mid
\vphi(Q)^{(*)}\psi(P) \in {\ra},\; \forall \psi \in \co(X^*)\}.
\end{equation}
\end{theorem}
Note that, since $\ra$ is stable under taking adjoints, if we
replace $U_xA$ by $AU_x$ and $(U_x-1)A$ by $A(U_x-1)$ in the second
condition above we get an equivalent condition. If each element $A$
of a $C^*$-subalgebra $\ra\subset\rb(X)$ verifies the two conditions
of the theorem, we shall say that $\ra$  \emph{satisfies Landstad's
conditions}.

\medskip

The following reformulation of the second Landstad condition is
useful. 
\begin{lemma}\label{lm:fact} 
If $T \in {\rb}(X)$ then the next three assertions are equivalent:
\begin{itemize}
\item
$\lim_{x\rarrow 0}\|(U_x - 1) T\|= 0$,
\item 
$T=\psi(P)T_0$ for some $\psi\in\co (X^*)$ and $T_0\in\rb(X)$,
\item 
$\forall\varepsilon>0$ $\exists F\subset X^*$ with $X^*\setminus F$
compact and $\|\ind_F(P)T\|<\varepsilon$.
\end{itemize}
\end{lemma} 
\proof
It suffices to consider only the first two conditions.
If $T=\psi(P)T_0$ then
$$
\|(U_x - 1) T\|\leq\|(U_x - 1)\psi(P) \|\|T_0\|\leq\|T_0\|\sup_k
|(k(x)-1)\psi(k)|\rarrow0 \hspace{1mm}\mbox{as }x\rarrow0.
$$ To prove the converse assertion, let
$\rb_0=\{T\in\rb\mid\lim_{x\rarrow 0}\|(U_x - 1) T\|= 0\}$.  This is
clearly a closed subspace of $\rb$ such that
$\psi(P)\rb_0\subset\rb_0$ if $\psi\in\co(X^*)$.  By taking
$\what\psi(k)=|K|^{-1}\ind_K$ in (\ref{eq:fr}), where $K$ runs over
the family of compact neighborhoods of the origin in $X^*$, we
easily see that each $T\in\rb_0$ is a norm limit of operators of the
form $\psi(P)T$. Now the Cohen-Hewitt factorization theorem
\cite[Th.\ V.9.2]{FeD} shows that each $T\in\rb_0$ can be written
as $T=\psi(P)T_0$ with $\psi\in\co(X^*)$ and $T_0\in\rb_0$.  \qed

\begin{corollary}\label{co:crnd}
If $\ra$ is a crossed product then each $A\in\ra$ can be factorized
as $A=A_1\psi_1(P)=\psi_2(P)A_2$ with $A_i\in\ra$ and
$\psi_i\in\co(X^*)$. In particular, if $A\in\ra$ and $\psi$ is a
bounded continuous function on $X^*$ then $A\psi(P)$ and $\psi(P)A$
belong to $\ra$.
\end{corollary}

Theorem \ref{th:Land} allows us to give an intrinsic description of
some crossed products. By ``intrinsic'' we mean a description which
makes no reference to the crossed product operation.  Examples may
be found in Section \ref{s:exloc}, here we give the description of
the largest crossed product $\rc(X)$ which makes the connection with
the definition (\ref{eq:01}).
\begin{theorem}\label{th:dg}
The crossed product $\rc(X)=\cc(X)\rtimes X$ is given by
(\ref{eq:01}). 
\end{theorem}
For the proof, it suffices to note that the right hand side of
(\ref{eq:01}) is a $C^*$-algebra and to apply Theorem \ref{th:Land}.
It is useful to view the last condition in (\ref{eq:01}) from the
perspective of Lemma \ref{lm:fact}: this gives a precise meaning to
the fact that the operators from $\rc(X)$ tend to zero as
$P\rarrow\infty$.

\begin{remark}\label{re:ux}{\rm
    If $X=\mbr^n$ we see that $\rc(X)$ is the norm closed linear
    subspace of $\rb(X)$ generated by the operators
    $\varphi(Q)\psi(P)$ with $\varphi$ in the space of $C^\infty$
    functions which are bounded together with all their derivatives
    and $\psi$ in the space of $C^\infty$ functions with compact
    support.  So $\rc(X)$ is generated by a rather restricted class
    of pseudo-differential operators. In particular, $\rc(X)$ is the
    norm closure of the set of pseudo-differential operators with
    symbols of class $S^m$ if $m<0$ (see \cite[Definition 18.1.1]{H}
    and use \cite[Theorem 18.1.6]{H}).  From Proposition \ref{pr:da}
    it also follows that $\rc(X)$ is generated by a rather small
    class of elliptic operators.
}\end{remark}

As a consequence, we get an intrinsic description of the algebras of
quantum Hamiltonians, in the sense of Definition \ref{df:aqh}.

\begin{proposition}\label{pr:aqh}
A $C^*$-subalgebra $\ra\subset\rb(X)$ is a $C^*$-algebra of quantum
Hamiltonians if and only if $\ra\supset\rt(X)$ and
\begin{itemize}
\item
$
x\in X,k\in X^*, A\in\ra\Longrightarrow V_k^*AV_k 
\mbox{ and }U_xA \mbox{ belong to }\ra,
$
\item
$
\lim_{ k\rarrow 0}\|[A, V_k]\|=
\lim_{x\rarrow 0}\|(U_x - 1) A\|=0.
$
\end{itemize}
\end{proposition}

\begin{remark}\label{re:rk}{\rm
Observe that the classical Riesz-Kolmogorov compactness criterion
\begin{eqnarray*}
\rk(X)\hspace{-3mm} &=&\hspace{-3mm} \{T\in {\rb}(X)\mid \lim_{
  k\rarrow 0}\|(V_k-1)T\|=0 
\mbox{{ \rm  and} }
\lim_{ x\rarrow 0}\|(U_x - 1) T\|=0\}\nonumber\\
 &=&\hspace{-3mm} \{T\in {\rb}(X) \mid
T=\varphi(Q)S=\psi(P)R \;\mbox{ with } \varphi\in\co(X),
\psi\in\co (X^*)\\
&&\hspace{18mm} \mbox{and } S,R\in\rb(X)\}\nonumber
\end{eqnarray*}
is also an intrinsic characterization of a crossed product and
follows easily from Theorem \ref{th:Land} and Lemma \ref{lm:fact}
together with a similar fact with the group $U_x$ replaced by $V_k$.
In more intuitive terms, the compact operators are characterized by
the fact that they vanish when $P\rarrow\infty$ and
$Q\rarrow\infty$.
}\end{remark}

Now we show that the set of $C^*$-subalgebras of $\rb(X)$ which are
crossed products is stable under arbitrary intersections and that
the $C^*$-algebra generated by an arbitrary family of crossed
products is again a crossed product. We denote by
$C^*\big(\ccup_\lambda\cb_\lambda\big)$ the $C^*$-subalgebra
generated by a family of $C^*$-subalgebras $\cb_\lambda$.
\begin{theorem}\label{th:main}
  If $(\ca_\lambda)$ is an arbitrary family of $X$-subalgebras of
  $\cc(X)$ then $\ccap_\lambda\ca_\lambda$ and
  $C^*\big(\ccup_\lambda\ca_\lambda\big)$ are $X$-subalgebras and:
\begin{eqnarray}
\ccap_\lambda(\ca_\lambda\rtimes X) &=&
(\ccap_\lambda\ca_\lambda)\rtimes X,\label{eq:int}\\
C^*\big(\ccup_\lambda(\ca_\lambda\rtimes X)\big) &=&
C^*\big(\ccup_\lambda\ca_\lambda\big)\rtimes X.\label{eq:uni}
\end{eqnarray} 
\end{theorem}
\proof The fact that $\ccap_\lambda\ca_\lambda$ and
$C^*\big(\ccup_\lambda\ca_\lambda\big)$ are $X$-subalgebras is easy
to prove and the inclusions $\supset$ in (\ref{eq:int}) and
$\subset$ in (\ref{eq:uni}) are obvious. The proof of $\supset$ in
(\ref{eq:uni}) is elementary. Indeed, it suffices to show that
$\varphi(Q)\psi(P)$ belongs to the left hand side of (\ref{eq:uni})
if $\varphi\in C^*\big(\ccup_\lambda\ca_\lambda\big)$. Then we may
assume that $\varphi=\varphi_L=\prod_{\lambda\in L}\varphi_\lambda$
with $\varphi_\lambda\in\ca_\lambda$ and $L$ a finite set.
Let $\lambda\in L$ and $M=L\setminus\{\lambda\}$. Then Corollary
\ref{co:crnd} applied to
$\varphi_\lambda(Q)\psi(P)\in\ca_\lambda\rtimes X$ gives
$$
\varphi_L(Q)\psi(P)=\varphi_M(Q)\varphi_\lambda(Q)\psi(P)=
\varphi_M(Q)\psi_\lambda(P)A_\lambda
$$
for some $\psi_\lambda\in\co(X^*)$ and
$A_\lambda\in\ca_\lambda\rtimes X$. Repeating the argument with
$\varphi_L$ replaced by $\varphi_M$ we see that
$\varphi_L(Q)\psi(P)$ can be written as a product of elements of
$\ca_\lambda\rtimes X$ with $\lambda\in L$. This proves
(\ref{eq:uni}).

The inclusion $\subset$ in (\ref{eq:int}) is a deeper fact, it
depends on Theorem \ref{th:Land}. Let
$\ra_\lambda=\ca_\lambda\rtimes X$ and
$\ra=\ccap_\lambda\ra_\lambda$. It is easy to check that $\ra$
satisfies the two conditions of Theorem \ref{th:Land}, so
$\ra=\ca\rtimes X$ where $\ca$ is defined by (\ref{eq:Land}). If
$\varphi\in\cc(X)$ has the property $\varphi(Q)^{(*)}\psi(P)\in\ra$
for all $\psi\in\co(X^*)$ then we also have
$\varphi(Q)^{(*)}\psi(P)\in\ra_\lambda$ for all such $\psi$, hence
$\varphi\in(\ra_\lambda)_\flat=\ca_\lambda$ for each $\lambda$. Thus
$\varphi\in\ccap_\lambda\ca_\lambda$, hence
$\ca\subset\ccap_\lambda\ca_\lambda$. 
\qed

\begin{proposition}\label{pr:var}
  If $\ca,\cj$ are $X$-subalgebras then $\cj$ is an ideal of $\ca$
  if and only if $\cj\rtimes X$ is an ideal of $\ca\rtimes X$.
\end{proposition}
\proof
The fact that
``$\cj\subset\ca$ ideal $\Rightarrow\cj\rtimes X\subset\ca\rtimes X$
ideal''  follows easily from Lemma \ref{lm:crp}. For the converse it
suffices to show that if $\rj,\ra$ are crossed products and if $\rj$
is an ideal of $\ra$, then $\rj_\flat$ is an ideal of $\ra_\flat$.
Let $\xi\in\rj_\flat$ and $\varphi\in\ra_\flat$, then by Corollary
\ref{co:crnd}, for each $\psi\in\co(X^*)$ we can factorize
$\varphi(Q)\psi(P)=\psi_0(P)S$ for some $\psi_0\in\co(X^*)$ and
$S\in\ra$. Thus $(\xi\varphi)(Q)\psi(P)=\xi(Q)\psi_0(P)S\in\cj$
because $\xi(Q)\psi_0(P)\in\rj$ and $\rj$ is an ideal of $\ra$,
hence $\xi\varphi\in\rj_\flat$.
\qed

\begin{proposition}\label{pr:ds}
  Assume that $\ca,\cb,\cj$ are $X$-subalgebras of $\cc(X)$ such
  that $\ca=\cb+\cj$ and that $\cj$ is an ideal in $\ca$. Then
  $\cj\rtimes X$ is an ideal in $\ca\rtimes X$ and $\ca\rtimes
  X=\cb\rtimes X+\cj\rtimes X$. If $\ca=\cb+\cj$ is a linear direct
  sum, then $\ca\rtimes X=\cb\rtimes X+\cj\rtimes X$ is a linear
  direct sum.
\end{proposition}
\proof We know that $\cj\rtimes X$ is an ideal in $\ca\rtimes X$ and
that $\cb\rtimes X\subset\ca\rtimes X$ is a $C^*$-subalgebra. From
\cite[Corollary 1.8.4]{Dix} we see that $\cb\rtimes X+\cj\rtimes X$
is closed in $\ca\rtimes X$, and since it is clearly dense in
$\ca\rtimes X$, we have $\ca\rtimes X=\cb\rtimes X+\cj\rtimes X$.
Finally,
$$
(\cb\rtimes X)\cap(\cj\rtimes X)=
\big(\cb\cap\cj\big)\rtimes X
$$
because of (\ref{eq:int}), and this is $\{0\}$ if
$\cb\cap\cj=\{0\}$.  . 
\qed

We mention a fact which is useful in the explicit computations of
$\ra_\flat$. 
\begin{remark}\label{re:nic}{\rm
It is clear that in (\ref{eq:Land}) it suffices to consider only
$\psi\in\Cc(X^*)$. Since, by Corollary \ref{co:crnd}, a crossed
product is a $\co(X^*)$-bimodule, we get the following simpler
description of $\ca$: \emph{if there is $\xi\in\co(X^*)$ such that
$\xi(k)\neq0$ for all $k\in X^*$, then}
\begin{equation}\label{eq:Landi}
\ca =\{\varphi\in\cc(X)\mid \vphi(Q)^{(*)}\xi(P) \in {\ra}\}.
\end{equation}
Such a $\xi$ exists if and only if $X^*$ is
$\sigma$-compact (i.e.\ a countable union of compact sets). 
}\end{remark}

\begin{remark}\label{re:gog}{\rm
    The following comment on the first Landstad condition is of some
    interest, although it does not play any r\^ole in our
    arguments. Let $C^{\rm u}(Q)$ be the set of $S\in\rb(X)$ which
    verify the first Landstad condition; this is clearly a
    $C^*$-algebra.  Let us say that an operator $S\in\rb(X)$ is of
    \emph{finite range} (not rank!)  if there is a compact
    neighborhood $\Lambda$ of the origin such that
    $S\ind_K(Q)=\ind_{K+\Lambda}(Q)S\ind_K(Q)$ for any Borel set
    $K$. Clearly, the set of finite range operators is a
    $*$-subalgebra of $\rb(X)$ and it can be shown that the set of
    finite range operators which belong to $C^{\rm u}(Q)$ is dense
    in $C^{\rm u}(Q)$. Moreover, under quite general conditions on
    $X$ it can be shown that a finite range operator belongs to
    $C^{\rm u}(Q)$ (this is probably always true). Thus, if
    $X=\mbr^n$ or if $X$ is a discrete group for example, then
    $C^{\rm u}(Q)$ is exactly the norm closure of the set of finite
    range operators.  These questions are treated in
    \cite[Propositions 4.11 and 4.12]{GG2}.  }\end{remark}

\section{Affiliation to \mbox{\secrond C $\,(X)$}} 
\label{s:opaf} 
\protect\setcounter{equation}{0}
Theorem  \ref{th:introo} shows that the essential spectrum of the
operators affiliated to $\rc(X)$ is determined by their
localizations at infinity, so it is important to show that the class
of operators affiliated to $\rc(X)$ is large. We show in this
section that this is indeed the case: singular perturbations of
hypoelliptic self-adjoint pseudo-differential operators are
affiliated to $\rc(X)$. If one thinks of $\rc(X)$ as the
$C^*$-algebra generated by the operators of the form
$\varphi(Q)\psi(P)$ with $\varphi\in\cc(X),\;\psi\in \co (X)$, this
is far from obvious.

\medskip

In the rest of the section we fix a finite dimensional Hilbert space
$\bfe$, we set ${\ch}=L^2(X;\bfe)$ and define $\rc=\rc(X)$ as in
(\ref{eq:01}).  Since the adjoint space\,\symbolfootnote[2]{\ The
adjoint space (space of antilinear continuous forms) of a Hilbert
space $\cg$ is denoted $\cg^*$ and if $u\in\cg$ and $v\in\cg^*$ then
we set $v(u)=\langle u,v\rangle$.}  ${\ch}^*$ is identified with
${\ch}$ by using the Riesz isomorphism, if ${\cg}$ is a Hilbert
space with $\cg\subset\ch$ continuously and densely then we get a
similar embedding ${\ch}\subset{\cg}^*$.

Let $H$ be a self-adjoint operator on ${\ch}$ and let
$z\in\rho(H)$. As we saw in (\ref{eq:intro1}), $H$ is affiliated to
$\rc$ if and only if
\begin{equation}\label{e:3.0} 
\lim_{x\rarrow0}\|(U_x-1)(H-z)^{-1}\|=0 \mbox{\ \ and\ \ } 
\lim_{k\rarrow0}\|[V_k,(H-z)^{-1}]\|=0.
\end{equation}
In the next subsection we make an abstract analysis of these
relations and in Subsection \ref{ss:af2} we give concrete examples. 

\PAR \label{ss:af1}
A function $\theta:X^*\rarrow\mbr$ such that
$\lim_{k\rarrow\infty}\theta(k)=+\infty$ will be called {\em
  divergent}.  Lemma \ref{lm:fact} and an interpolation argument
give:

\begin{lemma}\label{l:3.0} 
  The first condition in (\ref{e:3.0}) is fulfilled if and only
  there are $s>0$ and a continuous divergent function $\theta$ such
  that $D(|H|^s)\subset D(\theta(P))$. And then this property holds
  for all real numbers $s>0$.
\end{lemma}

Let $S(\bfe)$ be the space of symmetric operators on $\bfe$. If
$h:X^*\rarrow S(\bfe)$ is Borel, then $h(P)$ is the self-adjoint
operator on $\ch$ such that $\cf h(P)\cf^*$ is the operator of
multiplication by $h$ in $L^2(X^*;\bfe)$.  If
$\lim_{k\rarrow\infty}{\rm dist}(0,\sigma(h(k)))=\infty$ then we
write $\lim_{k\rarrow\infty}h(k)=\infty$. This property is
equivalent to $\lim_{k\rarrow\infty}\|(h(k)+i)^{-1}\|=0$ and implies
$\lim_{k\rarrow\infty}\|\vf(h(k))\|=0$ for all $\vf\in\co(\mbr)$.
If $\bfe=\mbc$ this means $\lim_{k\rarrow\infty}|h(k)|=\infty$.

\begin{corollary}\label{c:3.1} 
  If $h:X^*\rarrow S(\bfe)$ is a continuous function on $X^*$ then
  $h(P)$ is affiliated to $\rc$ if and only if
  $\lim_{k\rarrow\infty}h(k)=\infty$.
\end{corollary} 

In particular, if $X=\mbr^2$ then the operator
$H=\partial_1^2-\partial_2^2$ is not affiliated to $\rc$.
A second interesting operator not affiliated to $\rc$ is
$H=(\partial_1+ix_2)^2+(\partial_2+ix_1)^2$.

\medskip

We now give the simplest affiliation criterion.

\begin{proposition}\label{pr:sil}
Assume that $H_0$ is a self-adjoint operator affiliated to $\rc$ and
that $V$ is a bounded symmetric operator such that
$\lim_{k\rarrow0}\|[V_k,V]\|=0$. Then $H=H_0+V$ is a self-adjoint
operator affiliated to $\rc$.
\end{proposition}
\proof Let $R=(H+i)^{-1}$ and $R_0=(H_0+i)^{-1}$.  Since $H$ and
$H_0$ have the same domain and $R[1+VR_0]=R_0$, the operator
$1+VR_0$ is invertible. On the other hand, $1+VR_0$ clearly
satisfies the second condition in (\ref{e:3.0}), hence its inverse
verifies it too. From $R=R_0[1+VR_0]^{-1}$ we see that both
conditions in (\ref{e:3.0}) are satisfied.  \qed

From now on we consider only situations when $V$ is not bounded.

\begin{proposition}\label{pr:bt}
  Let $H$ be a self-adjoint operator such that $V_k D(H)\subset
  D(H)$ for all $k$. Then $H$ is affiliated to $\rc$ if and only if
  $D(H)\subset D(\theta(P))$ for some continuous divergent function
  $\theta$ and
\begin{equation}\label{eq:bt}
\lim_{k\rarrow0}\|[V_k,H]\|_{D(H)\rarrow D(H)^*}=0.
\end{equation}
\end{proposition}
\proof It is clear that $V_k D(H)\subset D(H)$ for all $k$ if and
only if $V_k$ extends to a continuous map $D(H)^*\rarrow D(H)^*$ for
each $k$, and then we have in $B(\ch)$:
\begin{equation}\label{e:3.1}
[V_k,(H-z)^{-1}]=(H-z)^{-1}[H,V_k](H-z)^{-1}.
\end{equation}
The operator $[H,V_k]$ belongs to $B(D(H),\ch)$ and so we can
consider it as a map $D(H)\rarrow D(H)^*$. But $(H-z)^{-1}$ is an
isomorphism ${\ch}\rarrow D(H)$ and $D(H)^*\rarrow{\ch}$. To end the
proof it suffices to use Lemma \ref{l:3.0}.
\qed

We shall give below three perturbative criteria of affiliation: we
add to an operator affiliated to $\rc$ an operator which is not
necessarily affiliated to it.  Note that functions of $Q$ are never
affiliated to $\rc$. First we consider operator bounded
perturbations.

\begin{corollary}\label{co:bt}
  Let $H_0$ be a self-adjoint operator affiliated to $\rc$ such that
  $V_k D(H_0)\subset D(H_0)$ for all $k$.  Let $V$ be a symmetric
  operator with domain $D(H_0)$ and such that $H=H_0+V$ is
  self-adjoint. Then $H$ is affiliated to $\rc$ if and only if 
\begin{equation}\label{eq:btc}
\lim_{k\rarrow0}\|[V_k,V]\|_{D(H_0)\rarrow D(H_0)^*}=0.
\end{equation}
\end{corollary}

Now we want to consider form bounded perturbations in a generalized
sense (in order to cover not semibounded operators).  Let $H$ be a
self-adjoint operator on ${\ch}$. We say that a Hilbert space
${\cg}$ is {\em adapted} to $H$ if $D(H)\subset{\cg}\subset {\ch}$
continuously and densely and $H-z$ extends to an isomorphism
${\cg}\rarrow{\cg}^*$ for some (hence for all) $z\in\CC$ outside the
spectrum of $H$.  Then $H$ extends to a continuous operator
${\cg}\rarrow{\cg}^*$ and we keep the notation $H$ for the extended
map. It is not difficult to show that \emph{if $H$ is a semibounded
operator then ${\cg}$ is adapted to $H$ if and only if ${\cg}=
D(|H|^{1/2})$ as topological vector spaces}, see \cite[page 47]{GG2}.
But in general, for example in the case of Dirac operators, this is
not the case.  Observe that
$$D(H)\subset{\cg}\subset{\ch}\subset{\cg}^*\subset D(H)^*$$
continuously and densely, in particular $B({\cg},{\cg}^*)\subset
B(D(H),D(H)^*)$.  It is then clear that one has $V_k{\cg}\subset
{\cg}$ for all $k$ if and only if $V_k$ extends to a continuous map
${\cg}^*\rarrow{\cg}^*$ for each $k$, and in this case the identity
(\ref{e:3.1}) is valid in $B({\cg}^*,{\cg})$.  The operator
$[H,V_k]$ belongs to $B({\cg},{\cg}^*)$ and so we can consider it as
a map $D(H)\rarrow D(H)^*$. But $(H-z)^{-1}$ is an isomorphism
${\ch}\rarrow D(H)$ and $D(H)^*\rarrow{\ch}$.  Thus:

\begin{proposition}\label{l:3.2}
  Let $H$ be a self-adjoint operator on ${\ch}$ such that
  $D(H)\subset D(\theta(P))$ for some continuous divergent function
  $\theta$. Assume that ${\cg}$ is a Hilbert space adapted to $H$
  and that $V_k{\cg}\subset {\cg}$ for all $k$.  Then $H$ is
  affiliated to $\rc$ if and only if
\begin{equation}\label{e:3.2}
\lim_{k\rarrow0}\|[V_k,H]\|_{D(H)\rarrow D(H)^*}=0.
\end{equation}
\end{proposition}
In many situations of interest in quantum mechanics the domain of
the Hamiltonian is difficult to determine while its form domain is
quite explicit. For this reason the following condition stronger
than (\ref{e:3.2}) is often more convenient:
\begin{equation}\label{e:3.22}
\lim_{k\rarrow0}\|[V_k,H]\|_{{\cg}\rarrow {\cg}^*}=0.
\end{equation}
We shall use this in the following context.

\begin{definition}\label{df:std}
Let $H_0$ be a self-adjoint operator on $\ch$ and let $\cg$ be a
Hilbert space adapted to it. We say that $V$ is a {\em standard form
perturbation} of $H_0$ if $V$ is a continuous symmetric sesquilinear
form on $\cg$ and there are numbers $\mu \in [0, 1)$ and $\nu\geq0$
such that one of the following conditions is satisfied:\\
(1) $\pm V \leq  \mu |H_0|+ \nu$ as forms on ${\cg}$\\
(2) $H_0$ is bounded from below and 
$V \geq - \mu H_0- \nu$ as forms on ${\cg}$.
\end{definition}
Then the operator $H=H_0+V:\cg\rarrow\cg^*$ is such that its
restriction to $D(H)=\{u\in\cg\mid Hu\in\ch\}$ \label{dirac} is a
self-adjoint operator on $\ch$ (and will also be denoted $H$) and
$\cg$ is adapted to $H$ too (see \cite{DaG3}).  Note that $V$ is
seen as a continuous operator ${\cg}\rarrow{\cg}^*$.

\begin{corollary}\label{co:ts}
  Let $H_0$ and $V$ as above. We assume that $\cg\subset
  D(\theta(P))$ for some continuous divergent function $\theta$,
  that $V_k\cg\subset\cg$ for all $k$, and
$\lim_{k\rarrow0}\|[V_k,H]\|_{{\cg}\rarrow {\cg}^*}$.
Then $H$ is affiliated to $\rc$.
\end{corollary}

The next result covers perturbations of $H_0$ which are not
dominated by $H_0$.

\begin{proposition}\label{p:3.1}
Let $H_1,H_2$ be bounded from below self-adjoint operators
and let us denote ${\cg}_i=D(|H_i|^{1/2})$.
Assume that ${\cg}\equiv{\cg}_1 \cap {\cg}_2$ is dense in ${\ch}$
and let $H=H_1+H_2$, the sum being defined in form sense.
Let us suppose that
${\cg}\subset D(\theta(P))$ for some continuous divergent function
$\theta$ and that for $i=1,2$ we have
$V_k {\cg}_i\subset {\cg}_i$  and
$\lim_{k\rarrow0}\|[V_k,H_i]\|_{B({\cg}_i,{\cg}_i^*)}=0$.
Then $H$ is affiliated to $\rc$.
\end{proposition} 
\proof Let us recall that the form sum $H=H_1+H_2$ is defined as the
unique self-adjoint operator such that $ D(|H|^{1/2})={\cg}$ and
$\langle u,Hu\rangle=\langle u,H_1u\rangle+\langle u,H_2u\rangle$
for all $u\in{\cg}$. The topology of ${\cg}$ is the intersection
topology of ${\cg}_1$ and ${\cg}_2$, so thinking in terms of
sesquilinear forms we see that
$$
\|[V_k,H]\|_{B({\cg},{\cg}^*)}\leq
C\|[V_k,H_1]\|_{B({\cg}_1,{\cg}_1^*)}+
C\|[V_k,H_2]\|_{B({\cg}_2,{\cg}_2^*)}
$$
for some constant $C$. Hence (\ref{e:3.22}) is satisfied.
\qed
\PAR \label{ss:af2}
If $w$ is a continuous divergent function on $X^*$ let
${\ch}^w\equiv{\ch}^w(X)=D(w(P))$ equipped with the graph norm.  We
saw in Lemma \ref{l:3.0} that if $H$ is affiliated to $\rc$ then
$D(|H|^{1/2})\subset {\ch}^w$ for such a $w$.  We consider now
operators whose form domain is {\em equal} to some ${\ch}^w$.

We say that $w$ is a {\em weight}\,\symbolfootnote[2]{\ The
  terminology is suggested by that from \cite[Section 10.1]{H}, cf.\
  the remark after Theorem 10.1.5.}  if $w:X^*\rarrow]0,\infty[$ is
  continuous and $w(k+p)\leq \omega(k)w(p)$ for some function
  $\omega$ and all $k,p\in X^*$. If $\omega$ is the smallest
  function satisfying such an estimate, then
  $\omega(k+p)\leq\omega(k)\omega(p)$. From now on we shall assume
  that $\omega$ satisfies this submultiplicativity condition. We
  also say {\em $\omega$-weight} if we need to be more specific. If
  $X=\mbr^n$ then a standard choice is $w(k)=\braket{k}^s$ for some
  real $s$.

\begin{lemma}\label{l:3.3}
  A continuous divergent function $w$ on $X^*$ is an $\omega$-weight
  if and only if $V_k{\ch}^w\subset{\ch}^w$ and
  $\norm{V_k}_{B({\ch}^w)}\leq\omega(k)$ for all $k$.
\end{lemma}
\proof We may take $\|w(P)u\|$ as norm on ${\ch}^w$.  From
$V_{k}^*w(P)V_k=w(P+k)$ we see that we have $V_k{\ch}^w\subset
{\ch}^w$ and $\norm{V_k}_{B({\ch}^w)}\leq \omega(k)$ if and only if
$\|w(P+k)u\|\leq \omega(k)\|w(P)u\|$ for all $u$, which is
equivalent to $w(k+p)\leq \omega(k)w(p)$ for all $k,p$.  \qed

\begin{proposition}\label{p:3.2}
A self-adjoint operator on ${\ch}$ with
$D(|H|^{1/2})={\ch}^w$ for some divergent weight $w$ and such that 
$\lim_{k\rarrow0}\|[V_k,H]\|_{B({\ch}^w,{\ch}^{w*})}=0$
is affiliated to $\rc$.
\end{proposition}
This is an immediate consequence of Proposition \ref{l:3.2}. 

\begin{proposition}\label{pr:3.3}
  Let $H$ be as in Proposition \ref{p:3.2} and bounded from below.
  Let $V\in L^1_{\rm loc}(X)$ be a real function whose negative part
  is form bounded with respect to $H$ with relative bound strictly
  less than 1. Then the self-adjoint operator $H+V(Q)$ (form sum) is
  affiliated to $\rc$.
\end{proposition}
\proof Let $V_+,V_-$ be the positive and negative parts of $V$, then
we define the sum as $(H-V_-)+V_+$ and apply successively
Propositions \ref{l:3.2} and \ref{p:3.1}.  \qed

\begin{example}\label{ex:ell}{\rm
    The most common situation is $X=\mbr^n$ and $w(k)=\braket{k}^s$
    for some real $s>0$. Then ${\ch}^w$ is the usual Sobolev space
    ${\ch}^s$ and typical operators satisfying the conditions of the
    Proposition \ref{p:3.2} are the uniformly elliptic operators of
    order $2s$.  For example, let $s=m\geq1$ integer and
    $$L=\textstyle\sum_{|\alpha|,|\beta|\leq m} P^\alpha
    a_{\alpha\beta}P^\beta$$ for some measurable functions
    $a_{\alpha\beta}:X\rarrow B(\bfe)$ such that the operator of
    multiplication by $a_{\alpha\beta}$ is a continuous map
    $\ch^{m-|\beta|}\rarrow\ch^{|\alpha|-m}$ (this is a very general
    assumption which allows one to give a meaning to the
    differential expression $L$). Then $L:\ch^m\rarrow\ch^{-m}$ is a
    continuous map and $V^*_kLV_k$ is a polynomial in $k$. If
    $\langle u,Lu\rangle\geq\mu\|u\|^2_{\ch^m}-\nu\|u\|^2$ for some
    $\mu,\nu>0$, then $L$ induces a self-adjoint operator in $\ch$
    which is affiliated to $\rc$.  }\end{example}

\begin{example}\label{ex:fiz}{\rm
    We give an explicit example of physical interest in the case
    $s=1$. Let
\begin{equation}\label{e:3.}
H=\textstyle\sum_{i,j}(P_i-A_i)G_{ij}(P_j-A_j)+V\equiv (P-A)G(P-A)+V
\end{equation}
where $G_{ij},A_i,V$ are (the operators of multiplication by)
locally integrable real functions having the following properties
($\|\cdot\|_1$ is the norm of ${\ch}^1$):\\
(1) $G_{ij}\in L^\infty(X)$, the matrix $G(x)=(G_{ij}(x))$ is
  symmetric and $G(x)\geq\nu>0$,\\
(2) for each $\ve>0$ there is $\delta\in\mbr$ such that
  $\|A_ju\|\leq\ve\|u\|_1+\delta\|u\|$ for all $u\in{\ch}^1$,\\
(3) if $V_-$ is the negative part of $V$ then for each $\ve>0$
  there is a real number $\delta$ such that $\langle
  u,V_-u\rangle\leq\ve\|u\|_1^2+\delta\|u\|^2$ for all
  $u\in{\ch}^1$.\\
Note that the conditions on $A_j$ and $V_-$ are satisfied if there
is $s<1$ such that $\|A_ju\|\leq C\|u\|_s$ and $\langle
u,V_-u\rangle\leq C\|u\|_s^2$.  Then $H$ is affiliated to $\rc$.
Indeed, observe first that $H_0\equiv(P-A)G(P-A)$ is a self-adjoint
operator with form domain equal to ${\ch}^1$, because there is
$\delta$ such that: 
$$
\langle
u,H_0u\rangle\geq\nu\|(P-A)u\|^2\geq\frac{\nu}{2}\|Pu\|^2-
\nu\|Au\|^2\geq\frac{\nu}{4}\|Pu\|^2-\delta\|u\|^2
$$ 
Hence, according to Proposition \ref{pr:3.3}, it suffices
to prove that $H_0$ is affiliated to $\rc$. But
\begin{eqnarray*}
V_k^*H_0V_k &=&(P-A+k)G(P-A+k)\\
&=&H_0+kG(P-A)+(P-A)Gk+kGk.
\end{eqnarray*}
Thus $\|V_k^*H_0V_k-H_0\|_{B({\ch}^1,{\ch}^{-1})}\leq C(|k|+|k|^2)$
so we can use Proposition \ref{p:3.2}.
  }\end{example}

\begin{remark}\label{r:3.2}{\rm
    Let us consider the operator $H_0$ under the more general
    condition $A_j\in L^2_{\rm loc}(X)$. More precisely, $H_0$ is
    the positive self-adjoint operator associated to the closed
    quadratic form $\|(P-A)u\|^2$ whose domain is the set ${\cg}$ of
    $u\in{\ch}$ such that the distributions $(P_j-A_j)u$ belong to
    ${\ch}$. The preceding computation shows that
    $V_k{\cg}\subset{\cg}$ and that (\ref{e:3.22}) is satisfied.
    Hence $H_0$ is affiliated to $\rc$ if and only if ${\cg}\subset
    \theta(P)$ for some continuous divergent function $\theta$. But
    this cannot be true without some boundedness conditions on $A$
    at infinity.  }\end{remark}

As a final example we consider singular perturbations of $h(P)$,
where $h:X^*\rarrow\mbr$ is a continuous divergent function and $X$
is an arbitrary group.  Let ${\cg}=D(|h(P)|^{1/2})$.  Two functions
$u,v$ on a neighborhood of infinity will be called {\em equivalent}
if they satisfy $c_1|u(k)|\leq |v(k)|\leq c_2 |u(k)|$ for all large
$k$ and some constants $c_1,c_2>0$.  It is clear that
${\cg}={\ch}^w$ if and only if $h$ is equivalent to $w^2$.  Then
Proposition \ref{pr:3.3} implies:
 
\begin{proposition}\label{p:3.3}
  Let $h:X^*\rarrow\mbr$ be a divergent function 
  equivalent to a weight and such that
\begin{equation}\label{e:3.x}
\lim_{k\rarrow0}\sup_p\frac{|h(p+k)-h(p)|}{1+|h(p)|}=0.
\end{equation}
Let $W$ be a standard form perturbation of $h(P)$ with
$\lim_{k\rarrow 0}\|[V_k,W]\|_{B({\cg},{\cg}^*)}=0$ and define
$H_0=h(P)+W$ as a form sum. Let $V\in L^1_{\rm loc}(X)$ real
and such that $V_-\leq\mu H_0+\nu$ on $\cg$ for some $\mu<1,\nu>0$.
Then the form sum $H=H_0+V(Q)$ is a self-adjoint operator affiliated
to  $\rc$.
\end{proposition}

\begin{example}\label{ex:hypoe}{\rm
    Let $X=\mbr^n$ and assume that $h$ is of class $C^1$ and
    satisfies $|h'(k)|\leq C(1+|h(k)|)$. Then (\ref{e:3.x}) is
    fulfilled because
$$
|h(p+k)-h(p)|\leq\sup_{0<\theta<1}|h'(p+\theta k)||k|\leq
C\big(1+\sup_{0<\theta<1}|h(p+\theta k)|\big)|k|
$$
which is $\leq C'(1+|h(p)|)|k|$ if $|k|\leq1$ because $h$ is a
equivalent to a weight. On the other hand, assume that $h$ is of
class $C^m$ for some integer $m\geq1$ and that we have: (1)
$\lim_{k\rarrow\infty}h(k)=+\infty$, (2) the derivatives of order
$m$ of $h$ are bounded, (3) $\sum_{|\alpha|\leq
  m}|h^{(\alpha)}(k)|\leq C(1+|h(k)|)$.  Then from \cite[p.\ 
342--343]{ABG} we get that $h$ is equivalent to a weight.
\emph{Any real hypoelliptic polynomial satisfies all these
conditions}, see Definition 11.1.2 and Theorem 11.1.3 in \cite{H}.
}\end{example}

\section{Localizations at infinity}
\label{s:LIF}
\protect\setcounter{equation}{0}
In this section we prove our main result, Theorem \ref{th:ess}, and
some easy consequences.
\PAR                      \label{ss:LIF2}
We define first the localizations at infinity for functions in
$\cc(X)$. We denote $\cc_s(X)$ the space $\cc(X)$ equipped with the
topology given by the seminorms $\|\vf\|_\theta=\|\vf\theta\|$ with
$\theta\in\co(X)$ (this is the strict topology associated to the
essential ideal $\co(X)$).

\begin{lemma}\label{lm:stop}
  If $\vf\in\cc(X)$ and $\vk\in\delta X$ then
  $\vk.\vf(y):=\lim_{x\rarrow\vk}\vf(x+y)$ exists locally uniformly
  in $y\in X$. Equivalently, we have $x.\vf\rarrow\vk.\vf$ in
  $\cc_s(X)$ if $x\rarrow\vk$ in $\gamma X$.  The function $\vk.\vf$
  belongs to $\cc(X)$ and we have $(\vk.\vf)(y)=\vk(y.\vf)$.
\end{lemma}
\proof 
Since $\vf$ is a bounded function, we have
$$
\lim_{x\rarrow\vk}\vf(x+y)=\lim_{x\rarrow\vk}(y.\vf)(x)=
\vk(y.\vf)
$$
by taking into account the two interpretations of $\vk$. Then we
use Lemma \ref{lm:easy}.
\qed

Thus $\vk.\vf\in\cc(X)$ is well defined for all $\vk\in \gamma X$
(if $\vk=x\in X$, see page \pageref{act}) and all $\vf\in\cc(X)$.
The next lemma is a slight improvement of Lemma \ref{lm:stop}, it
will allow us to give a completely elementary proof of Theorem
\ref{th:gn} (see the remark after the proof of the theorem). Note
that the relation $(\vk.\vf)(y)=\vk(y.\vf)$ remains true for all
$\vk\in\gamma X$ if we interpret $x\in X$ as a character of
$\ell_\infty(X)$. Since $y.\vf\in\cc(X)$ we see that $\vk.\vf$
depends in fact only on the class of $\vk$ in $\beta_{\rm u}X$, cf.\ 
Subsection \ref{ss:pd5}. We shall keep the notation $\vk.\vf$ even
if $\vk\in\beta_{\rm u}X$.  Recall that $X\subset\beta_{\rm u}X$ is
an open dense subset.

\begin{lemma}\label{lm:sflu}
  Let $\varphi \in \cc(X)$. Then $X\ni x\mapsto x.\vf\in\cc(X)$
  extends to a continuous function $\beta_{\rm u}X \ni \vk \mapsto
  \vk.\vf \in \cc_s(X)$.  We have $\vk.\varphi(y)=\vk(y.\vf)$ for
  all $y\in X$.
\end{lemma}
\proof For $\vk \in \beta_{\rm u}X = \sigma(\cc(X))$ the
function $\vk.\varphi$ is given by $\vk.\varphi(y)=\vk(y.\vf)$,
$y\in X$.  It is easy to check directly that $\vk.\vf$ so defined
belongs to $\cc(X)$: we have $|\vk(y.\vf)| \leq \|y.\vf\| =
\|\vf\|$ and
$$
|\vk(y.\vf) - \vk(z.\vf)| = |\vk(y.\vf -z.\vf)|
\leq
\|y.\vf - z.\vf\| = \|(y-z).\vf - \vf\|.
$$
It remains to prove that $\vk \mapsto \vk.\vf\,\theta \in
\cc_0(X)$ is continuous for any $\theta \in \cc_0(X)$, i.e. that for
each $\cchi \in \beta_{\rm u}X$, each $\ve > 0$ and each $\theta \in
\cc_0(X)$ there is a neighborhood $V$ of $\cchi$ in $\beta_{\rm
  u}X$ such that $\|(\vk.\vf - \cchi.\vf)\theta\|<\ve$ if $\vk \in
V$.  Since $\theta$ is $C_0$, it will suffice to prove that for each
$\cchi$ and $\ve$ as before and each compact set $K\subset X$ there
is a neighborhood $V$ of $\cchi$ such that $\vk\in V$ implies
$|\vk(y.\vf) - \cchi(y.\vf)|<\ve$ for $y\in K$.  But the map $y
\mapsto y.\vf \in \cc(X)$ is norm continuous, thus $\{y.\vf \mid
y\in K\}$ is a compact subset of $\cc(X)$. Hence there is a finite
subset $Z$ of $K$ such that for each $y\in K$ we have $\min_{z\in
  Z}\|y.\vf - z.\vf\|<\ve$.  Thus for each $y\in K$ and $z\in Z$ we
have
\begin{eqnarray*}
|\vk(y.\vf) - \cchi(y.\vf)|
&=&
|\vk(y.\vf - z.\vf) + \vk(z.\vf) - \cchi(z.\vf) + \cchi(z.\vf -
y.\vf )|\\ 
&<&
2\ve + |\vk(z.\vf)- \cchi(z.\vf)|.
\end{eqnarray*}
Now, if we take $V= \{\vk \in \beta_{\rm u}X \mid \sup_{z \in
  Z}|\vk(z.\vf)- \cchi(z.\vf)|<\ve \}$, then $V$ is a neighborhood
of $\cchi$ in $\beta_{\rm u}X$ because $Z$ is a finite set, and for
each $\vk \in V$ and each $y\in K$ we have $|\vk(y.\vf) -
\cchi(y.\vf)|<3\ve$.
\qed

\begin{lemma}\label{lm:linf}
If $\vf\in\cc(X)$ then $\vk.\vf=0$ for all $\vk\in\delta X$ if and
only if $\vf\in\co(X)$.
\end{lemma}
\proof If $\vk.\vf=0$ for all $\vk\in\delta X$ then
$\vk(\vf)=(\vk.\vf)(0)=0$ for such $\vk$.  If $\vf\nin\co(X)$ then
there is a number $a>0$ such that the set $U=\{x\mid|\vf(x)|>a\}$ is
not relatively compact.  Since $U\cap V\neq\emptyset$ for each $V$
with relatively compact complement, we see that the family of sets
$U\cap V$ is a filter basis and the filter $\gof$ it generates is
finer than Fr\'echet and contains $U$. Let $\vk$ be any ultrafilter
finer than $\gof$, then $\vk\in\delta X$ and
$\vk(\ind_U)=1$. Finally, from $|\vf|\geq a\ind_F$ we get
$|\vk(\vf)|^2=\vk(|\vf|^2)\geq a^2\vk(\ind_V)=a^2$, so we cannot
have $\vk(\vf)=0$.  \qed

\begin{definition}\label{df:linf}
If $\vf\in\cc(X)$ and $\vk\in\delta X$ then the function
$\vk.\vf\in\cc(X)$ is the \emph{localization of $\vf$ at $\vk$}.
And $\ell(\vf):=\{\vk.\vf\mid\vk\in\delta X\}\subset\cc(X)$ is
the set of \emph{localizations of $\vf$ at infinity}.
\end{definition}

For each $\vk\in\delta X$ let $\tau_\vk:\cc(X)\rarrow\cc(X)$ be
given by $\tau_\vk(\vf)=\vk.\vf$. Clearly this is a unital
morphism and, since the property $x.(\vk.\vf)=\vk.(x.\vf)$ is easy
to check,  $\tau_\vk$ is in fact an $X$-morphism.
By Lemma \ref{lm:linf} we have
\begin{equation}\label{eq:gid}
\ccap_{\vk\in\delta X}\ker\tau_\vk=\co(X).
\end{equation}
Note that
\emph{$\ker\tau_\vk$ is the maximal $X$-ideal included in the maximal
ideal $\ker\vk$ of $\cc(X)$}.

\begin{remark}\label{re:com}{\rm
In general $\tau_\vk\tau_\chi\neq\tau_\chi\tau_\vk$.
}\end{remark}

\PAR                      \label{ss:LIF3}
In this subsection we extend the notion of localization to operators
in $\rc(X)$.

\begin{definition}\label{df:stop}
Let $\rc_s(X)$ be the space $\rc(X)$ equipped with
the topology defined by the family of seminorms
$\|T\|_\theta=\|T\theta(Q)\|+\|\theta(Q)T\|$ with $\theta\in\co(X)$.
\end{definition}
Note that if $X$ is $\sigma$-compact then there is $\theta\in\co(X)$
with $\theta(x)> 0$ for all $x\in X$ and then $\|\cdot\|_\theta$ is
a norm on $\rc(X)$ which induces on bounded subsets of $\rc(X)$ the
topology of $\rc_s(X)$. In any case, the topology of $\rc_s(X)$ is
finer than the strong operator topology induced by $\rb(X)$. Note
also that the topology of $\rc_s(X)$ does not depend on any Hilbert
space realization of $\rc(X)$ because $\rc(X)$ is a
$\cc(X)$-bimodule and $\co(X)$ is an ideal of $\cc(X)$. Finally,
observe that we could consider on $\rc(X)$ the (intrinsically
defined) strict topology associated to the ideal $\rk(X)$; this is
weaker than that of $\rc_s(X)$ and finer than the strong operator
topology (but coincides with it on bounded sets).

\begin{remark}\label{re:simon}{\rm
    That this is the natural topology in our context should have
    been clear for us a long time ago, since it is induced by the
    strict topology of $\cc(X)$, cf.\ \cite[p.\ 31]{GI*} and
    \cite[p.\ 148]{GIc}. However, we did not realize it until
    B.\,Simon, in a private communication, emphasized its
    importance, in relation with Proposition 3.11 and Theorem 4.5
    from \cite{LaS}. We are indebted to him for this remark. On the
    other hand, note that this topology does not play any r\^ole in
    our paper, the strong operator topology on $\rc(X)$ (used in
    \cite{GI*,GIc}) suffices.
}\end{remark}

We now describe some topological properties of $\rc_s(X)$.
\begin{lemma}\label{lm:sop}
  The map $T\mapsto T^*$ is continuous on $\rc_s(X)$ and the
  operation of multiplication is continuous on bounded sets. If
  $T\in\rc(X)$ the map $x\mapsto U_xTU_x^*\in\rc(X)$ is norm
  continuous and the set $\{U_xTU_x^*\mid x\in X\}$ is relatively
  compact in $\rc_s(X)$.
\end{lemma}
\proof The first assertion is obvious. To prove the second one, note
first that \emph{if $S\in\rc(X)$ and $\theta\in\co(X)$ then the
operators $S\theta(Q)$ and $\theta(Q)S$ are compact}. Indeed, it
suffices to show this for $S$ of the form $\vf(Q)\psi(P)$ and then
the assertion is obvious. In particular, from the Remark \ref{re:rk}
it follows that there are $K\in\rk(X)$ and $\theta'\in\co(X)$ such
that $S\theta(Q)=\theta'(Q)K$, and similarly for $\theta(Q)S$. Thus
for $A,B,S,T\in\rc(X)$ we have
$$
\|(BA-TS)\theta(Q)\|\leq\|B\|\|(A-S)\theta(Q)\|+\|(B-T)\theta'(Q)\|
\|K\| 
$$ from which the continuity of multiplication follows.  The norm
continuity of $x\mapsto U_xTU_x^*$ is obvious by (\ref{eq:01}).
Finally, the last assertion of the lemma says that $x\mapsto
U_xTU_x^*\theta(Q)$ has relatively compact range and similarly when
$\theta$ is on the left side. Clearly it suffices to take
$T=\vf(Q)\psi(P)$ and then
$U_xTU_x^*\theta(Q)=\vf(Q+x)\psi(P)\theta(Q)$ and $\psi(P)\theta(Q)$
is a compact operator. Now the assertion follows from the
Riesz-Kolmogorov criterion (Remark \ref{re:rk}) which clearly
implies: if $K$ is a compact operator and $\vf\in\cc(X)$ then
$\vf(Q+x)K$ is a norm relatively compact family of operators.  \qed

\begin{proposition}\label{pr:l8}
  If $T\in\rc(X)$ and $\vk\in\delta X$ then
  $\vk.T:=\lim_{x\rarrow\vk}U_xTU_x^*$ exists in the topological
  space $\rc_s(X)$. The map $\tau_\vk:\rc(X)\rarrow\rc(X)$ defined
  by $\tau_\vk(T)=\vk.T$ is a morphism uniquely determined by the
  property:
\begin{equation}\label{eq:vk-d}
\vf\in\cc(X),\psi\in\co(X^*)\Longrightarrow
\tau_\vk\big(\vf(Q)\psi(P)\big)=(\vk.\vf)(Q)\psi(P). 
\end{equation}
If $T\in\rc(X)$ and $\psi:X^*\rarrow\mbc$ is a bounded continuous
function, then 
\begin{equation}\label{eq:cbm}
\tau_\vk\big(T\psi(P)\big)=\tau_\vk(T)\psi(P)
\hspace{2mm}\mbox{and}\hspace{2mm} 
\tau_\vk\big(\psi(P)T\big)=\psi(P)\tau_\vk(T).
\end{equation}
For each $k\in X^*$ we have
$\tau_\vk\big(V_k^*TV_k\big)=V_k^*\tau_\vk(T)V_k$.
\end{proposition}
\proof We must show that there is an operator $\vk.T\in\rc(X)$ such
that
$$
\lim_{x\rarrow\vk}\|(U_xTU_x^*-\vk.T)\theta(Q)\|=
\lim_{x\rarrow\vk}\|\theta(Q)(U_xTU_x^*-\vk.T)\|=0
$$
for all $\theta\in\co(X)$.
It is clearly sufficient to consider $T=\vf(Q)\psi(P)$ with
$\vf\in\cc(X)$ and $\psi\in\co(X^*)$. Then we have
$$
U_xTU_x^*\theta(Q)=\vf(Q+x)\psi(P)\theta(Q)=\vf(Q+x)\theta'(Q)K
$$
for some $\theta'\in\co(X)$ and $K\in\rk(X)$. Indeed,
$\psi(P)\theta(Q)$ is a compact operator and so we can use the
Remark \ref{re:rk}. Now it suffices to use Lemma \ref{lm:stop}. The
argument for $\theta(Q)U_xTU_x^*$ is even simpler. The other
assertions are easy to prove, for example the last assertion follows
from $V_k^*U_xTU_x^*V_k=U_xV_k^*TV_kU_x^*$.  \qed

\begin{proposition}\label{pr:ker}
  Let $T\in \rc(X)$. Then $\vk.T=0$ for each $\vk\in \delta X$ if
  and only if $T \in \rk(X)$.
\end{proposition}
\proof In order to prove that $\vk.T=0$ if $T\in\rk(X)$ it suffices
to consider $T=\vf(Q)\psi(P)$ with $\vf\in \cc_0(X)$. Then
$\vk.T=(\vk.\vf)(Q)\psi(P)$ and $\vk.\vf=0$ if $\vf\in \cc_0(X)$.
Reciprocally, let $\rj = \{T\in \rc(X) \mid \vk.T=0,
\;\forall\,\vk\in \delta X\}$ and notice that $\rj$ is a
$C^*$-algebra and, moreover, it is a crossed product because of the
last assertions of Proposition \ref{pr:l8}.  Also, for each $S\in
\rc(X)$ we have $\vk.(ST)=(\vk.S)(\vk.T)=0$ so $\rj$ is an ideal.
Thus, by Proposition \ref{pr:var}, there is an ideal $\cj$ in
$\cc(X)$ such that $\rj = \cj \rtimes X$. Let us show that $\cj=
\cc_0(X)$.  This will finish the proof, because then $\rj = \cc_0(X)
\rtimes X =\rk(X)$. From (\ref{eq:Land}) we get
$$
\cj = \{\vf \in \cc(X) \mid \vk.(\vf(Q)\psi(P))=0\ \forall\, \psi
\in \cc_0(X^*) \,\mbox{ and }  \forall\,\vk\in \delta X\}.
$$
But $\vk.(\vf(Q)\psi(P)) = (\vk.\vf)(Q)\psi(P)$.  On the other
hand, if $\theta\in\cc(X)$ is such that $\theta(Q)\psi(P)=0$
$\forall \psi \in \cc_0(X^*)$, then $\theta = 0$.  Indeed,
$V_k^*\theta(Q)\psi(P)V_k = \theta(Q)\psi(P+k)$, so if $\psi \in
L^1(X^*)$ then we have in the weak operator topology 
$$
0= \int_{X^*}V_k^*\theta(Q)\psi(P)V_k\dd k =
\theta(Q)\int_{X^*}\psi(P+k)\dd k = \theta(Q)\int_{X^*}\psi\dd k.
$$
Thus it suffices to take $\psi$ such that $\int_{X^*}\psi\dd k
\neq 0$.  So we finally see that $\cj$ is the set of $\vf\in\cc(X)$
such that $\vk.\vf=0$ for all $\vk\in \delta X$, i.e.\ 
$\cj=\cc_0(X)$ by Lemma \ref{lm:linf}.
\qed

The next result follows easily from Propositions \ref{pr:l8} and
\ref{pr:ker}.

\begin{theorem}\label{th:ess}
  The map $T\mapsto(\vk.T)_{\vk\in\delta X}$ is a morphism
  $\rc(X)\rightarrow\prod_{\vk\in\delta X}\rc(X)$ with $\rk(X)$ as
  kernel, so we have a canonical embedding
\begin{equation}\label{eq:can2} 
\rc(X)/\rk(X)\hookrightarrow\pprod_{\vk\in\delta X}\rc(X).
\end{equation}
\end{theorem}

Theorem \ref{th:intro} is an immediate consequence.  As explained in
Subsection \ref{ss:pd1}, the morphism $\tau_\vk$ extends to
observables affiliated to $\ra$ and Theorem \ref{th:introo} follows
easily. 

\begin{remark}\label{re:roch}{\rm
It has been brought to our attention by Steffen Roch that it is not
possible to deduce Theorem \ref{th:intro} for not normal operators
from Theorem \ref{th:ess}, as we stated in an earlier version of
this paper, because the spectrum of a general element of an
infinite product of $C^*$-algebras is not so simply related to the
spectra of its components. We could have stated a version of Theorem
\ref{th:intro} valid for not normal operators in the spirit of
\cite[Theorem 2.2.1]{RRS2} but we did not do it because the only
applications we have in mind refer to quantum Hamiltonians, which
are self-adjoint operators. We mention, however, that for some
algebras $\ca$ the Theorem \ref{th:imp} remains true (without
closure) for non normal operators, see Sections 2.4 and 2.5 in
\cite{RRS2}. 
}\end{remark}

\begin{definition}\label{df:linfo}
  If $H$ is an observable affiliated to $\rc(X)$ and if
  $\vk\in\delta X$ then the observable $\vk.H$ affiliated to
  $\rc(X)$ is called \emph{localization of $H$ at $\vk$}.  The set
  of operators $\ell(H):=\{\vk.H\mid\vk\in\delta X\}$ is the set
  of \emph{localizations of $H$ at infinity}.
\end{definition}
Then we can write the relation (\ref{eq:introo}) as follows:
\begin{equation}\label{eq:ess}
\se(H)=\overline\ccup_{\vk\in\delta X}\sigma(\vk.H)=
\overline\ccup_{K\in \ell(H)}\sigma(K).
\end{equation}

\begin{remark}\label{re:mg}{\rm
By using the universal property of the Stone-\v{C}ech
compactification $\gamma X$ (cf.\,page \pageref{up}) we see that for
$T\in\rb(X)$ the following two assertions are equivalent:\\ 
(1) the set $\{x.T\mid x\in X\}$ is strongly relatively compact in
$\rb(X)$;\\ 
(2) $X\ni x\mapsto x.T$ extends to a strongly
continuous map $\gamma X\ni\vk\mapsto\vk.T\in\rb(X)$.\\ 
The set of operators having these properties is a norm closed
subalgebra of $\rb(X)$ (quite large, it contains $\rc(X),
L^\infty(X),L^\infty(X^*)$ and much more). It is easy to check that
$\sigma(\vk.T)\subset\se(T)$ if $\vk\in \delta X$, but in most cases
the operators $\vk.T$ do not suffice to determine the essential
spectrum of $T$. This fact extends to observables affiliated to this
algebra. For example, if $H$ is the Hamiltonian of a particle in 2
dimensions in a constant non-zero magnetic field, then $T=\vf(H)$
has the property (1) and $\vk.T=0$ if $\vf\in\co(\mbr)$, i.e.\
$\vk.H=\infty$ for all $\vk\in \delta X$. But $\se(H)\neq\emptyset$.
}\end{remark}

\PAR                      \label{ss:LIT}
We fix now an algebra of interactions $\ca$ on $X$, and set 
$\ra=\ca\rtimes X\subset\rc$. Theorem \ref{th:ess} gives a
description of $\ra/\rk(X)$ but we can make it more precise because
many ultrafilters give the same character of $\ca$. 

\begin{definition}\label{df:l8a}
  If $\vk\in\delta X$ the $C^*$-algebras $\ca_\vk=\tau_\vk(\ca)$ and
  $\ra_\vk=\tau_\vk(\ra)=\ca_\vk\rtimes X$ are the
  \emph{localizations at $\vk$} of the algebras $\ca$ and $\ra$
  respectively.
\end{definition}
As explained in Subsection \ref{ss:pd5}, and taking into account the
relation $(\vk.\vf)(y)=\vk(y.\vf)$ (see Lemma \ref{lm:stop}) and
Lemma \ref{lm:delta}, we see that $\ca_\vk$ and $\ra_\vk$ depend
only on the restriction to $\ca$ of the character $\vk$.  In other
terms, we have for example $\ca_\vk=\ca_\chi$ if
$\delta(\vk)=\delta(\cchi)$, where $\delta:\delta X\rarrow
\delta(\ca)$ is the canonical surjection, cf.\ (\ref{eq:lca}).
According to the convention made in Subsection \ref{ss:pd5} (see
page \pageref{dist}) we shall use the same notations $\ca_\vk$ and
$\ra_\vk$ if $\vk\in\delta(\ca)$. 

In the statement of the next theorem we use the canonical
identification of $X$ (as topological space) with an open dense
subset of $\sigma(\ca)$

\begin{theorem}\label{th:gn}
  If $T\in \ra$ the norm continuous map $X\ni x\mapsto
  x.T\in\ra\subset\rc$ extends to a continuous map
  $\sigma(\ca)\ni\vk\mapsto\vk.T\in\rc_s(X)$. For each
  $\vk\in\delta(\ca)$ the map $\tau_\vk:\ra\rarrow\rc$ defined by
  $\tau_\vk(T)=\vk.T$ is a morphism with $\ra_\vk$ as range.  One
  has $\vk.T=0$ for all $\vk\in\delta(\ra)$ if and only if
  $T\in\rk(X)$ which gives a canonical embedding
\begin{equation}\label{eq:emb}
\ra/\rk(X)\hookrightarrow\pprod_{\vk\in\delta(\ca)}\ra_\vk.
\end{equation}
\end{theorem}
\proof Consider for each $T\in \ra$ the map $F_T:\gamma
X\rarrow\rc_s(X)$ defined by $F_T(\vk)=\vk.T$. From Lemma
\ref{lm:sflu} it follows that $F_T$ is continuous: indeed, it
suffices to assume that $T=\vf(Q)\psi(P)$ and to argue as in the
proof of Lemma \ref{lm:sop}.  Notice that if the characters
$\vk,\cchi\in \gamma X$ are equal on $\ca$, then $F_T(\vk) =
F_T(\cchi)$.  Indeed, for $T$ as above we have $\vk.T=
(\vk.\vf)(Q)\psi(P)=(\cchi.\vf)(Q)\psi(P)=\cchi.T$.  Thus, as
explained on page \pageref{fact}, if $\pi:\gamma
X\rarrow\sigma(\ra)$ is the canonical surjection, we shall have
$F_T=f_T\circ\pi$, where $f_T:\sigma(\ra)\rarrow\rc_s(X)$ is
continuous. If $x\in X$ then $\pi(x)=x$ so $f_T(x)=F_T(x)=x.T$.  We
have both $X\subset \sigma(\ra)$ and $X\subset\gamma X$ and since
the restriction of $\pi$ to $X$ is the identity mapping, $\pi$ acts
non-trivially only on the boundary.  Let $\delta$ be the restriction
of the map $\pi$ to $\delta X$, hence $\delta:\delta X\rarrow
\delta(\ra)$ is a canonical surjection. Thus $f_T(\vk)=0$ for all
$\vk \in \delta(\ra)$ is equivalent to $F_T(\vk)=0$ for all $\vk \in
\delta X$ which means that $T \in \rk(X)$.  \qed

\noindent{\bf Remark:} By using the last assertion of Lemma
\ref{lm:sop} and the universal property of the space $\gamma
X$, cf.\ page \pageref{up}, one may avoid the use of Lemma
\ref{lm:sflu}.   
 
\begin{remark}\label{re:cm}{\rm
In nice situations, the localization at infinity $\ca_\vk$ is
simpler than $\ca$, and $(\ca_\vk)_\chi$ is still simpler, and so
on, but this is not always the case. Note also that in general
$\ca_\vk\not\subset\ca$. If, however, this holds for each
$\vk\in\delta(\ca)$, then it is natural to ask whether we have
$\tau_\vk\tau_\chi\vf=\tau_\chi\tau_\vk\vf$ for all $\vf\in\ca$ and
all $\vk,\chi\in\delta(\ca)$. Although this is not true if
$\ca=\cc(X)$, in several non-trivial and physically interesting
situations this property is satisfied. See  Examples \ref{ex:linf}
and \ref{ex:ggyr} and Section \ref{s:exloc}.  }\end{remark}

\begin{example}\label{ex:linf}{\rm
    We shall consider here the localizations at infinity of the
    simplest algebras. If $\ca=\cc_\infty(X)$ then
    $\sigma(\ca)=X\cup\{\infty\}$ is the Alexandroff
    compactification of $X$, we have $\delta(\ca)=\{\infty\}$, and
    the localization of $\vf\in\ca$ at $\infty$ is the constant
    function which takes the value
    $\vf(\infty)=\lim_{x\rarrow\infty}\vf(x)$. If $X=\mbr$ and $\ca$
    is the set of bounded continuous functions which have limits as
    $x\rarrow \pm\infty$ then $\sigma(\ca)=[-\infty,+\infty]$,
    $\delta(\ca)=\{-\infty,+\infty\}$, and the localization of
    $\vf\in\ca$ at $+\infty$ is again the constant function which
    takes the value $\vf(+\infty)=\lim_{x\rarrow+\infty}\vf(x)$ and
    similarly for the localization at $-\infty$. Thus in both
    examples we have $\ca_\vk=\mbc$ for all $\vk\in\delta(\ca)$.  In
    Subsection \ref{ss:LI4} we shall describe explicitly the largest
    $X$-subalgebra $\ca\subset\cc(X)$ such that $\ca_\vk=\mbc$ for
    all $\vk\in\delta(\ca)$.  }\end{example}

\begin{example}\label{ex:ggyr}{\rm
The next example is due to Gilles Godefroy (we thank him for
answering to our questions) and is relevant in the context of Remark
\ref{re:cm}.  Let $X=\mbz\times\mbz$ and let $\ca$ be the set of
$\vf\in\ell_\infty(X)$ such that $\lim_{k\rarrow\infty}\vf(j,k)=0$
for all $j\in\mbz$. Let $\theta\in\ell_\infty(\mbz)$ and set
$\vf(j,k)=\theta(k)$ if $|k|\leq j$ and $=0$ otherwise. Then
$\vf\in\ca$ and $\lim_{a\rarrow+\infty} \vf(a+j,k)=\theta(k)$ for
each $j,k$. It is clear now that we may construct an ultrafilter
$\vk\in\delta X$ such that $\vk.\vf=1\otimes\theta$ so
$\vk.\vf\nin\ca$ in general.  
}\end{example}

Theorem \ref{th:imp} is a corollary of Theorem \ref{th:gn}. Thus, if
$H$ is a normal element of $\ra$ or an observable affiliated to
$\ra$ and if we set $\vk.H=\tau_\vk(H)$, then
\begin{equation}\label{eq:imp*}
\se(H)=\overline\ccup_{\vk\in\delta(\ca)}\sigma(\vk.H).
\end{equation}
This representation of the essential spectrum of $H$, although more
precise than (\ref{eq:ess}), is still quite redundant, cf.\ page
\pageref{red}, and can be improved in many situations (the most
interesting one being the $N$-body case). To explain this, for
$\vk\in\delta(\ca)$ let us denote \label{disc}
\begin{equation}\label{eq:tiid}
\cj_\vk=\ker\tau_\vk=\{\vf\in\ca\mid
\vk(x.\vf)=0\quad\forall x\in X\}.
\end{equation}
This is is the maximal $X$-ideal included in the maximal ideal
$\ker\vk$ of $\ca$. Although the ideals $\ker\vk$ for different
$\vk$ are not comparable, it often happens that the $\cj_\vk$ are
comparable, i.e.\,we may have $\cj_\vk\subset\cj_\chi$ for
$\vk\neq\cchi$. 

\begin{lemma}\label{lm:tiid}
  If $\cj_\vk\subset\cj_\chi$ then
  $\sigma(H_\chi)\subset\sigma(H_\vk)$. In particular,
  (\ref{eq:imp*}) remains true if we restrict the union to the $\vk$
  such that the ideal $\cj_\vk$ is minimal in
  $\{\cj_\vk\mid\vk\in\delta(\ca)\}$.
\end{lemma}
\proof Here we use more abstract algebraic tools, as in
\cite{GI*,GI0}. The morphism $\tau_\vk:\ca\rarrow\ca_\vk$ is
surjective and has $\cj_\vk$ as kernel, hence induces an isomorphism
$\ca/\cj_\vk\cong\ca_\vk$. If $T\in\ca$ and if $T/\cj_\vk$ is its
projection in the quotient $\ca/\cj_\vk$, then $T/\cj_\vk$ is sent
by this isomorphism into $\vk.T$, hence
$\sigma(T/\cj_\vk)=\sigma(\vk.T)$. From $\cj_\vk\subset\cj_\chi$ we
get a canonical surjective morphism $\ca/\cj_\vk\rarrow\ca/\cj_\chi$
which sends $T/\cj_\vk$ into $T/\cj_\chi$. Finally, we recall that
if $\Phi$ is a morphism then $\sigma(\Phi(S))\subset\sigma(S)$.
\qed

\begin{example}\label{ex:tr}{\rm
    If, for $x\in X$ and $\vk\in\delta(\ca)$, we denote $x+\vk$ the
    character $\vk\circ\tau_x$, then clearly $\cj_{x+\vk}=\cj_\vk$ ,
    hence $\sigma((x+\vk).H)=\sigma(\vk.H)$. However, this case is
    trivial because clearly $(x+\vk).H=U_x(\vk.H)U_x^*$.
  }\end{example}

One further simplification may be obtained as follows.

\begin{lemma}\label{lm:kk}
Let $\ck\subset\delta(\ca)$ such that: if $\vf\in\ca$ and
$\vk(x.\vf)=0$ for all $\vk\in\ck$ and $x\in X$, then $\vf\in\co(X)$.
Then (\ref{eq:imp*}) remains valid if $\delta(\ca)$ is replaced by
$\ck$.  
\end{lemma}
\proof This is a consequence of the proof of Theorem \ref{th:gn} but
can also be proved directly as follows. One first notices that the
condition on $\ck$ is equivalent to the density in
$\delta(\ca)=\sigma(\ca/\co(X))$ of the set of characters of the
form $\vk\circ\tau_x$, with $\vk\in\ck$ and $x\in X$. Then one can
use the following easily proven fact: if $S_\alpha$ is a net of
operators such that $S_\alpha^{(*)}\rarrow S^{(*)}$ strongly, then
$\sigma(S)$ is included in the closure of
$\ccup_\alpha\sigma(S_\alpha)$.  \qed

\section{Applications} \label{s:exloc} 
\protect\setcounter{equation}{0}
After some preliminaries, we describe here three classes of
$C^*$-algebras of Hamiltonians which seem to us particularly
relevant and treat some more explicit examples.
\PAR                      \label{ss:ints}
{\bf Algebras associated to translation invariant filters.}  In this
preliminary subsection we give an intrinsic description of a class
of crossed products introduced in \cite{GI*,GI0}. Recall that a
filter $\gof$ is \emph{translation invariant} if: $x\in X,F\in\gof
\Rightarrow x+F\in\gof$. Note that $\gof^\circ$ will also be
translation invariant.  If $\gof$ is a translation invariant filter
let
\begin{equation}\label{eq:JF}
\cj(\gof)=\{\varphi\in\cc(X)\mid \textstyle\lim_\gof\varphi=0\}. 
\end{equation}
This is clearly an $X$-ideal in $\cc(X)$ and from Lemma
 \ref{lm:easy} we get:
\begin{equation}\label{eq:uf}
\cj(\gof)=\cj(\gof^\circ).
\end{equation}
Then $\cc(\gof)=\mbc+\cj(\gof)$ is the $X$-algebra consisting of
the bounded uniformly continuous functions $\varphi$ such that
$\lim_\gof\varphi$ exists. Observe that if $\gof$ is the Fr\'echet
filter then $\cj(\gof)=\co(X)$ and $\cc(\gof)=\cc_\infty(X)$.

Below we shall consider nets indexed by the filter $\gof$
equipped with the order relation $F\leq G\Leftrightarrow F\supset
G$. For example, $\lim_{F\in\gof}\|\ind_F(Q)T\|=0$ means that for
each $\varepsilon>0$ there is a Borel set $F\in\gof$ such that
$\|\ind_F(Q)T\|<\varepsilon$.

\begin{proposition}\label{pr:tif}
$\cj(\gof)\rtimes X
=\{T\in\rc(X)\mid\lim_{F\in\gof}\|\ind_F(Q)T^{(*)}\|=0\}$.
\end{proposition}
\proof Each $T\in\cj(\gof)\rtimes X$ has the property
$\lim_{F\in\gof}\|\ind_F(Q)T\|=0$.  Indeed, it suffices to consider
operators of the form $T=\varphi(Q)\psi(P)$ with
$\varphi\in\cj(\gof),\psi\in\co(X^*)$.  But then the set $F$ of
points $x$ such that $|\varphi(x)|<\varepsilon$ is open and belongs
to $\gof$, and so we have $\|\ind_F(Q)\varphi(Q)\|\leq\varepsilon$,
which is more than needed.

Conversely, let $\rj$ be the set of $T\in\rc(X)$ such that
$\lim_{F\in\gof}\|\ind_F(Q)T^{(*)}\|=0$. This is clearly a
$C^*$-subalgebra of $\rc(X)$ which is stable under the morphisms
$T\mapsto V_k^*TV_k$. By Theorem \ref{th:Land} we have
$\rj=\cj\rtimes X$ for a unique $X$-algebra $\cj$, namely the set of
$\varphi\in\cc(X)$ such that
$\lim_{F\in\gof}\|\ind_F(Q)\varphi(Q)^{(*)}\psi(P)\|=0$ for all
$\psi\in\co(X^*)$.

Thus it remains to prove the following assertion: if
$\varphi\in\cc(X)$ has the property
$\lim_{F\in\gof}\|\ind_F(Q)\varphi(Q)\psi(P)\|=0$ for
$\psi\in\co(X^*)$, then $\lim_{F\in\gof}\|\ind_F(Q)\varphi(Q)\|=0$.
Observe that, due to (\ref{eq:uf}) we may assume $\gof=\gof^\circ$.

Fix $f\in L^2(X),\psi\in\co(X^*)$ and let us set $\theta=\psi(P)f$
and $\theta_a(x)=(U^*_a\theta)(x)=\theta(x-a)$.  Clearly
$\lim_{F\in\gof}\|\ind_F(Q)\varphi(Q)U^*_a\theta\|=0$ uniformly in
$a\in X$. Thus, for any $\varepsilon>0$, there is $F\in\gof$ Borel
such that $\|\ind_F\varphi\theta_a\|<\varepsilon$ for all $a$,
hence
$$
|\varphi(a)|\|\ind_F\theta_a\|\leq
\|\ind_F(\varphi(a)-\varphi)\theta_a\|+
\|\ind_F\varphi\theta_a\|\leq
\|\ind_F(\varphi(a)-\varphi)\theta_a\|+\varepsilon.
$$
Since $\gof=\gof^\circ$ we may assume that $F=G+V$ where $G\in\gof$
and $V$ is a a compact neighborhood of the origin. Moreover, since
$\varphi$ is uniformly continuous and since we may choose $V$ as
small as we wish, we may assume that
$|\varphi(x)-\varphi(a)|<\varepsilon$ if $x-a\in V$. It is possible
to choose $f,\psi$ such that $\supp\theta\subset V$ and
$\|\theta\|=1$. Indeed, $\theta$ is equal to the convolution
product $\wtilde\psi*f$ where $\wtilde\psi(x)=\what\psi(-x)$ and it
suffices to choose $f,\wtilde\psi$ continuous, positive and not zero
and such that $\supp f+\supp\wtilde\psi\subset V$.
Then for $a\in G$ we clearly have $\supp\theta_a\subset F$ hence
$$
|\varphi(a)|=|\varphi(a)|\ \|\theta_a\|=
|\varphi(a)|\ \|\ind_F\theta_a\|\leq
\|(\varphi(a)-\varphi)\theta_a\|+\varepsilon\leq
2\varepsilon.
$$
This proves that $\lim_\gof\varphi=0$.
\qed

From Proposition \ref{pr:tif} we easily get:
$$
\cc(\gof)\rtimes X
=\{T\in\rc(X)\mid \exists S\in\co(X^*) \mbox{ such that }
\lim_{F\in\gof}\|\ind_F(Q)(T-S)^{(*)}\|=0\}.
$$
The $X$-algebras of the form $\ccap_\lambda\cc(\gof_\lambda)$ are of
some physical interest \cite{Ric}.  Indeed, one should think of a
filter finer than the Fr\'echet filter as the set of traces on $X$
of the filter of neighborhoods of some closed part of the boundary
of $X$ in a compactification of $X$. This explains the interest of
the algebras $\ccap_\lambda\cc(\gof_\lambda)$ in the present
context: they consist of ``potentials'' which have limits at
infinity when going in certain directions.  One may easily deduce
from Theorem \ref{th:main} and Proposition \ref{pr:tif} an intrinsic
description of the crossed products
$\ccap_\lambda\cc(\gof_\lambda)\rtimes X$.

\PAR \label{ss:LI4}
{\bf The $\rv(X)$ algebra.}
We shall consider now the simplest non-trivial functions in
$\cc(X)$, those all of whose localizations at infinity are
constants. Our purpose is to give a simple characterization of the
$X$-algebra $\ca$ defined by the condition $\ca_\vk=\mbc$ for all
$\vk\in\delta X$ and of the associated crossed product.
So we introduce the $X$-algebra:
\begin{equation}\label{eq:vo}
\cv(X):=\{\vf\in\cc(X)\mid
\vk.\vf\in\mbc,\ \forall\vk\in\delta X\}
\end{equation}
Observe that the relation $\vk.\vf\in\mbc$ is equivalent to
$\vk.\vf=\vk(\vf)$.
\begin{lemma}\label{lm:vo}
We have $\vf\in \cv(X)$ if and only if $\vf\in\cc(X)$ and
\begin{equation}\label{eq:voc}
\lim_{x\rarrow\infty}(\vf(x+y)-\vf(x))=0,\quad \forall y\in X.
\end{equation}
\end{lemma}
\proof The condition (\ref{eq:voc}) is equivalent to
$y.\vf-\vf\in\co(X)$ for all $y\in X$ and, by (\ref{eq:gid}), this
is equivalent to $\vk(y.\vf-\vf)=0$ for all $\vk\in\delta X$ and
all $y$, hence to $\vk.\vf(y)=\vk(\vf)$ for all $\vk,y$, which means
$\vf\in \cv(X)$.
\qed

It is easily shown that $\vf\in\cc(X)$ satisfies (\ref{eq:voc}) if
and only if $\vf$ is a bounded continuous function such that
$\lim_{x\rarrow\infty}(\vf(x+y)-\vf(x))=0$ uniformly in $y$ when $y$
runs over a compact neighborhood of the origin.  Thus the functions
from $\cv(X)$ are of \emph{vanishing oscillation at infinity} or
\emph{slowly oscillating}, and their r\^ole in the theory of
pseudo-differential operators was noticed a long time ago
due especially to a well known theorem of H.\ Cordes concerning the
compactness of the commutators $[\vf(Q),\psi(P)]$ (see \cite[p.\ 
176--177]{ABG} for a short presentation of the main ideas).
If $X=\mbr^n$ then $\cv(X)$ is just the norm closure of the set of
bounded functions of class $C^1$ whose derivative tends to zero at
infinity. Thus results of the same nature as the embedding
(\ref{eq:tay}) may be found already in \cite{Tay}.

The algebra $\cv(X)$ was systematically considered in the works
\cite{Rab,RRR,RRS,RRS2}; see especially \cite{RRS2} where one may
find references to other earlier papers. Although the authors
emphasize the case $X=\mbz^n$, it is clear for us that their methods
extend to many other groups. On the other hand, since they allow the
functions $\varphi$ to be Banach space valued, the applications of
their theory cover directly the case of operators on $L^2(\mbr^n)$
for example (this involves a certain discretization technique). In
particular, Theorems 2.4.2 and Corollary 2.4.28 from \cite{RRS2} are
much stronger than our next Proposition \ref{pr:voq} in the case
$X=\mbz^n$. Taking into account the wealth of informations and
applications in connection to these question which may be found in
\cite[Chapters 2,4,5]{RRS2}, we decided to keep this section to a
minimum, just to point out the special role of the algebra $\cv(X)$
in the crossed product formalism.

More recently, the relevance of $\cv(X)$ in questions
related to the computation of the essential spectrum has been
independently noticed in \cite{LaS,Ma2}. 

We mention that the compactification $\sigma(\cv(X))$ and the
boundary $\upsilon X=\delta(\cv(X))$ are called \emph{Higson
compactification} and \emph{Higson corona} of $X$ and play an
important r\^ole in recent questions of topology, $C^*$-algebras,
$K$-theory, etc.\ \cite{Ro1,Ro2}. 

Finally, we note that a non-abelian version of $\cv(X)$ appears in a
natural way in the spectral analysis of Schr\"odinger operators on a
tree $X$, see \cite{GG1}.

\medskip

We now give an intrinsic description of the crossed product
$\rv(X)=\cv(X)\rtimes X$ and a more specific decomposition of the
essential spectrum.

\begin{proposition}\label{pr:voq}
We have
\begin{equation}\label{eq:voq}
\rv(X)=\{T\in\rc(X)\mid \vk.T\in\co(X^*),\ \forall 
\vk\in\delta X\}.
\end{equation}
If $T\in\rv(X)$ then the map $\vk\mapsto\vk.T\in\co(X^*)$ is norm
continuous, hence (\ref{eq:emb}) takes the more precise form
\begin{equation}\label{eq:tay}
\rv(X)/\rk(X)\hookrightarrow \cc(\upsilon X;\co(X^*)).
\end{equation}
In particular, 
$\ell(T)=\{\vk.T\mid\vk\in\upsilon X\}\subset\co(X^*)$ 
is a compact set.  If $H$ is a normal element of
$\rv(X)$ or is an observable affiliated to $\rv(X)$ then:
\begin{equation}\label{eq:essv}
\se(H)=\ccup_{\vk\in\upsilon X}\sigma(\vk.H)=
\ccup_{K\in \ell(H)}\sigma(K).
\end{equation}
\end{proposition}
\proof To show the inclusion $\subset$ in (\ref{eq:voq}) and the
norm continuity of the map $\vk\mapsto\vk.T\in\co(X^*)$ it suffices
to consider $T=\vf(Q)\psi(P)$ with $\vf\in\cv(X)$ and
$\psi\in\co(X^*)$. But then $\vk.T=\vk(\vf)\psi(P)$ and these facts
become obvious. Note that the compactness of the set $\ell(T)$
implies that the union $\ccup_{T\in \ell(T)}\sigma(T)$ is closed,
hence (\ref{eq:essv}) is true. It remains to show the inclusion
$\supset$ in (\ref{eq:voq}). Since $\vk.(V_k^*TV_k)=V_k^*(\vk.T)V_k$
and $(\vk.T)U_x=\vk.(TU_x)$ and since $\co(X^*)$ is stable under the
automorphism generated by $V_k$ and under multiplication by $U_x$,
it is clear that the right hand side of (\ref{eq:voq}) satisfies
Landstad's conditions. Hence Theorem \ref{th:Land} shows that it
suffices to prove that if $\vf\in\cc(X)$ has the property
$(\vk.\vf)(Q)\psi(P)\in\co(X^*)$ for all $\psi\in\co(X^*)$ and all
$\vk\in\delta X$, then $\vf\in\cv(X)$. Thus it suffices to show
that if $\xi\in\cc(X)$ and $\xi(Q)\psi(P)\in\co(X^*)$ for all
$\psi\in\co(X^*)$, then $\xi$ is a constant. But we have
$$
\xi(Q)\psi(P)=U_x\xi(Q)\psi(P)U_x^*=\xi(x+Q)\psi(P)
$$ hence $(\xi(Q)-\xi(x+Q))\psi(P)=0$ for all $\psi$, so
$\xi(Q)=\xi(x+Q)$ for all $x$.
\qed 
\noindent{\bf Remark:} If the reader has any difficulty in proving
that the union in (\ref{eq:essv}) is closed, he should look at the
proof of \cite[Theorem 2.10]{DaG2}.

\begin{remark}\label{re:vo1}{\rm
    $\rv(X)$ is the largest crossed product $\ra$ such that
    $\ra/\rk(X)$ is abelian. Indeed,
    $\ra/\rk(X)\hookrightarrow\pprod_{\vk\in\delta(\ca)}\ra_\vk$ by
    (\ref{eq:emb}) and the $\ra_\vk$ are crossed products, so
    $\ra_\vk$ is abelian if and only if $\ra_\vk=\{0\}$ or
    $\ra_\vk=\co(X^*)$.
}\end{remark}

\begin{remark}\label{re:vo2}{\rm
The observables affiliated to $\co(X^*)$ are functions of momentum,
so that it is natural to call them \emph{free Hamiltonians}. Then we
may describe in physical terms $\rv(X)$ as the largest $C^*$-algebra
of energy observables such that if $H$ is affiliated to it then all
its localizations at infinity are free Hamiltonians. 
}\end{remark}

\begin{remark}\label{re:cm1}{\rm
We reconsider here the question of Remark \ref{re:cm} for
$\ca=\cv(X)$. If $\vk\in\delta X$ then $\tau_\vk:\cv(X)\rarrow\mbc$
is just the character associated to $\vk$ and so if $\cchi\in\delta
X$ then
$\tau_\chi\tau_\vk\vf=\tau_\vk\vf\neq\tau_\chi\vf
=\tau_\vk\tau_\chi\vf$  in general.
}\end{remark}

\PAR \label{ss:mf}
{\bf More remarks on filters.}  The following general remarks will
be useful in the next subsections. Let $Y$ be a closed subspace of
$X$ (thus $K\cap Y$ is compact for each compact $K\subset X$).  If
$\gof$ is a filter on $Y$ then $\gof$ can be seen as a filter basis
on $X$ and we shall denote (just for a moment) by $\gof^X$ the
filter on $X$ that it generates (this is the set of subsets of $X$
which contain a set from $\gof$).  The map $\gof\mapsto\gof^X$ is an
injective map from the set of filters on $Y$ onto the set of filters
on $X$ which contain $Y$.  Indeed, we have $\gof=\{F\cap
Y\mid F\in\gof^X\}$.  It is also clear that if $\vk$ is an
ultrafilter on $Y$ then $\gof^X$ is also an ultrafilter.  Finally,
if $\gof$ is finer than Fr\'echet on $Y$ then $\gof^X$ is finer than
Fr\'echet on $X$.

Since $Y\in\gof^X$, if $T:X\rarrow Z$ then $\lim_{\gof^X} T$ exists
if and only if $\lim_\gof T|_Y$ exists and then they are equal.

From now on we shall not distinguish $\gof^X$ from $\gof$, so we use
the same notation $\gof$ for both. In particular, we get natural
embeddings
\begin{equation}\label{eq:xy}
\gamma Y\subset\gamma X \quad \mbox{and} \quad 
\delta Y\subset\delta X.
\end{equation}
It is convenient to understand this when the ultrafilters are
interpreted as characters.  We have an obvious embedding
$\ell_\infty(Y)\subset\ell_\infty(X)$ so each character of
$\ell_\infty(X)$ gives a character of $\ell_\infty(Y)$ by
restriction, and reciprocally, each character of $\ell_\infty(Y)$
has a canonical extension to a character of $\ell_\infty(X)$, namely
$\vk(\vf):=\vk(\vf\ind_Y)$. Thus:
$$
\gamma Y=\{\vk\in\gamma X\mid \vk(Y)=1\} \quad
\mbox{and} \quad \delta Y=\gamma Y\cap\delta X.
$$
It is easy to see now that $\gamma Y$ is a \emph{clopen} subset of
$\gamma X$, equal to the closure of $Y$ in $\gamma X$.  

\medskip

One says that a filter on a topological space is \emph{convergent}
to some point $x$ if it is finer than the filter of neighborhoods
of $x$. \emph{Any ultrafilter on a compact space is convergent}.
\label{upc} This is easily seen to be equivalent to any of the usual
definitions of compactness \cite[Chapter 1, \S 9]{Bo}.

It is easy now to understand the universal property of $\gamma X$,
cf.\ page \pageref{up}. We first observe that $\gamma$ should be
considered as a functor from the category of sets into the category
of compact spaces. Indeed, if $X,Y$ are sets and $\theta:X\rarrow Y$
then it is obvious how to define $\gamma\theta:\gamma X\rarrow\gamma
Y$ if ultrafilters are thought as characters: note first that
$\vf\mapsto\vf\circ\theta$ is a morphism
$\theta^*:\ell_\infty(Y)\rarrow\ell_\infty(X)$ and then if
$\vk\in\gamma X$  define $\gamma\theta(\vk)$ as the character of
$\ell_\infty(Y)$ given by $\gamma\theta(\vk)=\vk\circ\theta^*$.
The continuity of $\gamma\theta$ is clear.

Now assume $Y$ is a compact topological space. The only thing we
need to accept is that $\sigma(\cc(Y))=Y$, this is not difficult to
prove directly. Then we have a natural continuous map 
$\gamma Y\ni\cchi\mapsto\cchi_\flat\in Y$ which associates to a
character $\cchi$ of $\ell_\infty(Y)$ its restriction to $\cc(Y)$.
In fact, the ultrafilter $\cchi$ is convergent and $\cchi_\flat$ is
just its limit. Finally, $\vk\mapsto\gamma\theta(\vk)_\flat$ is the
unique extension of $\theta$ to a continuous map $\gamma X\rarrow
Y$.

\PAR \label{ss:klaus}
{\bf Sparse sets.}  From the point of view of the
complexity of the interactions, the algebra of interactions that one
should consider next is
\begin{equation}\label{eq:cunk}
\ca=\{\vf\in\cc(X)\mid\vk.\vf\in\cc_\infty(X),\ \forall
\vk\in\delta X\}.
\end{equation}
The corresponding algebra of energy observables is
\begin{equation}\label{eq:qunk}
\ra=\ca\rtimes X=
\{T\in\rc(X)\mid \vk.T\in\rt(X),\ \forall \vk\in\delta X\}.
\end{equation}
Thus $\ra$ is the largest $C^*$-algebra of energy observables such
that all the localizations at infinity of a Hamiltonian $H$
affiliated to it are two-body Hamiltonians.
We shall leave for the second part of our work the study of the
algebra (\ref{eq:qunk}) and we shall consider here only subalgebras
corresponding to Klaus type potentials.

\begin{remark}\label{re:cm2}{\rm
    The algebra $\ca$ defined by (\ref{eq:cunk}) is characterized by
    $\ca_\vk=\cc_\infty(X)$ for each $\vk$, hence contains
    $\cc_\infty(X)$ and is stable under all the morphisms
    $\tau_\vk$. It is also clear that $\tau_\chi\tau_\vk\vf\in\mbc$
    and is distinct from $\tau_\vk\tau_\chi\vf$ in general,
    cf.\,Remark \ref{re:cm}.  }\end{remark}

M.\ Klaus discovered in \cite{Kla} the following class of
Hamiltonians with nontrivial essential spectrum. Let $L\subset\mbr$
be a discrete set such that the distance between two successive
points of $L$ tends to infinity when we approach infinity. For each
$l\in L$ let $V_l\in L^1(\mbr)$ real such that $\|V_l\|_{L^1}\leq A$
and $\supp V_l\subset[-A,A]$ for a fixed finite $A$.  Denote
$H=P^2+\sum_lV_l(Q-l)$ and $H_l=P^2+V_l(Q)$. Then the description of
$\se(H)$ given in \cite{Kla} is equivalent to:
\begin{equation}\label{eq:kla}
\se(H)=\ccap_{F\in\cf(L)}
\overline\ccup_{l \in F^c} \sigma(H_l)
\end{equation}
where $\cf(L)$ is the set of finite subsets of $L$ and
$F^c=L\setminus F$. One of the main examples in \cite{GI*,GI0}
consisted in an algebraic treatment of this example, treatment
based on the construction of a $C^*$-algebra to which operators like
$H$ are affiliated. We recall below the definition of this type of
algebras and then we shall give a description of $\se(H)$ for the
operators affiliated to them which is more in the spirit of Theorem~
\ref{th:intro} (description which also appears in \cite{GI*,GI0} but
which is deduced there by very different means). 

\medskip

If $L,\Lambda$ are subsets of $X$ we denote
$L_\Lambda=L+\Lambda$ and $L_\Lambda^c=X\setminus L_\Lambda$.
If $L$ has the property $L_\Lambda\neq X$ for each compact $\Lambda$
then we associate to it the filter
\begin{equation}\label{eq:fl}
\gof_L=\{A\subset X\mid A\supset L_\Lambda^c
\mbox{ for some compact }\Lambda\subset X\}. 
\end{equation}
This is clearly a translation invariant filter finer than the
Fr\'echet filter and such that $\gof_L^\circ=\gof_L$. Thus
\begin{equation}\label{eq:cl}
\cc_L(X)=\{\vf\in\cc(X)\mid
\textstyle\lim_{\gof_L}\!\vf \mbox{ exists }\}
\end{equation}
is an algebra of interactions on $X$. An intrinsic description of
the corresponding algebra of Hamiltonians $\rc_L(X)$ follows
immediately from the results of Subsection \ref{ss:ints}. Let
\begin{equation}\label{eq:LC}
\delta_L X=\delta(\cc_L(X))=\sigma(\cc_L(X))\setminus X
\end{equation}
be the boundary of $X$ in the compactification associated to
$\cc_L(X)$.  We recall that $\delta_LX$ is a quotient of $\delta X$.
We set
\begin{equation}\label{eq:8}
\infty_L=\{\vk\in\gamma X\mid \vk\supset\gof_L\}=
\{\vk\in\gamma X\mid L_\Lambda^c\in\vk
\mbox{ if }\Lambda\subset X \mbox{ is compact }\}.
\end{equation}
This is a compact subset of $\delta X$ and if $\vk\in\infty_L$ then
$\vk(\vf)=\lim_{\gof_L}\!\vf$, so that $\infty_L$ gives just a point
in $\delta_L X$.  The problem that remains to be solved is the
description of the other points of $\delta_LX$.

\medskip

In this subsection we consider only the case when $L$ is a
\emph{sparse} set, in the following sense: $L$ is locally finite and
for each compact set $\Lambda$ there is a co-finite set $M\subset L$
(i.e.\ such that $L\setminus M$ is finite) with the following
property: if $m\in M$ and $l\in L$, $l\neq m$, then
$(m+\Lambda)\cap(l+\Lambda)=\emptyset$.

With the conventions made in Subsection \ref{ss:mf}, we have $\delta
L\subset\delta X$, more explicitly for $\vk\in\delta L$ and
$\vf\in\ell_\infty(X)$ we set
$$
\vk(\vf)\equiv\vk(\vf\ind_L)=\lim_{l\rarrow\vk}\vf(l).
$$
Below we use the symbol $\amalg$ to denote disjoint union of sets.
\begin{lemma}\label{lm:spa}
  Let $\theta:X\times\delta L\rarrow\delta_LX$ be defined by
  $\theta(x,\vk)=\vk\circ\tau_x$. Then $\theta$ is injective and its
  range is $\delta_LX\setminus\{\infty_L\}$, which gives us an
  identification
\begin{equation}\label{eq:spa}
\delta_LX\cong (X\times\delta L)\amalg \{\infty_L\}.
\end{equation} 
\end{lemma}
\proof We set $\theta(x,\vk)=\theta_{x,\vk}$ and note the more
explicit formula
$$
\theta_{x,\vk}(\vf)=\lim_{l\rarrow\vk}\vf(l+x).
$$ We first prove that $\theta$ is injective. It is clearly
sufficient to show that if $x\in X$ and $\vk,\cchi\in\delta L$ are
such that $\vk(x.\vf)=\cchi(\vf)$ for all $\vf\in\cc_L$, then $x=0$
and $\vk=\cchi$. Let $M\subset L$ such that $\vk(M)=1$. Since $\vk$
is finer than the Fr\'echet filter, $M$ is infinite and $\vk(N)=1$
if $N$ is a co-finite subset of $M$. Let $\Lambda\subset X$ be
compact and such that $0,x\in\Lambda$. Eliminating if needed a
finite number of points from $M$, we may assume that
$(L\setminus M)\cap M_\Lambda=\emptyset$ 
and $M_\Lambda=\amalg_{l\in M}(l+\Lambda)$.  Choose
$\vf\in\co(X)$ with $\supp\vf\subset\Lambda$ and let us define
$\vf_M=\sum_{l\in M}\tau_{-l}\vf$.  Then:
$$
(\ind_Lx.\vf_M)(y)=
\textstyle\sum_{l\in M}\ind_M(y)(\tau_{x-l}\vf)(y)=
\textstyle\sum_{l\in M}\ind_M(y)\vf(x+y-l).
$$ In the sum from the right hand side the terms are zero unless
$l,y\in M$ and $x+y\in l+\Lambda$; but this implies $l=y$ because
$x\in\Lambda$. We get $\ind_Lx.\vf_M=\ind_M\vf(x)$ and so, by
choosing $\vf$ such that $\vf(x)\neq0$, we see that
$$
\vk(x.\vf_M)=\vk(\ind_Lx.\vf_M)=\vk(\ind_M)\vf(x)=\vf(x)\neq0.
$$
Similarly $\ind_L\vf_M=\ind_M\vf(0)$ and so
$\cchi(\vf_M)=\cchi(\ind_M\vf_M)=\cchi(\ind_M)\vf(0)$. 
If $x\neq0$ we may choose $\vf$ such that $\vf(0)=0$ and we see that
$\vk(x.\vf)\neq\cchi(\vf)$ for some $\vf\in\cc_L$. If $x=0$ but
$\vk\neq\cchi$ then $M$ can be chosen such that $\cchi(M)=0$ (because
$\vk$ and $\cchi$ are distinct ultrafilters) hence again 
$\vk(x.\vf)\neq\cchi(\vf)$ for some $\vf\in\cc_L$. This proves the
injectivity of the map $\theta$.

Now we show that for any $\cchi\in\delta X$ such that
$\cchi\nin\infty_L$ there is $(x,\vk)\in X\times\delta L$ such that
$\cchi(\vf)=\vk(x.\vf)$ for all $\vf\in\cc_L$.  Since $\cchi$ is not
finer than $\gof_L$, there is a compact set $\Lambda\subset X$ such
that $L_\Lambda^c\nin\cchi$. But $\cchi$ is an ultrafilter, so
$L_\Lambda\in\cchi$. Since $\cchi$ is finer than the Fr\'echet
filter, there is $M\subset L$ such that
$\cchi(M_\Lambda)=1$ and
$$M_\Lambda=\amalg_{l\in M}(l+\Lambda)\equiv M\times \Lambda.
$$
The sets $F\subset M_\Lambda$ with $\cchi(F)=1$ form a basis for
$\cchi$ and each such $F$ can be uniquely written as a disjoint
union $F=\amalg_{l\in N}(l+F(l))$ with $N\subset M$ and
$F(l)\subset\Lambda$ non empty sets.  We define surjective maps
$\pi_M:M_\Lambda\rarrow M$ and $\pi_\Lambda:M_\Lambda\rarrow
\Lambda$ with the help of the identification $M_\Lambda\equiv
M\times \Lambda$. The image $\vk=\pi_M(\cchi)$, i.e.\ the filter of
subsets of $M$ generated by the $\pi_M(F)$ with $F\in\cchi$, is
obviously an ultrafilter on $M$, hence on $L$, finer then the
Fr\'echet filter.  Similarly, $\pi_\Lambda(\cchi)$ is an ultrafilter
on $\Lambda$, which is a compact space, hence $\pi_\Lambda(\cchi)$
converges to some point $x\in\Lambda$. If $F$ is as above
then $\pi_M(F)=N$ and $\pi_\Lambda(F)=\bigcup_{l\in N}F(l)$ and
the families of these sets are bases for the filters $\vk$ and
$\pi_\Lambda(\cchi)$ respectively. In particular, since
$\pi_\Lambda(\cchi)$ is finer than the filter of neighborhoods of
$x$, for each neighborhood $V$ of $x$ there is $F$ such that 
$\bigcup_{l\in N}F(l)\subset V$.

We prove now that $\cchi(\vf)=\vk(x.\vf)$ if $\vf\in\cc_L$.  We have
$\cchi(\vf)=\lim_\chi\vf$, thus for each $\ve>0$ there is
$F\in\cchi$ as above such that $|\vf(y)-\cchi(\vf)|<\ve$ for all
$y\in F$.  Thus $|\vf(l+\lambda)-\cchi(\vf)|<\ve$ for all $l\in N$
and $\lambda\in F(l)$. On the other hand, $\vf$ being uniformly
continuous, there is a neighborhood $V$ of $x$ such that
$|\vf(l+\lambda)-\vf(l+x)|<\ve$ for all $l\in N$ and $\lambda\in V$.
By what we said above, the preceding $F$ may be chosen such that
$\bigcup_{l\in N}F(l)\subset V$. Hence we get 
$|\vf(l+x)-\cchi(\vf)|<2\ve$ for all $l\in N$. Since $N\in\vk$ and
$\ve>0$ is arbitrary, this shows that
$\lim_{l\rarrow\vk}\vf(l+x)=\cchi(\vf)$.
\qed

\noindent{\bf Remark:}
It is easy to show that $\delta_L X$ is homeomorphic with
$(X\times\delta L)\amalg \{\infty_L\}$, thought as the one point
compactification of $X\times\delta L$, but we do not need this.

\medskip

In the next lemma we use the notation of Definition \ref{df:l8a}.
Let $\vk$ be a point in $\delta X$.

\begin{lemma}\label{lm:L8}
If $\vk\in\infty_L$ then $\cc_L(X)_\vk=\mbc$. If $\vk\nin\infty_L$,
then $\cc_L(X)_\vk=\cc_\infty(X)$.
\end{lemma}
\proof If $\vk\in\infty_L$ then
$\vk.\vf(x)=\vk(x.\vf)=\lim_{\gof_L}x.\vf=\lim_{\gof_L}\vf$
because $\gof_L$ is translation invariant. Thus $\vk.\vf\in\mbc$ in
this case. Now let $\cchi\nin\infty_L$. It suffices then to show
that $\cchi.\vf\in\cc_0(X)$ if $\lim_{\gof_L}\vf=0$ and by an easy
density argument we see that it suffices to assume that
$\supp\vf\subset L_K$ for a compact subset $K$ of $X$. If
$\vk,x$ are such that $\theta(x,\vk)=\cchi$ then
$$
\cchi.\vf(y)=\cchi(y.\vf)=\vk(x.(y.\vf))=\vk((x+y).\vf)=
\lim_{l\rarrow\vk}\vf(l+x+y).
$$
But if $z\nin K$ then there is $M\subset L$ co-finite such that
$l+z\nin L_K$ if $l\in M$, and then $\vf(l+z)=0$ for all such $l$,
and so $\lim_{l\rarrow\vk}\vf(l+z)=0$. Hence
$\supp\cchi.\vf\subset K-x$.

To finish the proof it remains to show that if $\cchi\nin\infty_L$
and $\xi\in\cc_\infty(X)$, then there is $\vf\in\cc_L$ such that
$\cchi.\vf=\xi$. It suffices to show this under the assumption
that $\xi$ has compact support. Then it suffices to take
$\vf=\xi_L=\sum_{l\in L}\tau_{-l}\xi$
\qed

\begin{lemma}\label{lm:nc}
If $\vf\in\cc_L(X)$ the map
$\delta L\ni\vk\mapsto\vk.\vf\in\cc_\infty(X)$ is norm continuous.
\end{lemma}
\proof By a density argument, it suffices to show this for
$\supp\vf\subset M_\Lambda$, where $M\subset L$ is a co-finite set
and $\Lambda\subset X$ is a compact set such that
$M_\Lambda=\amalg_{l\in M}(l+\Lambda)$.  If $l\in M$ let $\vf_l$ be
the function defined by $\vf_l(x)=\vf(l+x)$ for $x\in\Lambda$ and
$\vf_l(x)=0$ otherwise.  Then clearly $\vf_l\in\co(X)$,
$\supp\vf_l\subset\Lambda$, and the family $\{\vf_l\}_{l\in M}$ is
equicontinuous. Thus the set $\{\vf_l\mid l\in M\}$ is relatively
compact in $\co(X)$. From the universal property of $\gamma M$,
cf.\ page \pageref{up}, there is a unique continuous map
$\gamma M\ni\vk\mapsto\vf_\vk\in\co(X)$ such that $\vf_\vk=\vf_l$
for all $l\in M$. Since $\delta M=\delta L$, it suffices to show
that  $\vf_\vk=\vk.\vf$ if $\vk\in\delta M$. But we have
$$
\vk.\vf(x)=\lim_{l\rarrow\vk}\vf(l+x)=\lim_{l\rarrow\vk}\vf_l(x)=
\vf_\vk(x)
$$
because $\gamma M\ni\vk\mapsto\vf_\vk(x)$ is continuous. 
\qed

Putting all this together we obtain, for the algebra $\ra=\rc_L(X)$,
an improvement of Theorem \ref{th:gn}. If $T\in\rc_L(X)$ then,
according to that theorem, we have a continuous map 
$\sigma(\cc_L(X))\ni\vk\mapsto \vk.T\in\rc_s(X)$ which induces an
embedding
\begin{equation}\label{eq:LC1}
\rc_L(X)/\rk(X)\hookrightarrow\pprod_{\vk\in\delta_LX}\rc_L(X)_\vk.
\end{equation}
From Lemma \ref{lm:L8} we see that the localization 
$\rc_L(X)_\vk=\cc_L(X)_\vk\rtimes X$ at $\vk$ is
\begin{equation}\label{eq:LC2}
\rc_L(X)_\vk=\left\{
\begin{array}{ll}
\cc_0(X^*)\quad \mbox{if}\quad \vk=\infty_L,\\[3mm]
\rt(X)\quad \mbox{if}\quad  \vk\neq\infty_L.
\end{array}
\right.
\end{equation}
Here $\delta_LX$ is represented as in (\ref{eq:spa}). We identify
$\delta L\equiv\{0\}\times\delta L\subset X\times\delta L$ and we
simplify the relation (\ref{eq:LC1}) by taking into account the
discussion made on page \pageref{disc}. First, since
$(\vk\circ\tau_x).T=U_xTU_x^*$, it suffices to restrict the product
to the set $\delta L\cup\{\infty_L\}$. Second, we note that the
contribution of the point $\infty_L$ is already covered by the other
ones. Indeed, this follows from the easy to check
relation\,\symbolfootnote[2]{\
Note that this is related to Remark \ref{re:cm}: we have
$\tau_\vk\tau_\chi=\tau_\chi\tau_\vk=\infty_L$ on $\cc_L$. 
} 
$$
\infty_L(\vf)=\infty_L(\tau_\vk(\vf))\quad
\mbox{for all}\quad \vf\in\cc_L(X),
$$ which implies $\cj_\vk\subset\cj_{\infty_L}$, and from Lemma
\ref{lm:tiid}.  Finally, we get:

\begin{theorem}\label{th:LCi}
If $T\in\rc_L(X)$ and $\vk\in\delta L$ then 
$\lim_{l\rarrow\vk}U_lTU_l^*=\vk.T\equiv\tau_\vk(T)$ exists in
$\rc_s(X)$ and belongs to $\rt(X)$. The map $\delta
L\ni\vk\mapsto\vk.T\in\rt(X)$ is norm continuous. The maps
$\tau_\vk:\rc_L(X)\rarrow\rt(X)$ are surjective  morphisms and the
intersection of their kernels is $\rk(X)$, which gives us a
canonical embedding
\begin{equation}\label{eq:LCi1}
\rc_L(X)/\rk(X)\hookrightarrow\cc(\delta L;\rt(X)).
\end{equation}
If $H$ a normal operator in $\rc_L(X)$ or an observable affiliated
to $\rc_L(X)$, then
\begin{equation}\label{eq:LCi2}
\se(H)=\ccup_{\vk\in\delta L}\sigma(\vk.H).
\end{equation}
\end{theorem}
The last assertion follows from the norm continuity of the map
$\vk\mapsto\vk.T$.

\begin{remark}\label{re:his1}{\rm
Theorem \ref{th:LCi} has been obtained by rather different methods
in \cite{GI*,GI0}, see for example Theorems 5.5 and 5.6 in
\cite{GI0}. The point is that in these references the quotient
$\rc_L(X)/\rk(X)$ was computed directly and the notion of
localization at infinity did not play a r\^ole. Our purpose here was
only to show that Theorem \ref{th:intro} can be effectively used
even in some rather complicated situations.  Our arguments in this
subsection may, in fact, serve as a model for other computations.
}\end{remark}

\begin{remark}\label{re:his2}{\rm
    That the class of operators affiliated to $\rc_L(X)$ is quite
    large can be seen from the following result \cite[Theorem
    6.1]{GI0}.  Let $X=\mbr^n$ and denote $\ch^t$ the Sobolev space
    of order $t$ and $\|\cdot\|_t$ the norm in
    $B(\ch^{t},\ch^{-t})$.  Let $h:X \rarrow\mbr$ be a continuous
    function such that $c^{-1}(1+|k|)^{2s}\leq |h(k)|\leq
    c(1+|k|)^{2s}$ for some constant $c$ and all large $k$, and
    denote $H_0=h(P)$.  Let $0\leq t<s$ reals, choose a sparse set
    $L\subset X$ and let $\{W_l\}_{l\in L}$ be a family of symmetric
    operators $W_l:\ch^{t}\rarrow\ch^{-t}$ with the property
    $\sup_{l\in L}\|(1+|Q|)^\lambda W_l \|_t< \infty$ for some
    $\lambda>2n$.  Then the series $\sum_{l\in L}U_l^*W_l U_l\equiv
    W$ converges in the strong topology of $B(\ch^{t},\ch^{-t})$.
    Let $H=H_0 +W$, $H_l=H_0 +W_l$ be the self-adjoint operators in
    $L^2(X)$ defined as form sums.  Then $H$ is affiliated to
    $\rc_L(X)$. If $\vk\in\delta L$ then we also have
    $\vk.H=\lim_{l\rarrow\vk}H_l$ in norm resolvent sense.
  }\end{remark}

\begin{remark}\label{re:exo}{\rm
The preceding arguments can be simplified and everything becomes an
elementary exercise for the subalgebras of $\rc_L(X)$ corresponding
to a finite number of types of bumps \cite[p.\,548]{GI*}. The
case of just one type is already interesting. More precisely, let
$\cl$ be a finite partition of $L$ consisting of $n$ infinite sets
and let $\cc_\cl$ be the set of $\vf\in\cc_L$ such that $\lim_{a\ni
l\rarrow\infty}\vf(l+x)\equiv a.\vf(x)$ exists for each $x$ and for
each $a\in\cl$. Then $\delta L$ is replaced by the finite set $\cl$
and the Hamiltonians affiliated to $\rc_\cl$ have (modulo
translations) exactly $n+1$ localizations at infinity: a free one
$H_0\in'\co(X^*)$ and a two body one $a.H\in'\rt(X)$ for each
$a\in\cl$. And $\se(H)=\ccup_{a\in\cl}\sigma(a.H)$.
}\end{remark}

We make a final comment on the algebra $\ca$ defined in
(\ref{eq:cunk}). We saw that for any sparse set $L$ we have
$\cc_L(X)\subset\ca$.  On the other hand, if $M$ is a second sparse
set, then $L\cup M$ is not sparse in general. However, the
$C^*$-algebra $\cc_{L,M}$ generated by $\cc_L\cup\cc_M$ is still
included in $\ca$.  Note that to each Hamiltonian affiliated to
$\rc_L$ one may associate in a canonical way a free Hamiltonian, this
is the localization of $H$ at the point $\infty_L$. But this is not
the case for Hamiltonians affiliated to $\ra$.

\PAR \label{ss:nbody}
{\bf Grassmann algebras.}  We shall construct here $C^*$-algebras
canonically associated to finite dimensional vector spaces and which
allow one to consider a very general version of $N$-body
Hamiltonians. This algebras have first been pointed out in
\cite{DaG1} and the spectral theory of the operators affiliated to
them (essential spectrum and the Mourre estimate) has been studied
in detail in \cite{DaG2}. Our approach here is rather different, the
graded algebra structure so important in the quoted works does not
play a big r\^ole anymore.

If $Y$ is a closed subgroup of a locally compact abelian group $X$
then $X/Y$ is also a locally compact abelian group and we have a
continuous surjective group morphism $\pi_Y:X\rarrow X/Y$. Then the
map defined by $\vf\mapsto\vf\circ\pi_Y$ gives us a natural
embedding $\cc(X/Y)\hookrightarrow\cc(X)$. In fact \label{csg}
\begin{equation}\label{eq:yper}
\cc(X/Y)=\{\vf\in\cc(X)\mid y.\vf=\vf\quad\forall y\in Y\}.
\end{equation}
Note that we shall denote just $0$ the group $\{0\}$ and then
$\cc(0)=\co(0)=\mbc$, hence $\cc(X/X)=\co(X/X)=\mbc$.  On the other
hand, if $0\subset Y\subset Z\subset X$ are closed subgroups then
$X/Z\cong (X/Y)/(Y/Z)$ and we have natural maps
\begin{equation}\label{eq:gr}
X\rarrow X/Y\rarrow X/Z\rarrow 0
\end{equation}
hence we get embeddings
\begin{equation}\label{eq:gr1}
\mbc\subset \cc(X/Z)\subset \cc(X/Y)\subset \cc(X).
\end{equation}
In the rest of this subsection \emph{we shall consider only finite
dimensional real vector spaces}, although much of the theory can be
extended to more general groups. We shall consider the algebra
generated by the $C^*$-subalgebras
$\co(X/Y)\subset\cc(X/Y)\subset\cc(X)$.
We recall that the \emph{Grassmannian} $\mbg(X)$ is the set of all
vector subspaces of $X$ and the \emph{projective space} $\mbp(X)$ is
the set of all one dimensional subspaces of $X$.

\begin{definition}\label{df:gr}
The \emph{(classical) Grassmann algebra} of the vector space $X$ is
the $X$-subalgebra $\cg(X)\subset\cc(X)$ defined by
\begin{equation}\label{eq:cgr}
\cg(X)=\mbox{\rm norm closure of }\textstyle\sum_Y\co(X/Y)
\end{equation}
where $Y$ runs over $\mbg(X)$. The \emph{quantum Grassmann algebra}
of $X$ is the $C^*$-algebra $\rg(X)\subset\rb(X)$ defined by
\begin{equation}\label{eq:qgr}
\rg(X)=\cg(X)\rtimes X=
\mbox{\rm norm closure of }\textstyle\sum_Y\co(X/Y)\rtimes X.
\end{equation}
\end{definition}
\begin{remarks}\label{re:gr}{\rm
The fact that $\cg(X)$ is a $C^*$-algebra follows from the obvious
relation
\begin{equation}\label{eq:grad}
\co(X/Y)\cdot\co(X/Z)\subset\co(X/(Y\cap Z)).
\end{equation}
The second equality from (\ref{eq:qgr}) follows from Theorem
\ref{th:main}. 
}\end{remarks}

Let $G(X)$ be the set of finite unions of strict vector subspaces of
$X$:
$$
G(X)=\{L\subset X\mid \exists\,\mbf\subset\mbg(X)\setminus\{X\}
\mbox{ finite such that }  L=\ccup_{Y\in \mbf}Y\}.
$$
If $L$ is as above and $\Lambda\subset X$ is compact then
$L_\Lambda=L+\Lambda=\ccup_{Y\in \mbf}(Y+\Lambda)$.
Thus $L_\Lambda$ is a closed set, $L_\Lambda\neq X$, and we
have $L_\Lambda\cup M_\Lambda=(L\cup M)_\Lambda$ and
$L_{\Lambda'}\subset L_{\Lambda''}$ if
$\Lambda'\subset\Lambda''$. 
This justifies the next definition.

\begin{definition}\label{df:grf}
  The \emph{Grassmann filter} $\gog=\gog_X$ on $X$ is the filter
  generated by the family of open sets $L_\Lambda^c=X\setminus
  L_\Lambda$ where $L$ runs over $G(X)$ and $\Lambda$ over the set
  of compact subsets of $X$. If $Y$ is a subspace of $X$, then we
  denote also by $\gog_Y$ the filter on $X$ generated by the
  Grassmann filter of $Y$.
\end{definition}
Clearly, $\gog_X$ is translation invariant, finer than the Fr\'echet
filter, and $\gog_X^\circ=\gog_X$. If $X$ is one dimensional, then
$\gog_X$ is just the Fr\'echet filter.

\begin{remark}\label{re:flg}{\rm
    For $L\in G(X)$ we may consider the filter $\gof_L$ defined as
    in (\ref{eq:fl}).  Then $\gog_X$ is just the filter generated by
    $\bigcup_L\gof_L$. This can be expressed in other terms as
    follows: (1) $\gog_X$ is the upper bound of the set of filters
    $\gof_L$; (2) when seen as compact subsets of $\delta X$ (cf.\ 
    page \pageref{filclo}), $\gog_X$ is the intersection of the
    compact sets $\gof_L$.  }\end{remark}

\begin{remark}\label{re:euc}{\rm
If we equip $X$ with an Euclidean norm $|\cdot|$ and denote $\pi_Y$
the orthogonal projection onto $Y^\bot\cong X/Y$, then
$\delta_L(x)\equiv\mbox{dist}(x,L)=\min_{Y\in \mbf}|\pi_Yx|$ (with
$L,\mbf$ as before). Then the sets $L^c_r=\{ x\in X\mid
\delta_L(x)>r\}$, with $L\in G(X)$ and $r>0$ real, form a basis of
the filter $\gog_X$. Note that $L$ has empty interior and if $x$ is
outside it then $\delta_L(tx)=|t|\delta_L(x)\rarrow\infty$ as
$t\rarrow\infty$.  }\end{remark}

If $\gof$ is a filter on a set $S$ and $\pi$ is a map from $S$ to a
locally compact space $T$ then $\lim_{\gof}\pi=\infty$ means: for
each compact $K\subset T$ there is $F\in\gof$ such that
$\pi(F)\cap K=\emptyset$.

\begin{lemma}\label{lm:gf}
Let $Y,Z\in\mbg(X)$. If $Y\subset Z$ then $\lim_{\gog_Y}\pi_Z=0$.
If $Y\not\subset Z$ then $\lim_{\gog_Y}\pi_Z=\infty$.
\end{lemma}
\proof
Since $Y\in\gog_Y$ the above limits involve only the restriction of
$\pi_Z$ to $Y$. In the first case, if $y\in Y$ then $\pi_Z(y)=0$, so
the assertion is clear. If $Y\not\subset Z$ then $E=Y\cap Z$ is a
strict subspace of $Y$. Let $E'$ be a supplementary subspace for $E$
in $Y$. Then $\pi_Z:E'\rarrow X/Z$ is injective, hence if $K\subset
X/Z$ is compact then the set $\Lambda$ of $y\in E'$ such that
$\pi_Zy\in K$ is a compact in $E'$ and thus in $Y$. If
$y\in Y\setminus E_\Lambda\in\gog_Y$ then $y=e+e'$ with $e\in E$ and
$e'\in E'\setminus\Lambda$ so $\pi_Zy=\pi_Ze'\nin K$.
\qed

\begin{corollary}\label{co:gf}
  If $\vf\in\cg(X)$ and $Y\in\mbg(X)$ then $\lim_{\gog_Y}\vf$
  exists.  If $\vf=\sum_Z\vf^Z$ is a finite sum of
  $\vf^Z\in\co(X/Z)$ , then $\lim_{\gog_Y}\vf=\sum_{Z\supset
    Y}\vf^Z(0)$.
\end{corollary}

We see that each filter $\gog_Y$ defines a character of $\cg(X)$
and we could proceed as in the proof of Lemma \ref{lm:spa} and
describe $\delta_\mbg X\equiv\delta(\cg(X))$ in terms of couples
$(Y,y)$ with $y\in X/Y$. We shall not do it explicitly, but this is
hidden in what follows. We only note that the r\^ole of $\infty_L$ is
now played by $\gog_X$. 

\begin{proposition}\label{pr:gf}
For each $\vf\in\cg(X)$ the limit
\begin{equation}\label{eq:gf}
\tau_Y\vf=\lim_{y\rarrow\gog_Y}y.\vf 
\end{equation}
exists locally uniformly on $X$.  If $\vf$ is a finite sum
$\vf=\sum_Z\vf^Z$ with $\vf^Z\in\co(X/Z)$, then
$\tau_Y\vf=\sum_{Z\supset Y}\vf^Z$.  If $Y\in\mbp(X)$, $y\in
Y\setminus{0}$, then $\tau_Y\vf(x)=\lim_{t\rarrow\infty}\vf(x+ty)$.
\end{proposition}
\proof  We have to show that $\lim_{y\rarrow\gog_Y}\vf(x+y)$
exists locally uniformly in $x$. But this is an immediate
consequence of the Corollary \ref{co:gf} and Lemma \ref{lm:easy}.
\qed

According to the conventions we made at the beginning of this
subsection, we have $\cg(X/Y)\subset\cc(X/Y)\subset\cc(X)$ if $Y$
is a subspace of $X$.

\begin{proposition}\label{pr:cgr}
  We have $\cg(X/Y)\subset\cg(X)$. Moreover, there is a unique
  morphism $\tau_Y:\cg(X)\rarrow\cg(X/Y)$ such that $\tau_Y$ is a
  projection (in the sense of linear spaces). The map $\tau_Y$ is
  given by (\ref{eq:gf}). If\, $Y\subset Z\subset X$ then
\begin{equation}\label{eq:gre}
\mbc=\cg(0)\subset \cg(X/Z)\subset \cg(X/Y)\subset \cg(X)
\end{equation}
and $\tau_Y\tau_Z=\tau_Z\tau_Y=\tau_Z$. More generally, for any
$Y,Z\in\mbg(X)$ we have
\begin{equation}\label{eq:grs}
\cg(X/Y)\ccap\cg(X/Z)=\cg(X/(Y+Z))
\end{equation}
and $\tau_Y\tau_Z=\tau_Z\tau_Y=\tau_{Y+Z}$.
\end{proposition}
\proof
The algebra $\cg(X/Y)$ is generated the $\co((X/Y)/E)$ with
$E\subset X/Y$ subspace. If $Z=\pi_Y^{-1}(E)$ then $Y\subset
Z\subset X$, $E=Z/Y$ and  $(X/Y)/E\cong X/Z$ allows us to identify
$\co((X/Y)/E)=\co(X/Z)$ and thus to get the first assertion of the
proposition. Observe that
\begin{eqnarray}\label{eq:grt}
\cg(X/Y) &=& \mbox{\rm norm closure of }
\textstyle\sum_{Z\supset Y}\co(X/Z)\\
&=& \{\vf\in\cg(X)\mid y.\vf=\vf\quad\forall y\in Y\}.\nonumber
\end{eqnarray}
The other assertions of the proposition are easy to check.
\qed

\begin{proposition}\label{pr:dg}
If $\vf\in\cg(X)$ and $\tau_Y\vf=0$ for all $Y\in\mbp(X)$, then
$\vf\in\co(X)$.
\end{proposition}
\proof This follows from Theorem 3.2 and Lemma 4.1 of \cite{DaG1},
but we give a self-contained proof here.  Consider first a finite
set $\mbf\subset\mbg(X)$ which is stable under intersections and
such that $0\in\mbf$ and let $\ca=\sum_{Y\in\mbf}\co(X/Y)$. Then
$\ca$ is a $*$-algebra because of (\ref{eq:grad}) and
$\co(X)\subset\ca$. Clearly $\|\tau_Y\vf\|\leq\|\vf\|$ for all
$Y\in\mbf, \vf\in\ca$. Let us write $\vf=\sum_Y\vf^Y$ with
$\vf^Y\in\co(X/Y)$. From Proposition \ref{pr:gf} we get
$\tau_Y\vf=\sum_{Z\supset Y}\vf^Z$, so if $Y$ is a maximal element of
$\mbf$ then $\tau_Y\vf=\vf^Y$, hence $\|\vf^Y\|\leq\|\vf\|$. By
induction,we easily see that there is a constant  $c$ such that
$$
\|\vf^Y\|\leq c\|\vf\| \quad \mbox{for all} \quad 
Y\in\mbf, \vf\in\ca.
$$
This clearly implies that $\ca$ is a $C^*$-algebra and that
$\sum_{Y\in\mbf}\co(X/Y)$ is a topological direct sum. If $\vf$ is
as above and $\tau_Y\vf=0$ for all $Y\neq0$ then
$\sum_{Z\supset Y}\vf^Z=0$ if $Y\neq0$ hence, the sum being direct,
we get $\vf^Z=0$ for all $Z\neq0$, thus $\vf\in\co(X)$.

It follows that the map $\vf\mapsto(\tau_Y\vf)_{Y\neq0}$ is a
morphism from $\ca$ into $\prod_{Y\neq0}\cg(X/Y)$ with kernel equal
to $\co(X)$. In particular, the induced map
$\ca/\co(x)\rarrow\prod_{Y\neq0}\cg(X/Y)$ is an isometry, so that if
$\psi\in\ca$ is such that $\|\tau_Y\psi\|\leq\ve$ for all $Y\neq0$
then there is $\psi_0\in\co(X)$ such that $\|\psi-\psi_0\|\leq2\ve$
(just by definition of the quotient norm). 

Let now $\vf\in\cg(X)$ such that $\tau_Y\vf=0$ for all
$Y\in\mbp(X)$. From Proposition \ref{pr:cgr} it follows that this
property remains true for all $Y\in\mbg(X),Y\neq0$. From the
definition (\ref{eq:cgr}) it follows easily that for each $\ve>0$
there is $\ca$ as above and there is $\psi\in\ca$ such that
$\|\vf-\psi\|\leq\ve$. Then clearly we have $\|\tau_Y\psi\|\leq\ve$
for all $Y\neq0$, so by what we proved above there is  
$\psi_0\in\co(X)$ such that $\|\psi-\psi_0\|\leq2\ve$, and hence
$\|\vf-\psi_0\|\leq3\ve$. This clearly implies $\vf\in\co(X)$.
\qed

The next theorem is now an immediate consequence of Theorem
\ref{th:gn}, Propositions \ref{pr:cgr} and \ref{pr:dg}, and of Lemma
\ref{lm:kk}. We denote 
\begin{equation}\label{eq:qf}
\rg_Y(X)=\rg(X/Y)\rtimes X=\mbox{\rm norm closure of
}\textstyle\sum_{Z\supset Y}\co(X/Z)\rtimes X. 
\end{equation}
We mention that we have non canonical isomorphisms
$\rg_Y(X)\simeq\rg(X/Y)\otimes\cc_0(Y^*)$.

\begin{theorem}\label{th:dga}
If $T\in\rg(X)$ and $Y\in\mbg(X)$ then
$\tau_YT=\lim_{y\rarrow\gog_Y}U_xTU_x^*$ exists in $\rc_s(X)$ and
belongs to $\rg_Y(X)$. The map $\tau_Y:\rg(X)\rarrow\rg_Y(X)$ is a
morphism and a linear projection and is uniquely characterized by
these properties. We have $\tau_Y\tau_Z=\tau_Z\tau_Y=\tau_{Y+Z}$.
If $Y\in\mbp(X)$ and $y\in Y,y\neq0$ then
$\tau_YT=\lim_{t\rarrow\infty}U_{ty}TU^*_{ty}$. 
We have $T\in\rk(X)$ if and only if $\tau_YT=0$ for all
$Y\in\mbp(X)$, which gives us
\begin{equation}\label{eq:emg}
\rg(X)/\rk(X)\hookrightarrow\pprod_{Y\in\mbp(X)}\rg_Y(X).
\end{equation}
\end{theorem}

From (\ref{eq:emg}) we get that the the essential spectrum of an
observable $H$ affiliated to $\rg(X)$ is equal to the closure of the
union $\sigma(\tau_YH)$ with $Y\in\mbp(X)$. But now we can prove
more: as in the situations considered in Theorem \ref{th:LCi} and 
Proposition \ref{pr:voq}, the union is already closed (although it
is not finite, as in the usual $N$-body problem).

\begin{theorem}\label{th:ncl}
If $T\in\rg(X)$ then $\{\tau_YT\mid Y\in\mbp(X)\}$ is a compact set
in $\rg(X)$. In particular, if $H$ a normal operator in $\rg(X)$ or
an observable affiliated to $\rg(X)$, then
\begin{equation}\label{eq:ncl}
\se(H)=\ccup_{Y\in\mbp(X)}\sigma(\tau_YH).
\end{equation}
\end{theorem}
One should note that the map $Y\mapsto\tau_YT$ is not continuous:
if $T\in\co(X/Z)\rtimes X$ then $\tau_YT=T$ if $Y\subset Z$ and 
$\tau_YT=0$ if $Y\not\subset Z$.

Theorem \ref{th:ncl} is a corollary of Theorem 4.2 and Proposition
5.4 from \cite{DaG2}. We shall give below a slightly improved
proof. Note that only some general properties of the lattice
$\mbg(X)$ and of the graded algebra structure of $\rg(X)$ are really
needed. The next two lemmas imply the first assertion of Theorem
\ref{th:ncl} (hence the second). 

\begin{lemma}\label{lm:1}
If $T\in\rg(X)$ then for each $Z\in\mbg(X),Z\neq0$ there is
$Y\in\mbp(X)$ such that $\tau_ZT=\tau_YT$. 
\end{lemma}
\proof
Let $\mbe\subset\mbg(X)$ be countable. Then
$\{E\cap Z\mid E\in\mbe,E\cap Z\neq Z\}$ is a countable set of
strict subspaces of $Z$, so its union is not $Z$. Let $Y\in\mbp(Z)$
such that $Y\cap E\cap Z=0$ if $E$ is in the preceding set. Then
from $E\in\mbe$ and $E\supset Y$  we get $E\supset Z$. Now if
$T_\mbe$ is a finite sum $\sum_{E\in\mbe}T^E$ with
$T^E\in\co(X/E)\rtimes X$ then clearly $\tau_ZT=\tau_YT$. Finally,
if $T$ is arbitrary, then there is $\mbe$ as above such that $T$ be
a norm limit  of operators of the form $T_\mbe$, so we have
$\tau_ZT=\tau_YT$.
\qed

\begin{lemma}\label{lm:2}
Let $\{Y_n\}_{n\geq0}$ be a sequence of linear subspaces of $X$ and
let us define $Y=\ccap_{n\geq0}\sum_{m\geq n}Y_m$. If $k$ is the
dimension of $Y$, then there is $N$ such that for all $n\geq N$ and
all $T\in\rg(X)$:
$$
\|(\tau_{Y_n}-\tau_Y)T\|\leq k\sup_{m\geq n}
\|(\tau_{Y_n}-\tau_{Y_m})T\|.
$$
\end{lemma}
\proof
Since a decreasing sequence of subspaces is eventually constant,
there is $N$ such that $Y=\sum_{m\geq n}Y_m$ for all $n\geq N$. The
dimension of $Y$ being $k$, for each $n\geq N$ there are
$n<n_1<\dots<n_k$ such that $Y=Y_n+Y_{n_1}+\dots+Y_{n_k}$. From
Theorem \ref{th:dga} we get
$\tau_Y=\tau_{Y_n}\tau_{Y_{n_1}}\dots\tau_{Y_{n_k}}$. Let
$\cp=\tau_{Y_{n}}$, $\cp_i=\tau_{Y_{n_i}}$, and
$\cp_i'=1-\cp_i$. Then:
$$
\cp-\tau_Y=\cp[1-\cp_1\dots\cp_k]=
\sum_{i=1}^{k-1}\cp\cp_i'\cp_{i+1}\dots\cp_k +\cp\cp_k'.
$$
Since the morphisms $\cp_i$ commute, we get
$\|(\cp-\tau_Y)T\|\leq\sum_{i=1}^{k}\|\cp\cp_i'T\|$. Now it suffices
to note that $\cp\cp_i'=\cp(\cp-\cp_i)$.
\qed

\noindent{\bf Proof of Theorem \ref{th:ncl}:} If $\{\tau_{Y_n}T\}$
is a norm Cauchy sequence and $Y$ is as in the Lemma \ref{lm:2} then
$\|(\tau_{Y_n}-\tau_Y)T\|\rarrow0$. Observe that we do not have
$k=0$ because this would imply $Y_n=0$ for large $n$. Thus we can
use Lemma \ref{lm:1}  and find $E\in\mbp(X)$ such that
$\tau_YT=\tau_ET$, which proves the first assertion of the theorem.
\qed

\begin{remark}\label{re:hvz}{\rm
    The usual form of the HVZ theorem for $N$-body Hamiltonians
    follows easily from Theorem \ref{th:ncl}. Indeed, in the
    Agmon-Froese-Herbst formalism \cite{ABG} one is given a finite
    lattice $\cl$ and an injective map $\cl\ni a\mapsto
    X_a\in\mbg(X)$ such that $X_{a\wedge b}=X_a\cap X_b$,
    $X_{\max\cl}=X$ and $X_{\min\cl}=0$. The $N$-body Hamiltonians
    are observables $H$ affiliated to the $C^*$-algebra
    $\rc=\sum_{a\in\cl}\co(X^a)\rtimes X\subset\rg(X)$, where
    $X^a=X/X_a$. Let $\tau_a=\tau_{X_a}$, then $\tau_a$ is a
    morphism and a linear projection of $\rc$ onto the
    $C^*$-subalgebra $\rc_a=\sum_{b\geq a}\co(X^b)\rtimes X$. Let us
    set $H_a=\tau_aH$. Then (a generalized version of) the HVZ
    theorem says that
\begin{equation}\label{eq:hvz}
\se(H)=\ccup_{a\in\cm}\sigma(H_a),
\end{equation}
    where $\cm$ is the set of
    atoms of $\cl$. To get this from (\ref{eq:ncl}), note that for
    each $Y\in\mbp(X)$ there is a smallest $b$ in $\cl$ such that
    $Y\leq X_b$, so we have $Y\subset X_c$ if and only if $b\leq c$.
    Then for $T\in\rc$ we have $\tau_YT=\tau_bT$. On the other hand,
    there is an atom $a$ such that $a\leq b$, and then
    $\tau_bT=\tau_b\tau_aT$. Thus
    $\sigma(\tau_YT)\subset\sigma(\tau_aT)$. Reciprocally, if
    $Z\in\mbp(X)$ and $Z\subset X_a$ then $\tau_aT=\tau_a\tau_ZT$ and
    so $\sigma(\tau_aT)\subset\sigma(\tau_ZT)$.
}\end{remark}

\begin{example}\label{ex:nb1}{\rm
    The simplest application of Theorem \ref{th:ncl} is obtained by
    taking $X=\mbr^n$ and $H=\Delta+V(x)$ where $V\in\cg(X)$.
    Although simple, this situation is, however, not trivial because
    the union in (\ref{eq:ncl}) contains an infinite number of
    distinct terms in general. For example, the construction of $V$
    may involve an infinite number of subspaces $Y$ whose union is
    dense in $X$.
}\end{example}

\begin{example}\label{ex:nb2}{\rm
    We show here that in an $N$-body type situation (i.e.\ involving
    only a finite number of subspaces $Y$) the class of Hamiltonians
    for which (\ref{eq:hvz}) applies is very large. We use the
    setting of Remark \ref{re:hvz} and, to simplify notations, we
    equip $X$ with an Euclidean structure, so that $X$ is identified
    with $X^*$ and $X^a=X_a^\bot$. For real $s$ let $\ch^s$ be the
    usual Sobolev spaces, set $\ch=\ch^0=L^2(X)$, and embed as usual
    $\ch^s\subset\ch\subset\ch^{-s}$ if $s>0$.  Fix $s>0$ and denote
    $\|\cdot\|_s$ the norm in $B(\ch^s,\ch^{-s})$.  Let
    $h:X\rarrow\mbr$ be continuous and such that $c'(1+|k|)^{2s}\leq
    h(k)\leq c''(1+|k|)^{2s}$ outside a compact, for some constants
    $c',c''$.  Then $H(\max\cl):=h(P)$ is a self-adjoint operator
    with domain $\ch^{2s}$ and form domain $\ch^s$. Then for each
    $a\neq\max\cl$ let $H(a):\ch^s\rarrow\ch^{-s}$ be a symmetric
    continuous operator such that the following properties
    hold:\\
    (1) $U_xH(a)U_x^*=H(a)$ if $x\in X_a$,\\
    (2) $\|V_kH(a)V_k^*-H_a\|_s\rarrow0$ as $k\rarrow0$ in $X$,\\
    (3) $\|(V_k-1)H(a)\|_s\rarrow0$ as $k\rarrow0$ in $X^a$,\\
    (4) $H_a:=\sum_{b\geq a}H(b)\geq \mu h(P)-\nu$ as forms on
    $\ch^s$, for some  $\mu,\nu>0$.\\
    Then $H\equiv H(\min\cl)$ is affiliated to $\rc$ and
    $\tau_aH=H_a$, so  (\ref{eq:hvz}) holds. See \cite[Theorem
    4.6]{DaG2} for the details of the computation and for more
    general results.
  }\end{example}

\PAR \label{ss:exo}
{\bf On the operators $\vk.H$.}
We observed after Theorem \ref{th:introo} that if $H$ is a
self-adjoint operator affiliated to $\rc(X)$ then its localizations
at infinity $\vk.H$ are not necessarily densely defined. We shall
make in this subsection some comments on this question and we shall
give conditions which allow one to compute $\vk.H$ directly in terms
of $x.H$ and so to avoid considering the resolvent of $H$. This is
not possible for $H=h(P)+v(Q)$ if the operator $V=v(Q)$ is not
relatively bounded with respect to $h(P)$, so we shall consider here
only more elementary situations which are of some physical interest. 

To fix the ideas we consider here only the case $X=\mbr^n$ and take
${\ch}=L^2(X;\bfe)$, where $\bfe$ a finite dimensional Hilbert
space, cf.\ Section \ref{s:opaf}. For simplicity we consider only
operators whose form domain is a Sobolev space $\ch^s$ with $s>0$
(everything extends with no difficulty to hypoelliptic
operators). Set $\braket{k}=(1+|k|^2)^{1/2}$ and denote $S(\bfe)$
the space of symmetric operators on $\bfe$ and $|S|$ the
absolute value of $S\in S(\bfe)$.

Let $h:X\rarrow S(\bfe)$ be locally Lipschitz and such that
$c'\braket{k}^{2s}\leq |h(k)|\leq c''\braket{k}^{2s}$ and
$|h'(k)|\leq c\braket{k}^{2s}$ outside a compact, where $c,c',c''$
are constants.  We set $H_0=h(P)$ and observe that
$D(|H_0|^{1/2})=\ch^s$.

Let $v:X\rarrow S(\bfe)$ be a locally integrable function such that
the operator $V=v(Q)$ satisfies $V\ch^s\subset\ch^{-s}$ and $\pm
V\leq\mu |H_0|+\nu$ for some real numbers $\mu,\nu$ with
$\mu<1$. Then the self-adjoint operator $H=H_0+V$ (form sum) is
affiliated to $\rc(X)$, cf.\ Corollary \ref{co:ts}. Note that
$x.V=U_xVU_x^*=v(x+Q)$ satisfies the same estimates as $V$ and that
$x.H=H_0+x.V$. We mention that the next lemma is valid under the
much more general conditions of Definition \ref{df:std}.

\begin{lemma}\label{lm:pre}
  Let us assume that for each $\cc^\infty$ function with compact
  support $f$ the set $\{x.Vf\mid x\in X\}$ is relatively compact in
  $\ch^{-s}$. Then for each $\vk\in\delta X$ the limit
  $\lim_{x\rarrow\vk}x.V=\vk.V$ exists in the strong operator
  topology in $B(\ch^s,\ch^{-s})$, we have $\pm\vk.V\leq\mu
  |H_0|+\nu$ as forms on $\ch^s$, and we have $\vk.H=H_0+\vk.V$ if
  $\vk.H$ is defined as in Theorem \ref{th:introo}.
\end{lemma}
\proof Let $z\in\rho(H)$ and $R=(H-z)^{-1}\in\rc(X)$. Then $\vk.H$
is defined by the operator $\vk.R=\lim_{x\rarrow\vk}x.R$ where the
limit exists in $\rc_s(X)$. Note that we know that the limit exists
but we do not yet know whether $\vk.R$ is injective or not.  On the
other hand, the existence of $\vk.V$ follows from the fact that the
set of operators $x.V$ is bounded in $B(\ch^s,\ch^{-s})$: thus it
suffices to show the existence of the limit $\lim_{x\rarrow\vk}x.Vf$
in $\ch^{-s}$ for $f$ a $\cc^\infty$ function with compact support,
and this is obvious by the universal property of the Stone-\v{C}ech
compactification $\gamma X$ of the discrete space $X$ and our
assumption. Note that $\vk.V$ is the operator of multiplication by a
distribution which could not be a function, but clearly the estimate
verified by $V$ remains valid in the limit. Hence if
\emph{we define} $\vk.H=H_0+\vk.V$ as form sum, we get a densely
defined self-adjoint operator such that $\vk.H-z$ extends to an
isomorphism $\ch^s\rarrow\ch^{-s}$. Now it suffices to prove that
$\vk.R=(\vk.H-z)^{-1}$. Since $x.H-z:\ch^s\rarrow\ch^{-s}$ is also
an isomorphism, one can easily justify the equality
$$
(x.H-z)^{-1}-(\vk.H-z)^{-1}=(x.H-z)^{-1}(\vk.V-x.V)(\vk.H-z)^{-1}
$$
in $B(\ch^{-s},\ch^s)$. Then for $f\in\ch^{-s}$ we have
$$
\|[(x.H-z)^{-1}-(\vk.H-z)^{-1}]f \|_{\ch^{s}} \leq
C \|(\vk.V-x.V)(\vk.H-z)^{-1}f\|_{\ch^{-s}}
$$
where 
$$
C=\|(H-z)^{-1}\|_{B(\ch^{-s},\ch^s)}=
\|(x.H-z)^{-1}\|_{B(\ch^{-s},\ch^s)}.
$$
This clearly finishes the proof.
\qed

This lemma gives a rather concrete method of computing $\vk.H$ and
also shows that this operator is densely defined. The most elementary
way of checking the relative compacity assumption from the
lemma is described below.

\begin{proposition}\label{pr:suf}
  Assume that for each $\mu>0$ there is $\nu$ such that
  $|V|\leq\mu\braket{P}^{2s}+\nu$. Then
  $\lim_{x\rarrow\vk}x.V=\vk.V$ exists strongly in
  $B(\ch^s,\ch^{-s})$ for each $\vk\in\delta X$, for each $\mu>0$
  there is $\nu$ such that $\pm\vk.V\leq\mu |H_0|+\nu$ as forms on
  $\ch^s$, and $\vk.H=H_0+\vk.V$ if $\vk.H$ is defined as in Theorem
  \ref{th:introo}. In particular, we have
$$
\se(H)=\overline\ccup_{\vk\in\delta X}\sigma(\vk.H).
$$
\end{proposition}
\proof We only have to show that the set $\{x.Vf\mid x\in X\}$ is
relatively compact in $\ch^{-s}$ if $f\in\cc_c^\infty(X)$, i.e.\ if
$f$ is a $\cc^\infty$ function with compact support. This is
equivalent to the relative compactness in $\ch$ of the set
$\{\braket{P}^{-s}x.Vf\mid x\in X\}$.  Let
$\psi,\xi\in\cc_c^\infty(X)$ with $\xi(x)=1$ on $\supp f$ and let
$S=\braket{P}^{-s}\xi(Q)\braket{P}^{s}$ and
$T=\braket{P}^{-s}V\braket{P}^{-s}$. Then:
$$
\psi(P)\braket{P}^{-s}x.Vf=
\psi(P)\braket{P}^{-s}\xi(Q)x.Vf=
\psi(P)SU_xTU_x^*\braket{P}^{s}f\equiv \psi(P)Sf_x.
$$ The set $\{f_x\mid x\in X\}$ is bounded in $\ch$ and the operator
$\psi(P)S$ is compact in $\ch$, so the set
$\{\psi(P)\braket{P}^{-s}x.Vf\mid x\in X\}$ is relatively compact in
$\ch$. Thus it suffices to prove the following assertion: for each
$\ve>0$ there is $\psi\in\cc_c^\infty(X)$ such that
$\|\psi(P)^\bot\braket{P}^{-s}x.Vf\|\leq\ve$ for all $x\in X$, where
$\psi(P)^\bot=1-\psi(P)$. Let $V_\pm$be the positive and negative
parts of $V$, so that $V=V_+-V_-$ and $|V|=V_++V_-$, then it is
clearly sufficient to prove this assertion with $V$ replaced by
$V_\pm$. If $T_\pm=\braket{P}^{-s}V_\pm\braket{P}^{-s}$ then
$$
\|\psi(P)^\bot\braket{P}^{-s}x.V_\pm f\|=
\|\psi(P)^\bot U_xT_\pm U_x^*\braket{P}^{s}f\|\leq
\|\psi(P)^\bot T_\pm\|\|\braket{P}^{s}f\|
$$
and if we set $C_\pm=\|T_\pm\|^{1/2}$ then
$$
\|\psi(P)^\bot T_\pm\|\leq C_\pm\|\psi(P)^\bot T_\pm^{1/2}\|= 
C_\pm\|\psi(P)^\bot T_\pm\psi(P)^\bot\|^{1/2}.
$$
On the other hand, from $|V|\leq\mu\braket{P}^{2s}+\nu$ we get
$V_\pm\leq\mu\braket{P}^{2s}+\nu$ and then 
$T_\pm\leq\mu+\nu\braket{P}^{-2s}$ and so
$$
\psi(P)^\bot T_\pm\psi(P)^\bot \leq
[\mu+\nu\braket{P}^{-2s}](1-\psi(P))^2.
$$
Since $\mu$ can be chosen as small as we wish, it is clear that the
right hand side above can be made $\leq\ve$ for any $\ve>0$ by
choosing $\psi$ conveniently. Since the left hand side is positive
we then get $\|\psi(P)^\bot T_\pm\psi(P)^\bot\|\leq\ve$.
\qed

\begin{corollary}\label{co:ls}
If the conditions of Proposition \ref{pr:suf} are satisfied and if
for each $C^\infty$ function $f$ with support in the unit ball we
have $\lim_{x\rarrow\infty}\|x.Vf\|_{\ch^{-s}}=0$, then the
essential spectrum of $H$ is given by $\se(H)=\sigma(H_0)$. 
\end{corollary}

\begin{example}\label{ex:ls}{\rm
There are at least three physically interesting situations covered
by Proposition \ref{pr:suf}:\\ 
(1) The Schr\"odinger operator $H=P^2+V=\Delta+V(x)$. Then $s=1$ and
the assumptions of the proposition are satisfied if $V$ is of Kato
class, so we get Theorem 4.5 from \cite{LaS}. Corollary \ref{co:ls}
is similar to \cite[Theorem 4.3]{LaS}.\\ 
(2) The relativistic spin zero operator $H=(P^2+m^2)^{1/2}+V(x)$,
then  $s=1/2$. Here $m$ is any real number.\\
(3) The Dirac operator $H=D+V(x)$. Here $D$ is the free Dirac
operator of mass $m\geq0$, $s=1/2$, $\bfe=\mbc^N$ is
not trivial, and $V(x)$ is matrix valued. The last two situations
are also considered in \cite{Rab}.
}\end{example}

\PAR \label{ss:fin}
{\bf Cocompact subgroups.}  We consider now a situation similar to
that from \cite[Section 5]{LaS}. In a $C^*$-algebra setting such
examples and generalizations appear in \cite{Ma2}.

Throughout this subsection $X$ is an abelian locally compact non
compact group and $Y$ a closed subgroup such that $X/Y$ is compact.
Since $Y$ is fixed, we shall abbreviate $\pi_Y=\pi$.  We embed
$\cc(X/Y)\subset\cc(X)$ as explained on page \pageref{csg}, so we
think of $\cc(X/Y)$ as a translation invariant $C^*$-subalgebra of
$\cc(X)$ containing the constants, explicitly described by
(\ref{eq:yper}).  More generally, we identify any function $v$
defined on $X/Y$ with the function $v\circ\pi$ defined on $X$.

The full justification of the class of functions introduced below
will become clear later on, cf.\ Lemma \ref{lm:wad}.

\begin{definition}\label{df:wad}
If $\theta:X\rarrow X$ then write $\theta(a+x)\sim a+\theta(x)$ if
\begin{equation}\label{eq:wad}
\lim_{x\rarrow\infty}[\theta(a+x)-a-\theta(x)]=0
\quad\forall a\in X. 
\end{equation}
\end{definition}
If $\theta$ is uniformly continuous then this is equivalent to
having $\theta(x)=x+\xi(x)$ where $\xi:X\rarrow X$ is uniformly
continuous and slowly oscillating in a sense similar to
(\ref{eq:voc}).

The next proposition is completely elementary but we state it
separately because the main idea of the proof is very clear in this
context. 

\begin{proposition}\label{pr:wa1}
  Let $h:X^*\rarrow\mbr$ be a continuous function such that
  $|h(k)|\rarrow\infty$ as $k\rarrow\infty$. Assume that
  $\theta:X\rarrow X$ is uniformly continuous and $\theta(a+x)\sim
  a+\theta(x)$. If $v:X/Y\rarrow\mbr$ is continuous, $H=h(P)+v(Q)$
  and $H_\theta=h(P)+v\circ\theta(Q)$, then
  $\se(H_\theta)=\sigma(H)$.
\end{proposition}
\proof This will be a consequence of Theorem \ref{th:introo}. The
self-adjoint operator $H_\theta$ is affiliated to $\rc(X)$ because
of Proposition \ref{pr:sil}. It remains to compute the localizations
$\vk.H$ for $\vk\in\delta X$.  The image $\pi\circ\theta(\vk)$ is an
ultrafilter on the compact space $X/Y$, hence it converges to some
unique point $\what\vk\in X/Y$. Let $z\in X$  such that
$\what\vk=\pi(z)$. Since $v\equiv v\circ\pi$ we may define the
translated function $\what\vk.v(s)=v(\what\vk+s)=v\circ\pi(z+x)$ for
$s=\pi(x)\in X/Y$.  We shall prove that $\vk.H_\theta=\what\vk.H$
where $\what\vk.H=h(P)+\what\vk.v(Q)$. Note that
$\what\vk.H=U_zHU_z^*$, so $\sigma(\what\vk.H)=\sigma(H)$, which
finishes the proof.

Observe that $D(H)=D(h(P))$ is stable under translations, so it
suffices to prove the much stronger fact: $\Slim{x\rarrow\vk}
x.H_\theta f=\what\vk.Hf$ if $f\in D(H)$.  But this follows from
$\Slim{x\rarrow\vk}x.v\circ\theta(Q)=\what\vk.v(Q)$. This means that
for each $f\in L^2(X)$ we have
\begin{equation}\label{eq:bdd}
\lim_{x\rarrow\vk}\int_X
|v\circ\pi(\theta(x+y))-v\circ\pi(z+y)|^2|f(y)|^2\dd y=0
\end{equation}
where $z$ is as above.  Now for large $x$ we have
$\pi(\theta(x+y))\sim\pi(\theta(x))+\pi(y)$ and
$\lim_{x\rarrow\vk}\pi(\theta(x))=\what\vk=\pi(z)$ and since $v$ is
uniformly continuous we have
$$
\lim_{x\rarrow\vk}\sup_{y\in K}
|v\circ\pi(\theta(x+y))-v\circ\pi(z+y)|=0
$$
for each compact $K\subset X$. This clearly implies (\ref{eq:bdd}).
\qed

The extension of Proposition \ref{pr:wa1} to bounded measurable
functions $v$ seems to require further conditions. Indeed, one could
say that it suffices to use the dominated convergence theorem in
(\ref{eq:bdd}). But this requires some care because $\vk$ is a
filter, we did not assume $X$ separable, and the dominated
convergence theorem is not true if sequences are replaced by nets.
We indicate below two situations where these problems can be
avoided. 

\begin{proposition}\label{pr:wa2}
If $X/Y$ is separable or if $\theta$ is such that $\theta^{-1}(N)$ is
of measure zero whenever $N\subset X$ is of measure zero, then
Proposition \ref{pr:wa1} remains valid if $v$ is a bounded
measurable function.
\end{proposition}
\proof If $X/Y$ is separable then the point $\what\vk$ has a
countable fundamental system of neighborhoods $\{G_n\}$. For each
$n$ choose $F_n\in\vk$ such that $\pi(\theta(F_n))\subset G_n$ and
then choose points $x_n\in X$ such that $x_n\in F_n$. Clearly we
shall have $\pi(\theta(x_n))\rarrow\what\vk$ and 
$\Slim{n\rarrow\infty}x_n.H_\theta =\vk.H_\theta $
if the left hand side exists. Now the rest of the proof of
Proposition \ref{pr:wa1} works after replacing $x$ by $x_n$ and
$x\rarrow\vk$ by $n\rarrow\infty$, this time we can use the
dominated convergence theorem directly in (\ref{eq:bdd}). 

If $\theta$ has the property $|N|=0\Rightarrow|\theta^{-1}(N)|=0$,
we argue as follows. Since $v$ is bounded, it is sufficient to prove
(\ref{eq:bdd}) for $f$ the characteristic function of a compact
set. Then we approximate $v$ in $L^2(X/Y)$ by functions
$w\in\cc(X/Y)$, for such $w$ the relation (\ref{eq:bdd}) being
obvious. The only problem which appears is to estimate the term
$$
\int_K|v\circ\eta(x+y)-w\circ\eta(x+y)|^2\dd y
$$
where $K\subset X$ is a compact and $\eta=\pi\circ\theta$. The map
$\eta:X\rarrow X/Y$ is continuous and has the property 
$|N|=0\Rightarrow|\eta^{-1}(N)|=0$, by hypothesis and
\cite[Theorem 2.9]{F}. But this implies that there is an integrable
function $g\geq0$ on $X/Y$ such that the preceding integral be
$\leq\int_{X/Y}|v-w|^2 g\dd s$, and this can be made as small as we
wish. 
\qed

In the case $X=\mbr^n$ and under stronger assumptions on $\theta$ we
may extend Proposition \ref{pr:wa1} to unbounded functions $v$, in
particular we may recover Theorem 5.1 of the revised version of
\cite{LaS}. In order to be specific, \emph{we assume that $h$ is as
in Subsection \ref{ss:exo} and that $s\leq1$}. In particular our
assumptions below imply those of Proposition \ref{pr:suf}. Then we
easily obtain:

\begin{proposition}\label{pr:wa3}
Let $Y$ be the additive subgroup of $X=\mbr^n$ generated by $n$
linearly independent vectors. Let $\theta:X\rarrow X$ be a
homeomorphism such that $\theta$ and $\theta^{-1}$ are Lipschitz and
such that $\theta(a+x)\sim a+\theta(x)$. Assume that $v:X\rarrow
B(\bfe)$ is a locally integrable $Y$-periodic function and that
$V=v(Q)$ has the property: for each $\mu>0$ there is $\nu$ such that
$|V|\leq\mu\braket{P}^{2s}+\nu$. Then the operator
$V_\theta=v\circ\theta(Q)$ has the same property and if we set
$H=h(P)+V$ and $H_\theta=h(P)+V_\theta$ then
$\se(H_\theta)=\sigma(H)$.
\end{proposition}

\begin{example}\label{ex:wa3o}{\rm
We give an elementary example.
Let $X=\mbr,Y=\mbz$ and let $v$ be a real periodic locally
integrable function on $\mbr$. Then the form sum
$H=-\frac{d^2}{dx^2}+v(x)$ is a self-adjoint operator on $L^2(\mbr)$
and its spectrum is purely absolutely continuous. Let
$\theta:\mbr\to\mbr$ be of class $C^1$ with $\theta'(x)>0$ for all
$x$ and such that $\theta'(x)\to1$ as $|x|\to\infty$. Then the form
sum  $H_\theta=-\frac{d^2}{dx^2}+v(\theta(x))$ is a self-adjoint
operator and its essential spectrum is equal to the spectrum of
$H$. 
}\end{example}

We shall now consider the questions treated above in this subsection
from the point of view of Theorem \ref{th:imp}.  As in the proof of
Proposition \ref{pr:voq}, we get from (\ref{eq:yper}) and from
Theorem \ref{th:Land}:
$$
\cc(X/Y)\rtimes X=\{T\in\rc(X)\mid y.T=T\quad\forall y\in Y\}.
$$
Clearly $\co(X)\cap\cc(X/Y)=\{0\}$, from which it follows easily
that $\ca=\co(X)+\cc(X/Y)$ is an algebra of interactions and that we
are in the conditions of Proposition \ref{pr:ds}, hence we have a
topological direct sum
\begin{eqnarray}\label{eq:x/y}
\ra &\equiv& \ca\rtimes X=\rk(X)+\cc(X/Y)\rtimes X\\
&=& \{T\in\rc(X)\mid y.T-T\in\rk(X)\quad\forall y\in Y\}.\nonumber
\end{eqnarray}
This is a rather trivial algebra but things become less trivial
when we look at the image of $\ra$ under an automorphism of
$\rc(X)$.

If $\theta:X\rarrow X$ is a uniformly continuous homeomorphism then
$\theta^*:\cc(X)\rarrow\cc(X)$ is the injective morphism defined by
$\theta^*\vf=\vf\circ\theta$. Clearly $\theta^*\co(X)=\co(X)$ but in
nontrivial situations the image through $\theta^*$ of an $X$-algebra
is not an $X$-algebra. However, we are interested only in algebras
of interactions (which contain $\co(X)$), and the property of
$\theta$ isolated in Definition \ref{df:wad} will be sufficient.
The next lemma and its corollary are obvious.

\begin{lemma}\label{lm:wad}
  If $\theta:X\rarrow X$ is uniformly continuous and
  $\theta(a+x)\sim a+\theta(x)$, then for each $a\in X$ the map
  $\tau_a\theta^*-\theta^*\tau_a$ sends $\cc(X)$ into $\co(X)$.
\end{lemma}

\begin{corollary}\label{co:wad}
  Let $\theta:X\rarrow X$ be a uniformly continuous homeomorphism
  such that $\theta(a+x)\sim a+\theta(x)$. Then, if $\ca$ is an
  algebra of interactions, $\ca^\theta:=\theta^*\ca$ is also an
  algebra of interactions. Moreover, $\theta^*$ leaves $\co(X)$
  invariant and so it induces a canonical isomorphism of
  $X$-algebras $\ca/\co(X)\cong\ca^\theta/\co(X).$
\end{corollary}

We apply this to the situation (\ref{eq:x/y}).  Since $X/Y$ is
compact, we have
$$
\delta(\ca)=\sigma(\ca/\co(X))=\sigma(\cc(X/Y))=X/Y
$$
and thus we get $\delta(\ca^\theta)\cong X/Y$. Since $X$ acts
transitively on $X/Y$ we see that, modulo a unitary equivalence, we
have only one localization at infinity for an observable affiliated
to $\ra_\theta:=\ca^\theta\rtimes X$. This assertion can be made more
precise as follows. 

\begin{proposition}\label{pr:waa}
If $\theta:X\rarrow X$ is a uniformly continuous homeomorphism such
that $\theta(a+x)\sim a+\theta(x)$,  then there is a unique morphism
$\cp:\ra_\theta\rarrow\cc(X/Y)\rtimes X$ such that
\begin{equation}\label{eq:waa}
\cp(\vf\circ\theta(Q)\psi(P))=\left\{
\begin{array}{ll}
0\quad \mbox{if}\quad \vf\in\co(X),\\[3mm]
\vf(Q)\psi(P)\quad \mbox{if}\quad  \vf\in\cc(X/Y).
\end{array}
\right.
\end{equation}
This morphism is surjective and has $\rk(X)$ as kernel. If $\vk$ is
a filter on $X$ finer than the Fr\'echet filter and such that
$\lim_\vk\pi\circ\theta=0$, then
$\cp(T)=\lim_{x\rarrow\vk}U_xTU_x^*$, where the limit exists in 
$\rc_s(X)$.
\end{proposition}
\proof The uniqueness of $\cp$ follows from the fact that the
operators $\vf\circ\theta(Q)\psi(P)$ with $\vf\in\ca$ generate
$\ra_\theta$, and the surjectivity holds for a similar reason. A
filter $\vk$ as in the statement of the proposition exists because
$\pi\circ\theta:X\rarrow X/Y$ is surjective. If
$T=\vf\circ\theta(Q)\psi(P)$ then
$U_xTU_x^*=\vf\circ\theta(x+Q)\psi(P)$ and
$\vf\circ\theta(x+Q)\psi(P)\xi(Q)$ converges strongly to zero or to
$\vf(Q)\psi(P)\xi(Q)$ as $x\rarrow\vk$ if $\vf\in\co(X)$ or
$\vf\in\cc(X/Y)$ respectively. Here $\xi\in\co(X)$ and Remark
\ref{re:rk} is used. Thus we can define $\cp(T)$ by the last
assertion of the proposition. If $\cp(T)=0$ then 
the beginning of the proof of Proposition \ref{pr:wa1} shows that
$\tau_\chi T=0$ for all $\cchi\in\delta X$, so $T$ is compact.
\qed

\begin{remark}\label{re:wa}{\rm
Thus, if $H$ is an element of $\ra_\theta$ or an observable
affiliated to $\ra_\theta$, then
\begin{equation}\label{eq:waq}
\se(H)=\sigma(\cp(H)).
\end{equation}
One may get a large class of Hamiltonians $H$ affiliated to
$\ra_\theta$ by using Theorem 2.8 and Lemma 2.9 from\cite{DaG3}. For
example, let $H_0\geq0$ be self-adjoint operator 
strictly\symbolfootnote[2]{\
Strictly means $\|(1+\varepsilon H_0)^{-1}T-T\|\to0$ as
$\varepsilon\to0$ for all $T\in\ra_\theta$. For example, it suffices
that $H_0=h(P)$ where $h$ is a positive continuous function on $X^*$
which diverges at infinity.}  
affiliated to $\ra_\theta$. Let
$V$ be a quadratic form with $-\mu H_0-\nu\leq V\leq\nu(H_0+1)$ for
some $0<\mu<1$ and $\nu>0$ and such that 
$(H_0+1)^{-\alpha}V(H_0+1)^{-1/2}\in\ra_\theta$ for some
$\alpha>0$. Then the form sum $H=H_0+V$ is a self-adjoint operator
affiliated to $\ra_\theta$. For example, the last condition is
satisfied if $V\in\ra^\theta$ and then one gets singular functions
$V$ as limits of sequences $V_n$ such that 
$(H_0+1)^{-\alpha}V_n(H_0+1)^{-1/2}\in\ra_\theta$ is norm
convergent (this gives a class of $V$ larger than that from
Proposition \ref{pr:wa3}).  
}\end{remark}

\appendix
\section{Appendix}      \label{s:app}
\renewcommand{\theequation}{A.\arabic{equation}}
\protect\setcounter{equation}{0}
\PAR           \label{ss:land}
We give here a detailed proof of Theorem \ref{th:Land}.  We follow
rather closely Landstad's arguments, but we use the characterization
of $\ca\rtimes X$ taken as Definition \ref{df:crp}, which makes the
proof more transparent. We mention that the space $\rb^\circ_2(X)$
is suggested by Kato's theory of smooth operators, cf.\ \cite{RS}.
We shall not discuss the uniqueness of $\ca$ because the proof of
\cite[Lemma 3.1]{Lad} can hardly be simplified (if $X$ is discrete
we have $\ca=\ri(\ra)$ so uniqueness is trivial, see Remark
\ref{re:b2o}).

We begin with some heuristic comments which will make the rigorous
proof quite natural. The first question is, given $\ra$, how to
determine $\ca$. Observe that if we know $A\psi(P+k)$ for all $k$
then we can recuperate the operator $A$ by integrating over $k$,
because this operation will give $A\langle\psi\rangle$ with
$\langle\psi\rangle:=\int_{X^*}\psi(k)\dd k$. On the other hand,
$\psi(P+k)=V_k^*\psi(P)V_k$ so that if $A$ commutes with $V_k$ then
we get $A\langle\psi\rangle=\int_{X^*}V_k^*A\psi(P)V_k\dd k$. Thus
if $T=\sum_j\varphi_j (Q)\psi_j(P)$ then
$$
\sum_j\varphi_j (Q)\langle\psi_j\rangle=\int_{X^*}V_k^*TV_k\dd
k=:\ri(T) 
$$
If the group $X$ is discrete, so that $X^*$ is compact, this
formal argument can easily be made rigorous, the map $\ri$ is well
defined on all $\rb(X)$ and we have $\ca=\ri(\ra)$ (we strongly
advise the reader to first prove Landstad's theorem for discrete
$X$; this is a really pleasant exercise).  In general, one can give a
meaning to $\ri(T)$ for a sufficiently large class of $T$ for the
rest of the proof to work. Anyway, the preceding formula shows how
to extract the part in $\ca$ of the operator $T\in\ra$. The second
point is that one can reconstruct $T$ from such quantities by using
the formally obvious relation
$$
T=\int_X\ri(TU_x^*)U_x\dd x
$$
which is just the Fourier inversion formula, see (\ref{eq:ifou}).
But the right hand side here is, again formally, in $\ca\rtimes X$.

To make all this rigorous demands some preliminary constructions
that we expose in Subsection \ref{ss:land1} in a more general
context. We set $\ch=L^2(X)$ and we abbreviate $\rb=\rb(X)=B(\ch)$.
We recall that we have unitary representations $U_x$ and $V_k$ of
$X$ and $X^*$ in $\ch$ which satisfy the canonical commutation
relations
\begin{equation}\label{eq:ccr2}
U_xV_k = k(x)V_kU_x.
\end{equation}
Most of the next arguments do not depend on the explicit form of the
operators $U_x,V_k$.
\PAR           \label{ss:land1}
We first introduce the space of ``smooth operators'' with respect to
the unitary representation $V_k$:
\begin{equation}
{\rb}^\circ_2:=
\big\{ T\in {\rb} \mid \int_{X^*}\|TV_k f\|^2\dd k < \infty,\;
\forall f \in \ch\big\}.
\end{equation}
\begin{lemma} \label{l:3norm}
${\rb}^\circ_2$ is a left ideal (not closed in general) in ${\rb}$.
If $T \in {\rb}^\circ_2$ then
\begin{equation} \label{eq:3norm}
\nnorm{T} :=
\sup_{\|f\|=1}\left(\int_{X^*}\|TV_k f\|^2\dd k \right)^{1/2}
< \infty
\end{equation}
and $\big({\rb}^\circ_2, \nnorm{\cdot}\big)$ is a Banach space such
that $\nnorm{ST}\leq \|S\|\cdot\nnorm{T}$ for all $S \in {\rb}$.
Finally, if $x\in X$ and $T \in {\rb}^\circ_2$ then
$TU_x\in{\rb}^\circ_2$ and $\nnorm{TU_x}=\nnorm{T}$.
\end{lemma}
For the proof of (\ref{eq:3norm}) we have only to remark that the
map which sends $f\in\ch$ into $(TV_k f)_{k\in X^*} \in
L^2(X^*;\ch)$ is clearly closed and linear, hence it is
continuous. The last assertion of the lemma follows from
(\ref{eq:ccr2}).

The map $x \mapsto TU_x\in {\rb}_2$ is not norm continuous in
general. For this reason it will be convenient to consider the
following left ideal in $\rb$ and closed subspace of $\rb^\circ_2$
\begin{equation}\label{eq:goods}
\rb_2:=\{T\in\rb^\circ_2\mid \lim_{x\rarrow0}\nnorm{TU_x-T}=0\}.
\end{equation}
The following property of $\rb_2$ will be important in what follows:
if $S^*S\leq \sum T_j^*T_j$ for some $T_j\in\rb_2$, then
$S\in\rb_2$.

\begin{lemma}   \label{lm:appdoi}
  If $\psi\in L^\infty(X^*)$ then $\psi(P)\in{\rb}^\circ_2$ if and
  only if $\psi\in L^2(X^*)$. In this case we have
  $\psi(P)\in{\rb}_2$ and $\nnorm{\psi(P)} = \|\psi\|_{L^2}$.
\end{lemma}
\proof
We have
\begin{eqnarray*}
\int_{X^*}\!\!\!\|\psi(P)V_k f\|^2\dd k 
&=&
\int_{X^*}\|V_k^*\psi(P) V_k f\|^2\dd k =
\int_{X^*}\|\psi(P + k) f\|^2\dd k\\
&=&
\int_{X^*}\!\dd k
\int_{X^*}|\psi(p+k)|^2|\what{f}(p)|^2\dd p\\
&=&
\|\psi\|_{L^2(X^*)}\|\what{f}\|_{L^2(X^*)} =
\|\psi\|_{L^2(X^*)}\|f\|_{L^2(X)}.
\end{eqnarray*}
Then $\nnorm{\psi(P)U_x-\psi(P)}=\|x\psi-\psi\|_{L^2}$ where $x$ is
identified with the map $k\mapsto k(x)$ on $X^*$, we clearly
get $\psi\in{\rb}_2$.
\qed

\begin{definition}  \label{df:B-unu}
  ${\rb}_1$ is the linear subspace of ${\rb}$ generated by the
  operators of the form $S^*T$ with $S,T\in {\rb}_2$.
\end{definition}
The polarization identity
\begin{equation} \label{eq:polar}
4S^* T = \sum_{m=0}^3 i^m (i^m S + T)^* (i^m S + T)
\end{equation}
shows that ${\rb}_1$ is linearly generated by the operators of
the form $S^*S$ with $S\in {\rb}_2$.

We recall that a subset $\rc\subset\rb$ is called hereditary
if: $0\leq S\leq T\in \rc \Rightarrow  S\in \rc$. 

\begin{lemma}   \label{lm:ereditar}   
${\rb}_1$ is an hereditary $*$-subalgebra of ${\rb}$.
If $S\geq 0$ then $S\in\rb_1$ if and only if $\sqrt{S}\in\rb_2$.
If $S\in \rb_1$ then $U_xSU_y\in\rb_1$ for all $x,y\in X$.
\end{lemma}
\proof The fact that ${\rb}_1$ is a linear space and that $S^*\in
{\rb}_1$ if $S\in {\rb}_1$ is obvious. $\rb_1$ is stable under
multiplication because for $S,T\in\rb_2$ we have $S^*S T^*T=S^*\cdot
S T^*T\in\rb_1$ the space $\rb_2$ being a left ideal.

We prove now that if $S\geq 0$ and $S\in\rb_1$ then
$\sqrt{S}\in\rb_2$ (the reverse implication being obvious). Since
$S\in \rb_1$ we have $S=\sum_{j=1}^n \lambda_j S^*_jS_j$ with
$\lambda_j\in\CC$ and $S_j\in\rb_2$. If $S=S^*$ then by taking the
real parts we may assume that $\lambda_j\in \mbr$.  Then
$$
S=
\Big(\sum_{\lambda_j > 0 } + 
\sum_{\lambda_j < 0 }\Big)\lambda_j S^*_jS_j
\leq
\sum_{\lambda_j > 0 } \lambda_j S^*_jS_j,
$$
which implies $\sqrt{S}\in \rb_2$ by the property mentioned after
(\ref{eq:goods}).

Finally, if $0\leq S\leq T\in \rb_1$ then $\sqrt{T}\in\rb_2$,
so $S\in \rb_1$ by the same property.
\qed

Let $T\in\rb_1$ and let us write $T=\sum_{j=1}^n S^*_jT_j$ with
$S_j,T_j\in\rb_2$.  Then if $f,g\in \ch$:
\begin{eqnarray*}
\int_{X^*}\!|\langle V_kf,TV_kg\rangle|\dd k
&\leq&
\sum_{j=1}^n
\Big(\int_{X^*}\!\|S_jV_kf\|^2\dd k\Big)^{1/2}
\Big(\int_{X^*}\!\|T_jV_kg\|^2\dd k\Big)^{1/2}\\
&\leq&
\sum_{j=1}^n \nnorm{S_j}\nnorm{T_j}\|f\|\,\|g\|<\infty.
\end{eqnarray*}
From the operator version of the Riesz lemma it follows that there
is a unique operator $\ri (T) \in {\rb}(X)$ such that
$$
\langle f,\ri (T) g\rangle =
\int_{X^*}\!\langle V_kf,TV_kg\rangle\dd k
\;\mbox{ \rm for all } f,g\in \ch.
$$
In other terms, we see that the strongly continuous map
$k \mapsto V_k^*TV_k$ is such that the integral
\begin{equation}\label{eq:I}
\ri (T)=\int_{X^*}V_k^*TV_k \dd k
\end{equation}
exists in the weak operator topology of ${\rb}(X)$. It is clear that
for all $T\in\rb_2$ we have
\begin{equation}\label{eq:in}
\|\ri(T^*T)\|^{1/2}=\nnorm{T}.
\end{equation}
Moreover, the computation done above gives for $S,T \in \rb_2$:
\begin{equation}  \label{eq:Ipoln}
\|\ri(S^*T)\|\leq \nnorm{S}\,\nnorm{T}.
\end{equation}

\begin{example}\label{ex:lp}{\rm
If $S\in {\rb}(X)$ and $\xi, \eta \in  L^\infty(X^*)\cap L^2(X^*)$
then $\xi(P)S\eta(P) \in \rb_1$ and
\begin{equation}  \label{e:Inorm}
\|\,\ri\big(\xi(P)S\eta(P)\big)\|\leq
\|S\|\,\|\xi\|_{L^2(X^*)}\|\eta\|_{L^2(X^*)}.
\end{equation}
Indeed, we write $\xi(P)S\eta(P)
=\big(S^*\overline{\xi}(P)\big)^*\eta(P)$ and use (\ref{eq:Ipoln})
and Lemma \ref{lm:appdoi}.
}\end{example}

\begin{lemma} \label{lm:Ipol}
  If $T\in \rb_1$ then $(x,y)\mapsto\ri(U_xTU_y)$ is
  bounded and norm continuous.
\end{lemma}
\proof Due to (\ref{eq:Ipoln}) it suffices to assume that $T=S^*S$
for some $S\in\rb_2$ and to show continuity at $x=y=0$ of the map
$(x,y)\mapsto\ri(S_x^*S_y)$ with $S_x=SU_x$. Then
\begin{eqnarray*}
\|\ri(S_x^*S_y)- \ri(S^*S)\| &\leq& \|\ri\big((S_x-S)^*S_y\big)\|
+\|\ri\big(S^*(S_y-S)\big)\|\\ &\leq&
\nnorm{S_x-S}\nnorm{S}+\nnorm{S}\nnorm{S_y-S}
\end{eqnarray*}
because of the estimate (\ref{eq:Ipoln}).
\qed

\begin{proposition}\label{pr:Ip}
  If $T\in\rb_1$ then $\ri(T) \in \cc(X)$ and
  $U_x^*\ri(T)U_x=\ri(U_x^*TU_x)$ for all $x\in X$. The map
  $\ri:\rb_1\rarrow \cc(X)$ is linear and positive.
\end{proposition}
\proof We clearly have $V_k\ri(T)V_k^* = \ri(T)$ for all $k\in X^*$.
Since the Von Neumann algebra generated by $\{V_k\}_{k\in X^*}$ is
just $L^\infty(X)$, we get $\vphi(Q)\ri(T) = \ri(T)\vphi(Q)$ for all
$\vphi\in L^\infty(X)$. But $L^\infty(X)$ is maximal abelian in
$\rb$, thus $\ri(T)\in L^\infty(X)$.  From (\ref{eq:ccr2}) we get
$U_x^*V_k^*TV_kU_x = V_k^*U_x^*TU_xV_k$ hence $\ri(U_x^*TU_x) =
U_x^*\ri(T)U_x$. Since $\varphi\in\cc(X)$ if and only if $\varphi\in
L^\infty(X)$ and $x\mapsto U_x^*\varphi(Q)U_x$ is norm continuous,
we get $\ri(T)\in\cc(X)$. The last assertion of the proposition is
obvious.  
\qed

\begin{remark}\label{re:b2o}{\rm
    In a similar way we can associate an hereditary $*$-subalgebra
    $\rb^\circ_1$ to $\rb^\circ_2$ and define an extension of $\ri$
    to it, but then we only have $\ri:\rb^\circ_1\rarrow
    L^\infty(X)$. If $X$ is a discrete group, then $\rb_1=\rb$ and
    $\ri$ is a conditional expectation.  }\end{remark}

For $T\in\rb_1$ and for $x\in X$ we set
$\wtilde{T}(x):=\ri(TU_{x}^*)$, so that we associate to $T$ a
function $\wtilde{T}:X\rarrow \cc(X)$ which, by Lemma \ref{lm:Ipol},
is bounded and norm continuous. Let $\what T$ be the Fourier
transform of the function $k\mapsto V_k^*TV_k$, more precisely
$$
\what T(x)=\int_{X^*}\overline{k(x)} V_k^*TV_k \dd k
$$
where the integral exists in the weak operator topology. From
(\ref{eq:ccr2}) we get
\begin{equation}\label{eq:fou}
\wtilde T(x)=\what T(x)U_x^*.
\end{equation}
so that $\wtilde T$ is a kind of twisted Fourier transform.
Now the inversion formula for the Fourier transform gives us a
\emph{formal relation}
\begin{equation}\label{eq:ifou}
T=\int_X \wtilde T(x)U_x\dd x
\end{equation}
whose rigorous meaning is given below.

\begin{lemma} \label{lm:ifou}
For each $T\in\rb_1$ and $\theta \in L^1(X)$ we have
\begin{equation}  \label{eq:ifour}
\int_X \widetilde{T}(x)U_x\theta(x)\dd x =
\int_{X^*}V_k^*TV_k\what{\theta}(k) \dd k
\end{equation}
where both integrals exist in the weak operator topology.
\end{lemma}
\proof 
For each $f\in \ch$,
\begin{eqnarray*}
\int_X \langle f,\wtilde{T}(x)U_x f\rangle \theta(x)\dd x
&=&
\int_X \Big(
\int_{X^*} \langle V_k f,T\overline{k(x)}V_k f\rangle \dd k
\Big)\theta(x)\dd x\\
&=&
\int_X \Big(
\int_{X^*} \overline{k(x)}\langle V_k f,TV_k f\rangle \dd k
\Big)\theta(x)\dd x.
\end{eqnarray*}
Since $\theta \in L^1(X)$ and the function
$k\mapsto \langle V_k f,TV_k f\rangle$ is in
$L^1(X^*)$, we can apply the Fubini theorem and get
thus get (\ref{eq:ifour}).
\qed

Let us remark that the l.h.s.\ of the identity (\ref{eq:ifour})
always exists in the strong operator topology, and the same is true
for the r.h.s.\ if $\what{\theta}\in L^1(X^*)$.  We recall the
following result (see e.g.\ \cite[Lemma 4.19]{F}).
%
\begin{lemma}             \label{lm:harm}
  Let $\Lambda\subset X^*$ be a neighborhood of the neutral element
  in $X^*$ and let $\varepsilon>0$.  Then there is $\theta \in
  \Cc(X)$ such that $\what\theta \geq 0$, $\what\theta \in
  L^1(X^*)$, $\int_{X^*}\what\theta(k)\dd k = 1$, and $\int_{X^*
    \setminus \Lambda}\what\theta(k)\dd k\leq \varepsilon$.
\end{lemma}

The next version of the Fourier inversion formula is an easy
consequence of Lemmas \ref{lm:ifou} and \ref{lm:harm}:
\begin{proposition}    \label{pr:ifoun}
If $T\in\rb_1$ and $k\mapsto V_k^*TV_k$ is norm continuous, then
$T$ belongs to the norm closure of the set of operators of the form
$T_\theta=\!\int_X \widetilde{T}(x)U_x\theta(x) \dd x$ with
$\theta\! \in\Cc(X)$ and $\what\theta\! \in L^1(X^*)$.
\end{proposition}

\begin{lemma}    \label{lm:precis}
Let $\psi \in \co(X^*)$ and $T\in\rb_1$. Then if $\theta \in
L^1(X)$ and $\what\theta \in L^1(X^*)$ the integral $\int_X
\wtilde{T}(x)U_x \psi(P)\theta(x)\dd x$ exists in the norm operator
topology and $\ri(T)\psi(P)$ is a norm limit of such integrals.
\end{lemma}
\proof The map $x\mapsto U_x\psi(P)$ is norm continuous if
$\psi\in\co(X^*)$, hence the integrand is norm continuous.  The
last assertion follows by choosing $\theta$ as in
Lemma \ref{lm:harm} but with the r\^oles of $X$ and $X^*$ inverted.
\qed
%
\PAR               \label{ss:land2}
We are now ready to prove Landstad's theorem (Theorem
\ref{th:Land}). From now on, $\ra$ and $\ca$ are as in that theorem.

\begin{lemma}\label{lm:nd}
${\ra}$ is a non-degenerate $\co(X^*)$-bimodule. More precisely, if
  $A\in\ra$ and $\psi\in\co(X^*)$ then $A\psi(P)\in\ra$,
  $\psi(P)A\in\ra$ and $A$ is a limit of operators of the form
  $A\psi(P)$ and of operators of the form $\psi(P)A$.
\end{lemma}
\proof It is clearly sufficient to consider only the right action
 and, since each $\psi\in\co(X^*)$ is limit in the sup norm of
 functions whose Fourier transform is integrable, we may assume
 $\what\psi\in L^1(X)$. Then $A\psi(P)=\int_XAU_x\what\psi(x)\dd x$,
 we have $AU_x\in\ra$ and the integral converges in norm by the
 second assumption of Theorem \ref{th:Land}, so $A\psi(P)\in\ra$.
 By taking $\what\psi=|K|^{-1}\ind_K$, where $K$ runs over the set
 of compact neighborhoods of the origin in $X$, and by taking into
 account the norm continuity of the map $x\mapsto AU_x$, we
 see that $A$ is a norm limit of operators of the form $A\psi(P)$.
\qed 

\begin{lemma}\label{lm:ca}
$\ca$ is an $X$-subalgebra of $\cc(X)$.
\end{lemma}
\proof It is clear that $\ca$ is a norm closed subspace of $\cc(X)$
 stable under conjugation and stable under translations (note that
 $(\tau_x\varphi)(Q)=U_x\varphi(Q)U_x^*$).  To show that it is
 stable under multiplication, let $\varphi_1,\varphi_2 \in\ca$ and
 $\psi\in\co(X^*)$. Since $\varphi_2(Q)\psi(P)\in\ra$ we can write
 it as a norm limit of operators of the form $\xi(P)A$ with
 $\xi\in\co(X^*)$ and $A\in\ra$, so that
 $\varphi_1(Q)\varphi_2(Q)\psi(P)$ is a norm limit of operators of
 the form $\varphi_1(Q)\xi(P)A$ which belong to $\ra$.
\qed

Now we may consider the crossed product $\ca\rtimes X$, this is the
norm closed subspace of $\rb(X)$ generated by the operators
$\varphi(Q)\psi(P)$ with $\varphi\in\ca$ and $\psi\in\co(X^*)$.  We
clearly have $\ca \rtimes X\subset {\ra}$ and it remains to prove the
reverse inclusion.

\begin{lemma}\label{lm:ab}
If $T\in\ra\cap\rb_1$ then $\ri(T)\in\ca$
\end{lemma}
\proof Due to Proposition \ref{pr:Ip} it suffices to show that
$\ri(T)\psi(P)\in {\ra}$ if $\psi \in \co(X^*)$. Because of
Lemma \ref{lm:precis}, it is enough to prove that $\int_X
\widetilde{T}(x)U_x \psi(P)\theta(x)\dd x \in\ra$ if $\theta\in
L^1(X)$ and $\what\theta\in L^1(X^*)$.  But (\ref{eq:ifour}) implies:
\begin{eqnarray*}
\int_X \widetilde{T}(x)U_x \psi(P)\theta(x)\dd x =
\int_X V_k^*TV_k\psi(P)\what{\theta}(k)\dd k.
\end{eqnarray*}
Since $V_k^*TV_k\psi(P) \in {\ra}$ and is a norm continuous function 
of $k$  the last integral belongs to ${\ra}$.
\qed

\begin{lemma}\label{lm:fin}
If $T\in\ra\cap\rb_1$ then $T\psi(P)\in\ca\rtimes X$ for all
$\psi\in \co(X^*)$.
\end{lemma}
\proof
We shall have $TU^*_x\in\ra\cap\rb_1$, hence
$\widetilde{T}(x)=\ri(TU_x^*)\in\ca$, and thus the map $\widetilde
{T}:X\rarrow\ca$ is bounded and norm continuous.  On the other hand,
Proposition \ref{pr:ifoun} shows that for each $\psi \in \co(X^*)$
the operator $T\psi(P)$ is a norm limit of integrals $\int_X
\widetilde{T}(x) U_x\psi(P)\theta(x)\dd x$.  But $U_x\psi(P)\in
\co(X^*)$ and the map $x\mapsto U_x\psi(P)$ is norm continuous, thus
 the preceding integral converges in norm.  Also, we have
$\widetilde{T}(x) U_x\psi(P)\in\ca\rtimes X$ for each $x$, thus the
integral belongs to $\ca \rtimes X$.  \qed

Now we prove $\ra\subset\ca\rtimes X$. For this it suffices
to find a dense subset of $\ra$ which is included in $\ca\rtimes X$.
The Example \ref{ex:lp} and Lemma \ref{lm:nd} imply that
$\ra\cap\rb_1$ is a dense subspace of $\ra$. Thus \emph{it suffices
to show that $\ra\cap\rb_1\subset\ca\rtimes X$}.  But this follows
from Lemma \ref{lm:fin} because each $T\in\ra$ is a norm limit of
operators of the form $T\psi(P)$ with $\psi\in\co(X^*)$.

%
\addcontentsline{toc}{section}{
{References}
}
%

%
\end{document}